\documentclass[apj]{emulateapj}
\usepackage{verbatim}
\usepackage{graphics}
\usepackage{grffile}

\shorttitle{\psrpi: VLBI parallax distances for 57 pulsars}

\newcommand{\psrpi}{\ensuremath{\mathrm{PSR}\pi}}
\newcommand{\mspsrpi}{\ensuremath{\mathrm{MSPSR}\pi}}
\newcommand{\Gaia}{\textit{Gaia}}

\newcommand{\uas}{\ensuremath{\mu\mathrm{as}}}
\newcommand{\degrees}{\ensuremath{^\circ}}
\newcommand{\mjybm}{mJy~beam${}^{-1}$}

\begin{document}

\title{Microarcsecond VLBI pulsar astrometry with \psrpi\ II. parallax distances for 57 pulsars}

\author{
A. T. Deller\altaffilmark{1},
W. M. Goss\altaffilmark{2},
W. F. Brisken\altaffilmark{3},
S. Chatterjee\altaffilmark{4},
J. M. Cordes\altaffilmark{4},
G. H. Janssen\altaffilmark{5,6},
Y. Y. Kovalev\altaffilmark{7,8},
T. J. W. Lazio\altaffilmark{9},
L. Petrov\altaffilmark{10},
B. W. Stappers\altaffilmark{11},
A. Lyne\altaffilmark{11}
}
\altaffiltext{1}{Centre for Astrophysics and Supercomputing, Swinburne University of Technology, John St, Hawthorn, VIC 3122, Australia}
\altaffiltext{2}{National Radio Astronomy Observatory, Socorro, NM 87801, USA}
\altaffiltext{3}{Long Baseline Observatory, Socorro, NM 87801}
\altaffiltext{4}{Department of Astronomy and Cornell Center for Astrophysics and Planetary Science, Cornell University, Ithaca, NY 14853, USA}
\altaffiltext{5}{ASTRON, Netherlands Institute for Radio Astronomy, Oude Hoogeveensedijk 4, 7991 PD, Dwingeloo, The Netherlands}
\altaffiltext{6}{Department of Astrophysics/IMAPP, Radboud University, P.O. Box 9010, 6500 GL Nijmegen, The Netherlands}
\altaffiltext{7}{Astro Space Center of Lebedev Physical Institute, Profsoyuznaya 84/32, 117997 Moscow, Russia}
\altaffiltext{8}{Moscow Institute of Physics and Technology, Dolgoprudny, Institutsky per., 9, Moscow region, 141700, Russia}
\altaffiltext{9}{Jet Propulsion Laboratory, California Institute of Technology, Pasadena, CA 91109, USA}
\altaffiltext{10}{NASA Goddard Space Flight Center, 8800 Greenbelt Rd, Greenbelt, MD 20771, USA}
\altaffiltext{11}{University of Manchester, Jodrell Bank Centre for Astrophysics, Manchester M13 9PL, UK}

\begin{abstract}
We present the results of \psrpi, a large astrometric project targeting radio pulsars using the Very Long Baseline Array (VLBA).  From our astrometric database of 60 pulsars, we have obtained parallax-based distance measurements for all but 3, with a parallax precision that is typically $\sim$45 \uas\ and approaches 10 \uas\ in the best cases.  Our full sample doubles the number of radio pulsars with a reliable ($\gtrsim$5$\sigma$) model-independent distance constraint.  Importantly, many of the newly measured pulsars are well outside the solar neighborhood, and so \psrpi\ brings a near-tenfold increase in the number of pulsars with a reliable model-independent distance at $d>2$ kpc.  Our results show that both widely-used Galactic electron density distribution models contain significant shortcomings, particularly at high Galactic latitudes.  When comparing our results to pulsar timing, two of the four millisecond pulsars in our sample exhibit significant discrepancies in their proper motion estimates.  With additional VLBI observations that extend our sample and improve the absolute positional accuracy of our reference sources, we will be able to additionally compare pulsar absolute reference positions between VLBI and timing, which will provide a much more sensitive test of the correctness of the solar system ephemerides used for pulsar timing.  Finally, we use our large sample to estimate the typical accuracy attainable for differential VLBA astrometry of pulsars, showing that for sufficiently bright targets observed 8 times over 18 months, a parallax uncertainty of 4 \uas\ per arcminute of separation between the pulsar and calibrator can be expected.
\end{abstract}


\section{Introduction}
\label{sec:introduction}

With magnetic field strengths exceeding $10^{14}$~\hbox{G}, rotation rates approaching 1000~Hz, central densities exceeding $10^{14}$~g~cm${}^{3}$, and surface gravitational field potentials of order 40\% of that of a comparable mass black hole, neutron stars have proven to be powerful physical laboratories.  With their large moments of inertia, when detected as radio pulsars, their pulses provide a highly regular clock.
Studies of pulsars have placed strong constraints on the equation of state of neutron stars \citep{demorest10a}, provided the first detection of extrasolar planets \citep{wolszczan92a}, and provided the first observational evidence for the existence of gravitational waves \citep{taylor89a}.

In many cases, these results have been obtained despite considerable
uncertainty in the distance of the pulsar (or pulsars).  
It has not proven
possible to relate a pulsar's radio luminosity to any other intrinsic physical
quantity that would provide an independent distance estimate \citep[][]{szary14a}---however, it is instead
possible to make use of the pulsar's dispersion measure (DM) and a
model of the Galactic electron density distribution to provide this
distance estimate.  However, the difficulty of modeling all the
small-scale structure of the ionized component of the Milky Way means
that the fidelity of Galactic electron density distribution models is
generally rather low.  Accordingly, the reliability of DM-based
distance estimates for individual pulsars is generally quite low, and
errors of a factor of several are not rare
\citep{deller09b,chatterjee09a}.  While some pulsar science use cases are
relatively unaffected by such errors, there are others for which knowing the distance is vital and the distance uncertainty becomes the limiting factor in the measurement.  For instance, studies of the pulsar velocity distribution and hence supernova kicks can be biased by distance errors \citep[e.g.,][and references therein]{verbunt17a}, while studies of pulsar gamma-ray emission cannot build an accurate energy budget without a correct calibration of high-energy flux into luminosity \citep[e.g.,][]{abdo13a}.

Various methods exist to obtain non-DM--based estimates of pulsar
distances.  These include measurements of annual orbital parallax via
pulsar timing \citep[e.g.,][]{matthews16a}, visible wavelength observations \citep[e.g.,][]{2001ApJ...561..930C}, or Very Long
Baseline Interferometry \citep{chatterjee09a}, or via model-dependent approaches such
as \ion{H}{1} absorption limits
\citep[e.g.,][]{1969Natur.221..249G,2008ApJ...676.1189M}.  Of these,
VLBI astrometry is the most robust.  In addition to being dependent
upon a model for Galactic rotation, \ion{H}{1} absorption formally
provides only a lower limit; the spectra of pulsars are such that
few pulsars are detected at wavelengths shorter than radio and angular
resolutions are typically poorer than can be achieved with
\hbox{VLBI}; and pulsar timing parallaxes are generally only achieved
with millisecond pulsars.

The \psrpi\ campaign was conceived as a successor to previous
intensive VLBI campaigns \citep{brisken02a,chatterjee09a,deller09b}
that would treble the number of radio pulsars with a distance
measurement having a precision of better than 10\% and use the result
to constrain the characteristics of the radio pulsar population (e.g.,
velocity, luminosity) as well as improving models of the Galactic electron density distribution.  A subset of \psrpi\ results for two binary millisecond pulsars has been previously presented \citep{deller16a}, and in this paper we present the results for the full sample of~60 pulsars.  Section~\ref{sec:obsdataproc} describes the observations, data reduction, and position extraction, while Section~\ref{sec:results} describes the astrometric results and error analysis.  Section~\ref{sec:discussion} contains an analysis of both individual pulsars and parameters of the pulsar population, an evaluation of different Galactic electron density distribution models, a comparison of the VLBI results to pulsar timing, and a forward look to future observations for reference frame ties with radio pulsars.  Section~\ref{sec:conclusions} contains our conclusions.

\section{Observations and data processing}
\label{sec:obsdataproc}

\subsection{Calibrator search and sample selection}
\label{sec:searchobs}

\begin{figure}
\includegraphics[width=0.48\textwidth]{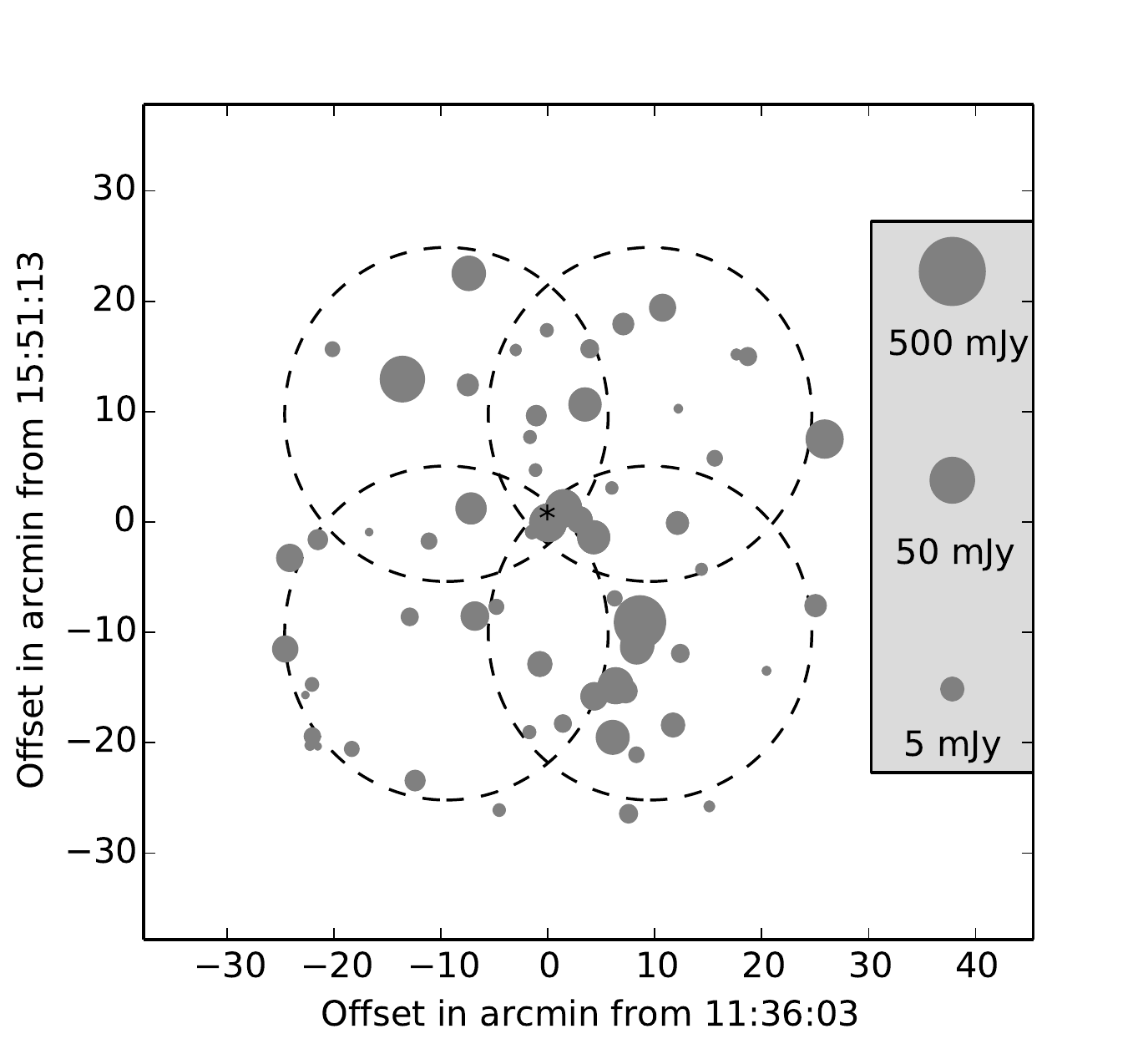}
% Figure generated on bunker with ~/svn_codebase/psrpi/plot_pulsar_searchpointings.py --pulsar=J1136+1551 --experiment=bd152a4 --plottype=eps
\caption{\label{fig:searchpointings}Layout of the calibrator search pointings for the target PSR J1136$+$1551.  Dashed lines show the 50\% response level of the primary beam, and candidate calibrator sources from the FIRST survey are shown in grayscale.}
\end{figure}

An initial list of target pulsars was produced consisting of sources located north of $-$20\degrees\ declination and with a ``gated equivalent" flux density\footnote{The gated equivalent flux density is the flux density of an unpulsed source that would provide an equivalent signal-to-noise to the pulsar when gating is applied in the correlator.  For a top-hat pulse shape and a perfectly placed pulsar gate, the gated equivalent flux density is given by the pulsar flux density divided by the square root of the duty cycle.} at 1400 MHz sufficient to obtain a detection exceeding 35$\sigma$ within a single \psrpi\ astrometric observation. This sample consisted of 225 pulsars with a gated equivalent flux density $>$3.2 mJy (bright enough for observations at the then--available data rate of 512 Mbps) and a further 55 sources with a gated equivalent flux density between 1.6 mJy and 3.2 mJy (bright enough for future observations at a data rate of 2 Gbps).
The first phase of \psrpi\ observations entailed the identification of suitable compact background sources close to the potential target pulsars on the sky, that could be used as secondary phase calibrators (``in-beam" calibrators).   The astrometric positions of the pulsars are ultimately measured relative to these sources.  A pilot program testing the observing strategy was undertaken between 2010 February and 2010 May (40 pulsars, 12 hours, VLBA project code BD148), which included some eventual \psrpi\ targets.  In-beam calibrator identification observations for the remaining potential \psrpi\ targets were undertaken in the main \psrpi\ observing program (240 pulsars, 85 hours, VLBA project code BD152) between 2010 November and 2011 December.  In all cases, all potentially useful candidates within $\sim$25\arcmin\ of the target pulsar were investigated using the multifield capability of the DiFX software correlator \citep{deller11a}.  The central observing frequency was 1660 MHz, and phase referencing was performed using a nearby calibrator to a grid of 4 pointing centers arrayed around the target pulsar, with right ascension and declination offsets of $\pm$10\arcmin\ in each direction.
Figure~\ref{fig:searchpointings} illustrates
an example pointing layout, for the target J1136$+$1551.

The candidate sources were taken from the Faint Images of the Radio Sky \citep[FIRST;][]{becker95a} catalog where available, and the NRAO VLA Sky Survey \citep[NVSS;][]{condon98a} in areas not covered by FIRST.  For future campaigns, the ongoing VLA Sky Survey (VLASS; Lacey et al. in prep.) will provide a deeper and higher resolution catalog covering the full NVSS footprint. Each pointing center was visited for approximately 3.5 minutes, which with an observing bandwidth of 64 MHz (dual polarization) gave a typical on-source root-mean-square (rms) noise of~0.3~\mjybm--0.8~\mjybm, depending on the location of the candidate in the pointing pattern.  In general, this was sufficient to identify (at $>$6.5$\sigma$) any candidates within $\sim$25\arcmin\ of the target pulsar brighter than 3 mJy (the completeness limit of NVSS) that could potentially serve as useful calibrators or astrometric check sources.  The calibration and source detection approach was essentially the same as that used by the mJIVE-20 project \citep{deller14a}, which was inspired by the procedure undertaken here.

These calibrator search observations for \psrpi\ served as a survey of over 200 square degrees at milliarcsecond resolution complete to $\sim$3 mJy, reliably detecting over 1500 sources.  Over 90\% of the 280 targeted pulsars were found to be located near at least one source suitable as a secondary phase calibrator for high-sensitivity observations (flux density $>$3 mJy contained within a component of maximum size several milliarcseconds, angular separation $<$25\arcmin).

From our initial sample of 280 pulsars, $\sim$110 met our requirements for astrometric observations with the then-available 512 Mbps recording system on the VLBA capable of recording dual polarization 64 MHz bandwidth.  These requirements were: pulsar gated equivalent flux density $>$3.2 mJy, and at least one compact secondary calibrator within 25\arcmin\ with flux density $>$6 mJy. We observed each of these 110 sources once using the astrometric scheme described in Section~\ref{sec:astrometricobs} below, before down-selecting to the final sample of 60 pulsars.  The initial observations were used to reduce the risk of selecting a target where the final astrometric precision would be insufficient to provide a useful distance constraint.  Some targets were rejected due to an unsuitable secondary calibrator, generally due to complicated source structure that was not identifed in the initial snapshot search observations due to limited $uv$ coverage.  Other targets were rejected because that pulsar's observed flux density was much fainter than the catalog value.  Finally, from the remaining viable targets, preference was given to sources that sampled a range of Galactic longitudes and latitudes (which meant primarily discarding sources in the Galactic plane and located towards to inner Galaxy), and sources that did not already have a high-precision parallax distance.   The final selection of 60 targets was therefore based on both logistical (range of right ascensions, strong calibrators close on the sky to the target, target flux density) and scientific (range of Galactic heights, predicted distances, individual objects where the distance was a high priority) considerations.  The 60 selected sources are summarized in Table~\ref{tab:targets}.

\begin{deluxetable*}{llrrcrr}
\tabletypesize{\scriptsize}
\tablecaption{\psrpi\ targets}
\tablehead{
\colhead{Pulsar} & \colhead{Pulsar} & \colhead{DM} & \colhead{S$_{1.4\mathrm{g}}$} & \colhead{Obs. Freq.} & \colhead{D$_{\rm NE2001}$} & \colhead{D$_{\rm YMW16}$} \\
 \colhead{Jname} & \colhead{Bname} & \colhead{(pc cm$^{-3}$)} & \colhead{(mJy)\tablenotemark{A}} & \colhead{(MHz)\tablenotemark{B}} & \colhead{(kpc)\tablenotemark{C}} & \colhead{(kpc)\tablenotemark{D}} }
\startdata
J0040+5716 &  B0037+56 &92.6 & 4.7 & 1660 & 3.05 & 2.42 \\
J0055+5117 &  B0052+51 &44.1 & 7.6 & 1660 & 1.90 & 1.94 \\
J0102+6537 &  B0059+65 &65.9 & 4.9 & 1660 & 2.29 & 1.98 \\
J0108+6608 &  B0105+65 &30.5 & 5.2 & 1660 & 1.42 & 1.46 \\
J0147+5922 &  B0144+59 &40.1 & 10.1 & 1660 & 2.22 & 1.58 \\
J0151--0635 & B0148--06 & 25.7 & 5.4 & 1660 & 1.22 & 25.00 \\
J0152--1637 & B0149--16 & 11.9 & 9.2 & 1660 & 0.51 & 0.92 \\
J0157+6212 &  B0154+61 &30.2 & 10.6 & 1660 & 1.71 & 1.39 \\
J0323+3944 &  B0320+39 &26.0 & 6.5 & 1660 & 1.01 & 1.20 \\
J0332+5434 &  B0329+54 &26.8 & 1244.9 & 1660 & 0.98 & 1.18 \\
J0335+4555 &  B0331+45 &47.2 & 4.6 & 1660 & 1.64 & 1.53 \\
J0357+5236 &  B0353+52 &103.7 & 6.1 & 1660 & 2.78 & 2.02 \\
J0406+6138 &  B0402+61 &65.3 & 13.5 & 1660 & 2.12 & 1.78 \\
J0601--0527 & B0559--05 & 80.5 & 11.8 & 1660 & 3.93 & 2.33 \\
J0614+2229 &  B0611+22 &96.9 & 12.5 & 1660 & 2.08 & 1.74 \\
J0629+2415 &  B0626+24 &84.2 & 17.9 & 1660 & 2.24 & 1.67 \\
J0729--1836 & B0727--18 & 61.3 & 8.0 & 1660 & 2.90 & 2.40 \\
J0823+0159 &  B0820+02 &23.7 & 8.3 & 1660 & 1.01 & 0.81 \\
J0826+2637 &  B0823+26 &19.5 & 76.4 & 1660 & 0.34 & 0.31 \\
J1022+1001 &   &10.3 & 9.5 & 1660 & 0.45 & 0.83 \\
J1136+1551 &  B1133+16 &4.9 & 181.9 & 1660 & 0.34 & 0.41 \\
J1257--1027 & B1254--10 & 29.6 & 7.3 & 1660 & 1.55 & 25.00 \\
J1321+8323 &  B1322+83 &13.3 & 4.3 & 1660 & 0.76 & 0.98 \\
J1532+2745 &  B1530+27 &14.7 & 4.8 & 1660 & 0.83 & 1.32 \\
J1543--0620 & B1540--06 & 18.4 & 15.2 & 1660 & 0.72 & 1.12 \\
J1607--0032 & B1604--00 & 10.7 & 26.2 & 1660 & 0.67 & 0.68 \\
J1623--0908 & B1620--09 & 68.2 & 4.7 & 1660 & 50.00 & 25.00 \\
J1645--0317 & B1642--03 & 35.7 & 167.4 & 1660 & 1.12 & 1.32 \\
J1650--1654 &  & 43.2 & 8.7 & 1660 & 1.47 & 1.05 \\
J1703--1846 & B1700--18 & 49.6 & 4.8 & 1660 & 1.48 & 1.69 \\
J1735--0724 & B1732--07 & 73.5 & 9.5 & 1660 & 2.26 & 0.21 \\
J1741--0840 & B1738--08 & 74.9 & 6.5 & 1660 & 2.17 & 0.22 \\
J1754+5201 &  B1753+52 &35.4 & 9.3 & 1660 & 2.18 & 4.17 \\
J1820--0427 & B1818--04 & 84.4 & 37.9 & 2267 & 1.94 & 2.92 \\
J1833--0338 & B1831--03 & 234.5 & 18.8 & 2267 & 5.14 & 5.17 \\
J1840+5640 &  B1839+56 &26.7 & 25.0 & 1660 & 1.68 & 2.19 \\
J1901--0906 &  & 72.7 & 20.4 & 1660 & 2.13 & 2.89 \\
J1912+2104 &  B1910+20 &88.3 & 7.4 & 1660 & 3.96 & 3.37 \\
J1917+1353 &  B1915+13 &94.5 & 10.4 & 2267 & 3.99 & 2.94 \\
J1913+1400 &  B1911+13 &145.1 & 7.3 & 2267 & 5.12 & 5.25 \\
J1919+0021 &  B1917+00 &90.3 & 5.7 & 1660 & 3.06 & 4.10 \\
J1937+2544 &  B1935+25 &53.2 & 7.7 & 1660 & 3.25 & 2.87 \\
J2006--0807 & B2003--08 & 32.4 & 8.8 & 1660 & 1.23 & 1.71 \\
J2010--1323 &  & 22.2 & 5.8 & 1660 & 1.02 & 1.16 \\
J2046--0421 & B2043--04 & 35.8 & 12.7 & 1660 & 1.75 & 3.27 \\
J2046+1540 &  B2044+15 &39.8 & 10.5 & 1660 & 2.42 & 3.34 \\
J2113+2754 &  B2110+27 &25.1 & 8.9 & 1660 & 2.03 & 1.87 \\
J2113+4644 &  B2111+46 &141.3 & 62.9 & 1660 & 4.53 & 4.12 \\
J2145--0750 &  & 9.0 & 21.2 & 1660 & 0.57 & 0.69 \\
J2149+6329 &  B2148+63 &128.0 & 11.0 & 1660 & 5.51 & 3.88 \\
J2150+5247 &  B2148+52 &148.9 & 8.6 & 1660 & 4.62 & 3.61 \\
J2212+2933 &  B2210+29 &74.5 & 3.8 & 1660 & 4.20 & 25.00 \\
J2225+6535 &  B2224+65 &36.1 & 10.0 & 1660 & 1.86 & 1.88 \\
J2248--0101 &  & 29.1 & 4.6 & 1660 & 1.65 & 25.00 \\
J2305+3100 &  B2303+30 &49.5 & 17.2 & 1660 & 3.66 & 25.00 \\
J2317+1439 &   &21.9 & 10.3 & 1660 & 0.83 & 2.16 \\
J2317+2149 &  B2315+21 &20.9 & 6.5 & 1660 & 0.95 & 1.80 \\
J2325+6316 &  B2323+63 &197.4 & 6.7 & 1660 & 8.26 & 4.86 \\
J2346--0609 &  & 22.5 & 8.2 & 1660 & 0.94 & 25.00 \\
J2354+6155 &  B2351+61 &94.7 & 31.6 & 1660 & 3.43 & 2.40 
\enddata
% Generated by ./makepsrpitable.py on bunker in /home/deller/svn_codebase/psrpi
% Note that mspsrpitable.py is on gygax
\tablenotetext{A}{Gated equivalent flux density; calculated from catalog 1.4 GHz flux density scaled by $\sqrt{\mathrm{duty\ cycle}}$.}
\tablenotetext{B}{Observing frequency used for the majority of astrometric observations (see Section~\ref{sec:obsdataproc}).}
\tablenotetext{C}{Distance estimated from the DM and the NE2001 Galactic electron density distribution \citep{cordes02a}.}
\tablenotetext{D}{Distance estimated from the DM and the YMW16 Galactic electron density distribution \citep{YMW17}.}
\label{tab:targets}
\end{deluxetable*}

\subsection{Astrometric observations}
\label{sec:astrometricobs}
Each of the 232 \psrpi\ astrometric epochs lasted $\sim$2.4 hours and targeted two pulsars located relatively close to each other on the sky, with typical angular separations of~10\degrees--20\degrees.  In each observation, 5 fields were observed: the two target fields (each of which encompassed both a pulsar and one or more in-beam calibrators), two out-of-beam phase reference sources (one near each pulsar), and one strong ``fringe finder" source used to calibrate the instrumental bandpass.  The observing sequence was as follows:
\begin{itemize}
\item Five 5.5 minute scans on the first target field interleaved and bracketed by 1.25 minute scans on the associated primary calibrator;
\item Five 5.5 minute scans on the second target field interleaved and bracketed by 1.25 minute scans on the associated primary calibrator;
\item A two-minute scan on the fringe finder;
\item Five 5.5 minute scans on the first target field interleaved and bracketed by 1.25 minute scans on the associated primary calibrator;
\item Five 5.5 minute scans on the second target field interleaved and bracketed by 1.25 minute scans on the associated primary calibrator;
\end{itemize}
This observing sequence was used to ensure that the $uv$ coverage was maximised while keeping the slewing overheads relatively low.

In some cases, a known and suitable VLBI calibrator was separated by less than 25\arcmin\ from the pulsar.  In these cases, no nodding calibration was performed, reducing the observing block for that pulsar to a single 27.5 minute long scan (repeated twice during the observation).

\begin{figure}
\includegraphics[width=0.48\textwidth]{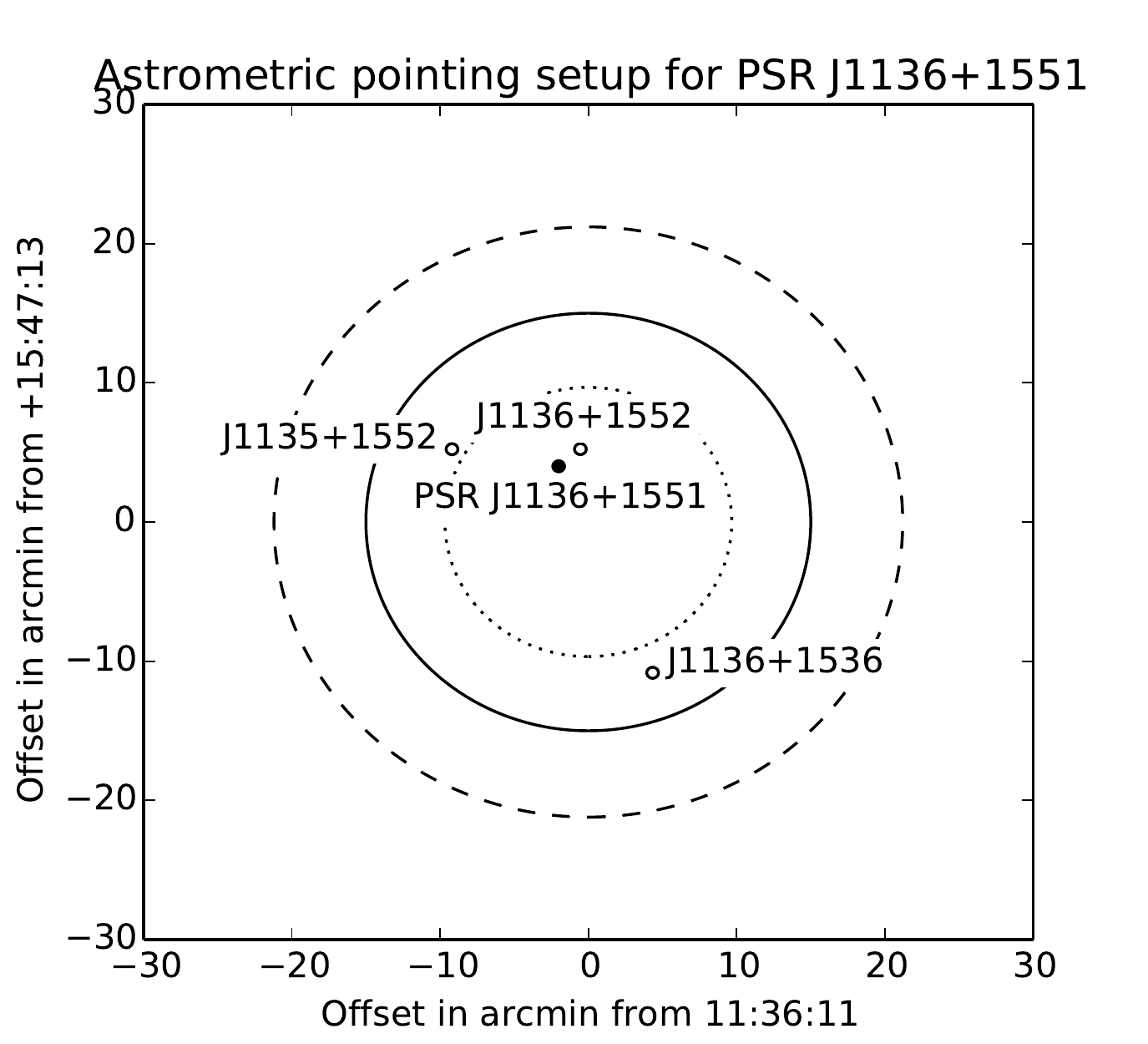}
\caption{\label{fig:astrometricpointing}The pointing layout of the target pulsar and in-beam calibrator sources for PSR J1136$+$1551.  The dotted, solid and dashed lines show the 75\%, 50\% and 25\% response contours of the primary beam at the center frequency of 1660 MHz.}
% Made on bunker in /home/deller/data/vlbi/psrpi/final/pointingplots/temp/, using make_oneoff_beamplot.py
\end{figure}

The pointing center for the target fields was typically chosen to be close to the midpoint between the pulsar and the primary in-beam calibrator, although adjustments were made based on the location of additional in-beam calibrator sources in some cases.  Figure~\ref{fig:astrometricpointing} shows an example astrometric pointing layout for the target PSR J1136$+$1551, while \url{https://safe.nrao.edu/vlba/psrpi/astrometric\_pointings.html} shows the pointing layout for all the target pulsars.

The astrometric observations were scheduled during the period 2011 January to 2013 December, and were optimized in time to provide maximum sensitivity to annual geometric parallax.  Since the VLBA is more extended in the east-west direction than north-south, the synthesized beam is narrower in right ascension than in declination.  Accordingly, we scheduled our astrometric observations around the time of the peak parallax signature in right ascension.  After the first ``check" observation, every successive parallax extremum was sampled with two observations within a $\sim$20 day period.  In total, each pulsar was observed 8 or 9 times spread over a $\sim$2 year period.

The observational setup consisted of four 16 MHz subbands, covering both circular polarizations and sampled at 2 bit precision for a total recording rate of 512 Mbps.  In the first astrometric observation of each pulsar, the frequency range chosen was 1624.49~MHz--1688.49~MHz, while for the remaining observations the frequency range was shifted slightly to~1627.49~MHz--1691.49~MHz, due to strong interference from the Iridium satellite constellation at $\sim$1625 MHz.  For four pulsars (PSRs~J1820$-$0427, J1833$-$0338, J1913$+$1400, and J1917$+$1353), strong scatter-broadening led us to choose observations at higher frequency, 2234.49~MHz--2298.49~MHz.  In the first observation of each pulsar, a significant portion of the data from the first subband was flagged due to this interference, while in the later observations, a lesser amount of data was flagged in the 4th subband due to interference at $\sim$1690 MHz.

Correlation was performed in Socorro using the DiFX software correlator \citep{deller11a}.  For each observation, a minimum of three and a maximum of six correlation passes were made, forming 3--6 separate visibility datasets.  The first pass correlated all sources, using the position of the primary in-beam calibrator for scans on the target field.  We refer to the resultant dataset henceforth as the ``calibrator" dataset.  All other passes correlated only the scans on the target fields, and differed in the location of the phase center and the presence or absence of special pulsar processing. The second pass used the position of the target pulsars and employed pulsar gating to boost the signal-to-noise (S/N) by down-weighting timeranges when the pulsar signal is weak or absent \citep{deller07a}.  The pulsar ephemeris and the pulsar gate parameters were obtained from timing observations with the Lovell telescope at Jodrell Bank Observatory.  We refer to the output as the ``gated" dataset.  The third pass also used the position of the target pulsars, but used no pulsar gating, to generate the ``ungated" dataset.  Comparison of the S/N of the pulsar detection in the gated and ungated datasets allowed us to check that the correct pulsar ephemeris and gate had been applied.  If present, the fourth, fifth, and sixth correlator passes used the position of the additional in-beam calibrator sources; we refer subsequently to these as the ``additional" datasets.

\subsection{Astrometric data calibration}
\label{sec:calibration}

In order to generate artifact-free images of the target pulsars that are located in a stable reference frame, careful calibration is needed to remove time- and frequency-dependent corruption of the measured visibilities by instrumental and propagation effects.  All calibration was performed using AIPS \citep{greisen03a}, facilitated by the ParselTongue python interface \citep{kettenis06a}.  AIPS version 31DEC15 was used for the final data processing, and each pulsar was processed independently.  Calibration was script-based, with configurable options set using a markup-language control file for traceability.  The cumulative calibration derived in the preceding stages is always applied before solving for the next stage of the incremental calibration.  We now describe the calibration script stages in detail.

\begin{enumerate}
\item {\bf Load:} The visibility datasets were loaded into AIPS using the task FITLD.
\item {\bf {\it A priori} flagging:} Logged time ranges when the antennas are slewing, settling, or have otherwise known pointing or recording problems are already recorded in the flag table accompanying the dataset.  In addition to these existing flags, we also flagged baselines during times when the natural fringe rate was low using the AIPS task UVFLG, as these are susceptible to corruption by radio frequency interference (RFI) and instrumental effects.  We also flagged baselines when one or both antennas was pointing below 20\degrees\ elevation (AIPS task UVFLG).  Finally, we applied any user-defined flags that were generated after inspection of final data products of an earlier pipeline run (AIPS task UVFLG).
\item {\bf Source shifting:} If required, the phase center for one or more sources was shifted using the AIPS task CLCOR.  This was typically only needed for the first observation, where the pulsar position was sometimes poorly known, and in some cases the in-beam calibrator position was also only poorly constrained after the initial snap-shot observation.
\item {\bf {\it A priori} ionosphere correction:} The delay model applied at the correlator does not include any ionsospheric contribution.  To correct for ionospheric propogation delays, we used the AIPS task TECOR, which makes use of a low-resolution global ionosphere model.  While these global models are unable to remove rapid and/or small-scale variations, they do account for bulk ionospheric effects.  We used the the final combined analysis models of the International GNSS Service (analysis center code {\tt igsg}) available from \url{ftp://cddis.gsfc.nasa.gov/gps/products/ionex/}.
\item {\bf EOP corrections:} The Earth Orientation Parameters (EOPs) used in the correlator model are often refined after the time of correlation.  To update the visibilities and make use of the most accurate available EOPs in order to minimise residual position offsets after phase referencing, we used the AIPS task CLCOR.
\item {\bf {\it A priori} amplitude calibration:} The visibilities in the correlator dataset are scaled to take the form of pseudo-correlation coefficients.  To convert these to an approximate flux density scale in Janskys, the following steps are taken:
  \begin{enumerate}
  \item {\bf Quantization correction:} Imperfect level-setting in the quantizers bias the amplitude scale; this effect can be detected and corrected by analysis of the station autocorrelations.  We used the AIPS task ACCOR to make these corrections.
  \item {\bf System temperature correction:} A continuously operating switched noise diode at each VLBA antenna records the system temperature at that antenna.  In combination with an {\it a priori} gain curve (which, at low frequencies such as used here, is quite accurate), this can be used to convert the pseudo-correlation coefficients produced by the correlator into Janskys.  We use the AIPS task APCAL to generate these corrections.
  \item {\bf Primary beam correction:} In the target pointing, the pulsar and in-beam calibrator(s) are not centered in the primary beam.  Accordingly, the amplitude response for each of these sources is attenuated by the primary beam fall-off.  We apply a correction based on a simplified model of a uniformly illuminated antenna scaled by the measured parameters for  beam width and beam squint of VLBA antennas at our observing band, using a custom ParselTongue script described in \citet{deller14a}.
  \end{enumerate}
\item{\bf Instrumental phase calibration:} Any instrumental phase variations due to changing propagation through the signal chain are tracked by an injected pulse train, and the measured phases are stored in a table that accompanies the visibilities.  We applied these corrections using the AIPS task PCCOR.
\item{\bf Time-independent delay calibration:} Using the AIPS task FRING, we measured the single-band delays for each subband independently on the fringe finder source.  A model of the fringe finder source derived from imaging the concatenated data for the source from all 8--9 \psrpi\ epochs was supplied to FRING.  The AIPS task SNSMO was used to apply a median window filter to the resultant delays and automatically exclude any solutions that differed by more than 10 nanoseconds from the median for that subband of that antenna.  Rates were zeroed before the delays were applied using CLCAL.
\item{\bf Time-independent instrumental bandpass calibration:}  The AIPS task BPASS was used to derive the instrumental bandpass, using the fringe finder scan.  As with the preceding step, the model of the fringe finder source was supplied to BPASS.  The resultant amplitude corrections were normalized to leave the flux density scale unaffected.
\item{\bf Time-dependent delay calibration:} We used the AIPS task FRING to now derive single-band delays using the phase reference calibrator source.  Again, a source model (based on imaging of concatenated \psrpi\ datasets) was supplied in all cases.  In almost all cases, each subband was solved separately, but for several weak phase reference sources, we combined all subbands together to improve the S/N.  Solving for all subbands separately is preferred, because the residual ionospheric delays can lead to a (slightly) varying delay between subbands.  When this is the case, approximating the delay as constant over all subbands reduces the S/N improvement resulting from the larger bandwidth, and also leaves per-subband phase residuals that must be corrected later.  The solutions were median-window filtered using SNSMO to excise values where the delay or phase offsets exceeded 10~ns or 10~mHz, respectively.
\item{\bf Refinement of time-dependent amplitude calibration:} The AIPS task CALIB was used to compute amplitude self-calibration corrections on a per-subband basis for the phase reference calibrator source, usually with a timescale of 20 minutes (shorter in some cases where the calibrator source was particularly strong).  As in previous steps, the epoch-averaged calibrator model was employed, which in effect forces the absolute flux density scale to match this model.  If the calibrator exhibited substantial flux density variations over the observing period, this could affect the flux density scale of individual observations; however, the absolute flux density scale is not important for the astrometric observables, which depend only on position.  SNSMO was used to median-window filter and remove amplitude corrections differing by more than 20\% from the median over the whole observation.
\item{\bf Refinement of time-dependent phase calibration in target direction:}
  The following steps now made use of the strongest source(s) in the target pointing, with the results then being applied to all sources in the target pointing.  In most cases, a single in-beam calibrator source was used to derive solutions.  For some targets where no strong in-beam calibrator source was available, multiple in-beam sources were used together in a ``multi-source selfcal" \citep{middelberg13a,radcliffe16a}.  Finally, in several cases (PSR J0332+5434, PSR J1136+1551, and PSR J2113+4644), the pulsar itself was by far the strongest source in the field, and the gated pulsar dataset was used to derive these calibrations rather than one of the other in-beam sources.
  \begin{enumerate}
  \item {\bf Frequency independent:} The dataset(s) corresponding to the designated source(s) were split using the AIPS task SPLIT, to apply calibration and flagging and average the data in frequency for speed.  For non-pulsar sources, the split dataset was then divided by the corresponding source model using the AIPS task UVSUB.  If multiple in-beam sources were used in a ``multi-source selfcal,'' these normalized datasets were then combined with the AIPS task \hbox{DBCON}.  CALIB was then used to solve for phase corrections on this normalized (and possibly concatenated) dataset, summing all subbands and polarizations.  The solution interval ranged from 10 seconds for the brightest sources to 5 minutes for the weakest source, with a median value of 1.25 minutes.
  \item {\bf Frequency dependent:} The preceding step was repeated, but this time treating frequency subbands separately while still summing polarizations.  The reduced bandwidth per solution was compensated with a longer solution interval, typically 6~minutes--30~minutes.  This step compensates for the small residual dispersive delays due to the ionosphere between the phase reference calibrator direction and the target direction.
  \end{enumerate}
\item{\bf Refinement of time-dependent amplitude calibration in target direction:} For a handful of targets with sufficiently bright in-beam calibrators, we used CALIB to derive further self-calibration amplitude corrections using the in-beam calibrator data.  SNSMO was applied to filter out solutions differing by more than 20\% from the median.
\item{\bf Correction of pulsar scintillation:} For some nearby pulsars, the scintles produced by diffractive scintillation in the interstellar medium are sufficiently large compared to our visibility resolution (diffractive timescale $\gg$ 1 second, diffractive bandwidth $\gg$ 1 MHz) that significant amplitude variations are apparent in the pulsar data.  When significant diffractive scintillation was present, we derived time-dependent amplitude correction factors using a custom ParselTongue routine described in \citet{deller09b}.  The solution interval was empirically determined by inspection of the uncorrected pulsar data.
\item{\bf Writing calibrated data:} The fully calibrated datasets for the pulsar (both gated and ungated), the in-beam calibrator source(s), and the phase reference calibrator source were split and averaged to a single visibility point per frequency subband (excising the two edge channels).  The non-pulsar sources were divided by the epoch-averaged source structure model, and all datasets (pulsar gated, pulsar ungated, calibrator sources, calibrator sources divided by model) were written to disk using the AIPS task FITTP.
\item{\bf Producing log files:} While the script was running, statistics on the failure rates of each calibration step were retained.  At the conclusion of the script, these are written into a summary webpage, along with any information from the VLBA operator's log and plots of the delay, amplitude, and phase calibration tables generated by the script.  Plots showing visibility amplitude as a function of baseline length are also generated for each source and included, to aid in the identification of unflagged RFI.  If any evidence of RFI or unsatisfactory calibration was evident, then additional flagging was undertaken, and the calibration script was re-run.
\end{enumerate}

\subsection{Imaging and position extraction}
\label{sec:posextraction}

After flagging, calibration, and averaging, the visibility
data are now in a suitable form for the extraction of our
astrometric observables. Given the calibration that has been applied, the
pulsar is effectively being determined
using the method of differential astrometry with respect to the
position of the in-beam calibrators. Among the 73 in-beam calibrators,
the positions of 14 were found in the Radio Fundamental Catalogue
(RFC\footnote{Available at \url{http://astrogeo.org/rfc}};
Petrov \& Kovalev, in preparation) derived using observations 
designed to improve the absolute positions of VLBI calibrator sources.
The absolute positions of the other in-beam calibrators
were determined with respect to the primary out-of-beam phase reference calibrators.
For the out-of-beam calibrators, the RFC2019a solution was used, and the 
absolute positions listed in the RFC have
coordinate uncertainties ranging from 0.1 to 0.9~mas (median: 0.18~mas).
J2000 coordinates are used for all sources.

Rather than these uncertainties transferred from the out-of-beam calibrators, the
absolute position uncertainties of the in-beam calibrators, and hence of 
the pulsars, is generally dominated by one of the following two factors:

\begin{enumerate}
\item A substantial angular separation from the out-of-beam
calibrators to the target field, (up to $3.5^\circ$), leading
to biases in position estimates when the calibration
is extrapolated. At an observing frequency of 1600~MHz,
the uncertainty depends strongly on the ionospheric
conditions during the observation(s) as well as the size of the angular separation and the
median observing elevation; a few mas is
typical \citep{deller16a}.  We estimate this value for each pulsar field by examining the scatter in the
in-beam positions obtained when imaging using only phase referencing from the
out-of-beam calibrator (i.e., no self-calibration on the in-beam calibrator).  
\item The frequency dependence of the core position of calibrator sources.
The apparent core of calibrator sources, which defines their reference position, is determined from the region of peak brightness in an image.
The true jet origin, the region at the jet apex, is invisible to an observer, as it is
opaque (optical depth $\tau\gg1$) due to synchrotron
self-absorption. The apparent core is located where the jet becomes visible further away from the origin, when
optical depth reaches $\tau\approx1$ at the apparent jet base. The higher the frequency, the closer the observed core is to the jet apex.
This effect is called the core-shift; more detailed descriptions can be found in, e.g., 
%\citet{marcaide84a,lobanov_cs1998,Kovalev_cs_2008,Sullivan09_coreshift,hada11a,Sokol_cs2011,pushkarev_etal12,fromm13a,kutkin_etal14,r:1030,r:3C273_lisakov,voitsik18a,kutkin18a,pushkarev19a}.
\citet{marcaide84a,lobanov_cs1998,Kovalev_cs_2008,Sullivan09_coreshift,voitsik18a,pushkarev19a}.
According to \citet{Sokol_cs2011} the core-shift at 1600~MHz ranges from 0.6 to 2.4~mas with median 1.1~mas for a specially pre-selected sample of active galactic nuclei (AGN) jets with significant shifts. \citet{pushkarev_etal12} has measured core-shifts for a large complete flux density limited sample beween 8 and 15~GHz and has found median values similar to \citet{Sokol_cs2011}.
Another complication arises with long-term core-shift variability, which can reach significant values of up to 1~mas between 2 and 8~GHz \citep{plavin19a}.
In this work we did not determine the core-shift, and thus the neglected core-shift introduces a $\sim$mas level bias in the chain of tying the pulsar position to the out-of-beam calibrator via an intermediate in-beam calibrator.
\end{enumerate}
 
 All three sources of uncertainty are added in quadrature when determining the absolute position uncertainty of the pulsar.
However, none of these issues affect any shifts {\em relative} to the in-beam calibrator, which are relevant for determining proper motion and parallax.  The contributions to the relative astrometric error budget are considered in more detail in Section~\ref{sec:errorbudget}.

\begin{figure}
\includegraphics[width=0.48\textwidth]{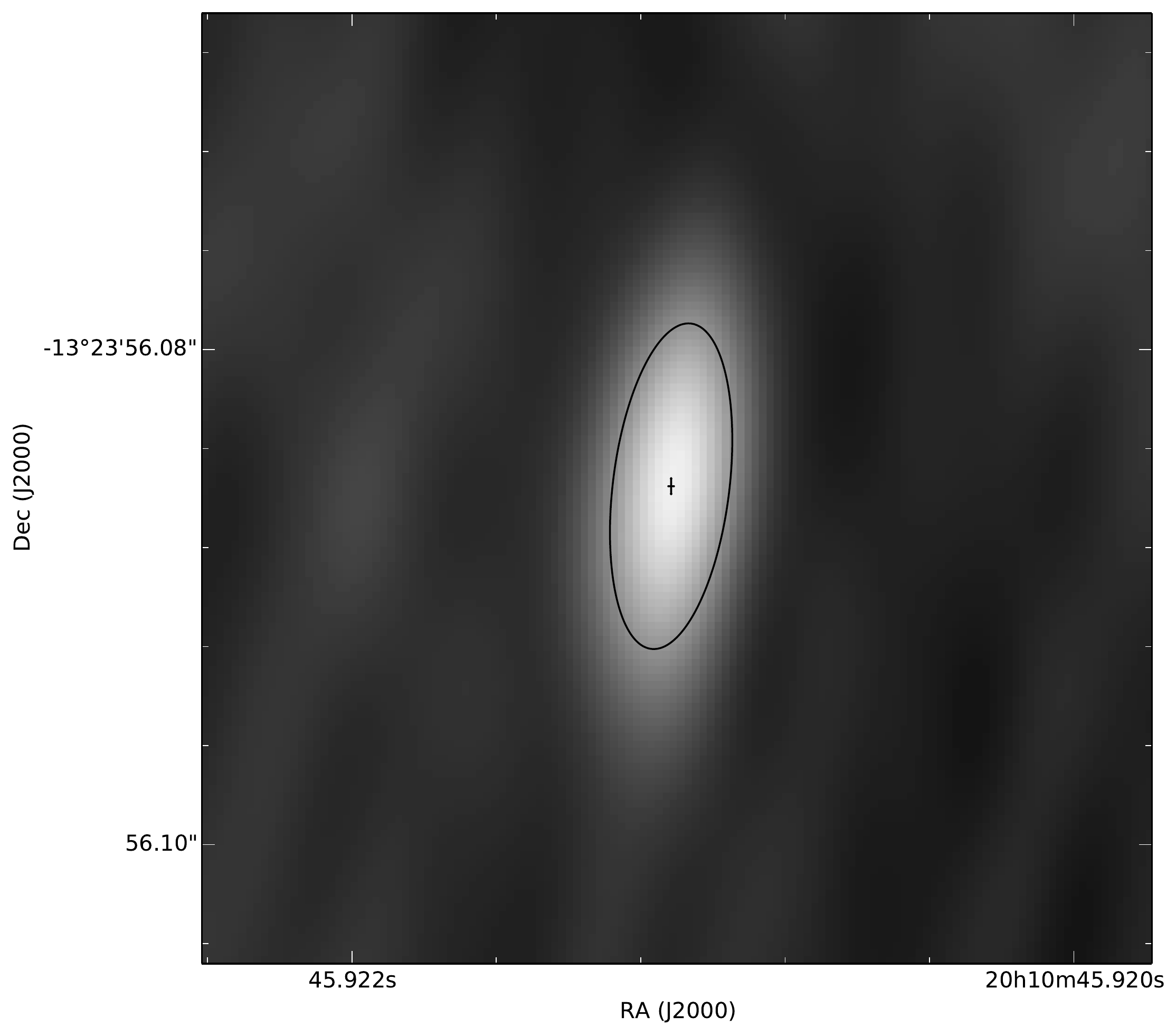}
% Figure generated on bunker, see bunker:data/vlbi/psrpi/plots/README
\caption{\label{fig:imageplanefit} A sample CLEAN image of a target pulsar, in this case PSR~J2010$-$1323.  Greyscale shows the sky brightness on a linear scale.  The solid ellipse shows the best-fit gaussian returned by JMFIT, while the cross shows the best-fit position and associated uncertainty for the pulsar at this epoch.  A detection with low signal-to-noise (S/N $\sim$ 16) was chosen to make the uncertainty bars visible.}
\end{figure}

For each visibility dataset (gated pulsar, ungated pulsar, calibrator source(s), and calibrator source(s) divided by their average model), we used the following procedure to extract a position, again implemented as a ParselTongue script.  We loaded the dataset into the difmap package \citep{shepherd97a} and inverted the Stokes I visibility data to form a dirty image, using natural weighting.  We then shifted the dataset to be approximately centered on the peak in the dirty image, and added a single point source component to the model at the location of the peak.  We ran a model fit for 20 iterations, and then wrote the resultant clean image (pixel size 0.75 milliarcseconds) to disk in FITS format.  This clean image was then loaded into AIPS and the position and position uncertainties were extracted with the task JMFIT to fit an elliptical gaussian, using a 20x20 pixel window centered on the peak.  An example image, showing the fitted gaussian, best-fit position, and uncertainty, is shown in Figure~\ref{fig:imageplanefit}.

In principle, position information can be extracted directly from the model fit, without the need to form an image and fit an elliptical gaussian.  However, while the best-fit position is easily accessible, extracting a position uncertainty from a model fit is highly dependent on the overall scaling of the visibility weights. In contrast, the position uncertainty resulting from an image plane fit, where the root mean square fluctuations of a residual image can easily be measured, is well defined and robust under conditions that are typically satisfied for radio interferometric images \citep{condon97a}.  It is for this reason that we use JMFIT to extract positions and position uncertainties, although we did cross-check our astrometric results using the model-fit positions and estimated uncertainties, finding results that typically agreed to well within 1$\sigma$ (where the uncertainty was taken from the image plane fit.)

% See bunker:data/vlbi/psrpi/plots/README for how to generate

\subsection{Astrometric fitting}

After the source modeling, calibration, and imaging described in the previous section was completed, we were left with a position time series for the target pulsar (taken from the higher S/N gated datasets) and one or more in-beam reference sources.  The reference frame in which these positions were measured has been defined by the assumed positions for the source(s) used in step~12 of Section~\ref{sec:calibration}.  This was not necessarily optimal for measuring the time-varying position of the target pulsar---the brightest in-beam calibrator might be located far from the pulsar, or might exhibit structure evolution that result in additional systematic errors.  In these cases, we could provide a more stable reference frame by utilizing a weaker and/or more stable background source located at a smaller angular separation to the target pulsar.

By subtracting the position residuals from a given in-beam source (or a weighted average of several in-beam sources), we could transform the reference frame into one defined by an in-beam source (or sources) of our choosing.  By selecting the source nearest to the target, it is possible to minimise the systematic position shifts at the target pulsar position introduced by the residual ionosphere.  However, this also necessitates adding the formal fit errors for the chosen position reference source(s) in quadrature to the pulsar position fit error, which may not be an acceptable trade-off for weak background sources where these errors are large.  Accordingly, for each pulsar we selected the position reference source considering both of these factors, and Table~\ref{tab:calibrators} shows the calibrator source(s) and the frame-defining source(s) for each target pulsar, along with the flux density as measured from the combined reference image for each source (at 1670 MHz for most sources, but at 2270 MHz for the calibrators of PSRs~J1820$-$0427, J1833$-$0338, J1913$+$1400, and J1917$+$1353).  Finally, any estimate of additional systematic sources of position uncertainty (see below) were added to the formal position errors; these were initially set to zero.

\begin{deluxetable*}{lrrrrrrrrr}
\tabletypesize{\scriptsize}
\tablecaption{Calibrator sources for \psrpi\ targets.}
\tablewidth{\textwidth}
\tablehead{
\colhead{Pulsar} & \colhead{Delay cal.} & \colhead{Sep.} & \colhead{Flux dens.} & \colhead{In-beam cal.} & \colhead{Sep.} & \colhead{Flux dens.} & \colhead{Position reference} & \colhead{Sep.} & \colhead{Flux dens.} \\
\colhead{} & \colhead{} & \colhead{(\degrees)} & \colhead{(mJy)} & \colhead{} & \colhead{(\arcmin)} & \colhead{(mJy)} & \colhead{} & \colhead{(\arcmin)} & \colhead{(mJy)} 
 }
\startdata
J0040$+$5716 & J0042$+$5708 & 0.3 & 910 & J004219$+$570836 & 16.5 & 914 & J004047$+$570321 & 13.2 & 7 \\
J0055$+$5117 & J0049$+$5128 & 1.0 & 180 & J005620$+$512226 & 7.5 & 84 & J005620$+$512226 & 7.5 & 84 \\
J0102$+$6537 & J0110$+$6805 & 2.6 & 400 & J010225$+$653553 & 1.5 & 16 & J010225$+$653553 & 1.5 & 16 \\
J0108$+$6608 & J0110$+$6805 & 2.0 & 400 & J010845$+$660807 & 2.4 & 16 & J010845$+$660807 & 2.4 & 16 \\
J0147$+$5922 & J0147$+$5840 & 0.7 & 210 & J014921$+$592512 & 12.8 & 45 & J014921$+$592512 & 12.8 & 45 \\
J0151$-$0635 & J0138$-$0540 & 3.2 & 270 & J015230$-$062955 & 17.5 & 57 & J015201$-$062904 & 11.4 & 12 \\
J0152$-$1637 & J0151$-$1732 & 0.9 & 180 & J015325$-$163113 & 19.0 & 20 & J015325$-$163113 & 19.0 & 20 \\
J0157$+$6212 & J0207$+$6246 & 1.2 & 1530 & J015553$+$620701 & 14.6 & 63 & J015553$+$620701 & 14.6 & 63 \\
J0323$+$3944 & J0322$+$3948 & 0.1 & 70 & J032251$+$394802 & 7.4 & 72 & J032251$+$394802 & 7.4 & 72 \\
J0332$+$5434 & J0346$+$5400 & 2.1 & 320 & J0332$+$5434 & 0.0 & 1244 & J033317$+$544011 & 6.1 & 7 \\
J0335$+$4555 & J0330$+$4656 & 1.3 & 220 & J033346$+$460819 & 20.0 & 86 & J033346$+$460819 & 20.0 & 86 \\
J0357$+$5236 & J0346$+$5400 & 2.2 & 320 & J035751$+$524922 & 12.5 & 11 & J035751$+$524922 & 12.5 & 11 \\
 &  &  &  & J035819$+$522936 & 9.0 & 12 &  &  &  \\
J0406$+$6138 & J0356$+$6043 & 1.5 & 180 & J040635$+$611543 & 23.0 & 17 & J040635$+$611543 & 23.0 & 17 \\
J0601$-$0527 & J0606$-$0724 & 2.3 & 410 & J060250$-$053757 & 16.3 & 26 & J060250$-$053757 & 16.3 & 26 \\
J0614$+$2229 & J0620$+$2102 & 2.0 & 870 & J061411$+$222204 & 8.0 & 15 & J061411$+$222204 & 8.0 & 15 \\
J0629$+$2415 & J0620$+$2102 & 3.8 & 870 & J062909$+$235751 & 17.9 & 15 & J062909$+$235751 & 17.9 & 15 \\
J0729$-$1836 & J0725$-$1904 & 1.0 & 230 & J072831$-$182206 & 20.5 & 105 & J072831$-$182206 & 20.5 & 105 \\
J0823$+$0159 & J0825$+$0309 & 1.3 & 420 & J082344$+$020257 & 9.5 & 11 & J082344$+$020257 & 9.5 & 11 \\
J0826$+$2637 & J0819$+$2747 & 2.0 & 240 & J082733$+$263715 & 9.4 & 34 & J082733$+$263715 & 9.4 & 34 \\
J1022$+$1001 & J1025$+$1253 & 3.0 & 470 & J102334$+$101200 & 13.5 & 213 & J102310$+$100126 & 3.2 & 18 \\
J1136$+$1551 & J1142$+$1547 & 1.5 & 160 & J1136$+$1551 & 0.0 & 181 & J113609$+$155228 & 1.9 & 15 \\
J1257$-$1027 & J1303$-$1051 & 1.6 & 270 & J125751$-$101040 & 20.1 & 65 & J125713$-$102403 & 3.8 & 4 \\
J1321$+$8323 & J1321$+$8316 & 0.1 & 390 & J132145$+$831613 & 7.4 & 392 & J132145$+$831613 & 7.4 & 392 \\
J1532$+$2745 & J1539$+$2744 & 1.7 & 140 & J153330$+$273502 & 20.8 & 18 & J153330$+$273502 & 20.8 & 18 \\
J1543$-$0620 & J1543$-$0757 & 1.6 & 1420 & J154416$-$064253 & 25.0 & 33 & J154416$-$064253 & 25.0 & 33 \\
J1607$-$0032 & J1557$-$0001 & 2.4 & 380 & J160533$-$003106 & 24.7 & 36 & J160533$-$003106 & 24.7 & 36 \\
J1623$-$0908 & J1624$-$0649 & 2.3 & 640 & J162431$-$090255 & 19.3 & 12 & J162431$-$090255 & 19.3 & 12 \\
 &  &  &  & J162414$-$092356 & 20.5 & 11 & J162414$-$092356 & 20.5 & 11 \\
J1645$-$0317 & J1638$-$0340 & 1.7 & 330 & J164410$-$031329 & 13.6 & 44 & J164410$-$031329 & 13.6 & 44 \\
J1650$-$1654 & J1642$-$2007 & 3.8 & 100 & J165133$-$170928 & 21.7 & 49 & J165015$-$165730 & 4.0 & 24 \\
J1703$-$1846 & J1709$-$1728 & 1.9 & 410 & J170441$-$185807 & 16.8 & 16 & J170441$-$185807 & 16.8 & 16 \\
 &  &  &  & J170429$-$190336 & 19.6 & 16 &  &  &  \\
J1735$-$0724 & J1735$-$0559 & 1.4 & 530 & J173401$-$071554 & 18.2 & 22 & J173500$-$073321 & 8.5 & 8 \\
 &  &  &  & J173500$-$073321 & 8.5 & 8 &  &  &  \\
J1741$-$0840 & J1740$-$0811 & 0.6 & 170 & J174002$-$083111 & 21.9 & 8 & J174002$-$083111 & 21.9 & 8 \\
J1754$+$5201 & J1740$+$5211 & 2.1 & 1570 & J175459$+$520114 & 5.7 & 19 & J175459$+$520114 & 5.7 & 19 \\
 &  &  &  & J175550$+$520506 & 14.0 & 35 &  &  &  \\
J1820$-$0427 & J1819$-$0258 & 1.5 & 1480 & J182043$-$042412 & 4.1 & 95 & J182103$-$042633 & 3.0 & 29 \\
 &  &  &  & J182103$-$042633 & 3.0 & 29 &  &  &  \\
J1833$-$0338 & J1827$-$0405 & 1.5 & 510 & J183323$-$032331 & 16.2 & 97 & J183323$-$032331 & 16.2 & 97 \\
J1840$+$5640 & J1824$+$5651 & 2.3 & 630 & J183849$+$564515 & 16.3 & 13 & J183849$+$564515 & 16.3 & 13 \\
J1901$-$0906 & J1855$-$1209 & 3.4 & 140 & J190252$-$085706 & 17.3 & 13 & J190252$-$085706 & 17.3 & 13 \\
 &  &  &  & J190230$-$085144 & 17.2 & 15 & J190230$-$085144 & 17.2 & 15 \\
J1912$+$2104 & J1908$+$2222 & 1.7 & 100 & J191255$+$210734 & 4.1 & 30 & J191255$+$210734 & 4.1 & 30 \\
 &  &  &  & J191326$+$205141 & 16.3 & 8 & J191326$+$205141 & 16.3 & 8 \\
J1917$+$1353 & J1911$+$1611 & 2.7 & 490 & J191718$+$140509 & 12.4 & 101 & J191718$+$140509 & 12.4 & 101 \\
J1913$+$1400 & J1911$+$1611 & 2.2 & 490 & J191324$+$140254 & 2.0 & 15 & J191324$+$140254 & 2.0 & 15 \\
J1919$+$0021 & J1920$-$0236 & 3.0 & 320 & J191851$+$002147 & 14.9 & 60 & J191851$+$002147 & 14.9 & 60 \\
J1937$+$2544 & J1929$+$2543 & 1.6 & 210 & J193805$+$253232 & 18.6 & 189 & J193805$+$253232 & 18.6 & 189 \\
J2006$-$0807 & J2011$-$0644 & 1.9 & 2040 & J200651$-$082625 & 21.3 & 78 & J200651$-$082625 & 21.3 & 78 \\
J2010$-$1323 & J2011$-$1546 & 2.4 & 540 & J201101$-$134359 & 20.4 & 34 & J201101$-$134359 & 20.4 & 34 \\
J2046$-$0421 & J2055$-$0416 & 2.5 & 350 & J204536$-$043534 & 15.4 & 56 & J204536$-$043534 & 15.4 & 56 \\
J2046$+$1540 & J2045$+$1547 & 0.2 & 140 & J204545$+$154727 & 14.7 & 135 & J204545$+$154727 & 14.7 & 135 \\
J2113$+$2754 & J2114$+$2832 & 0.8 & 360 & J211358$+$275059 & 12.4 & 68 & J211312$+$275002 & 4.4 & 17 \\
 &  &  &  & J211312$+$275002 & 4.4 & 17 &  &  &  \\
J2113$+$4644 & J2123$+$4614 & 1.8 & 120 & J2113$+$4644 & 0.0 & 62 & J211432$+$463439 & 15.1 & 52 \\
J2145$-$0750 & J2142$-$0437 & 3.3 & 380 & J214557$-$074748 & 3.1 & 20 & J214557$-$074748 & 3.1 & 20 \\
J2149$+$6329 & J2148$+$6107 & 2.4 & 1460 & J215159$+$633527 & 14.6 & 13 & J215159$+$633527 & 14.6 & 13 \\
J2150$+$5247 & J2201$+$5048 & 2.6 & 530 & J214842$+$525403 & 18.4 & 12 & J214842$+$525403 & 18.4 & 12 \\
J2212$+$2933 & J2205$+$2926 & 1.4 & 140 & J221207$+$293356 & 3.6 & 77 & J221207$+$293356 & 3.6 & 77 \\
J2225$+$6535 & J2238$+$6804 & 2.8 & 100 & J222346$+$654751 & 17.9 & 16 & J222346$+$654751 & 17.9 & 16 \\
 &  &  &  & J222417$+$652805 & 12.4 & 6 & J222417$+$652805 & 12.4 & 6 \\
J2248$-$0101 & J2247$+$0000 & 1.1 & 450 & J224808$-$011532 & 14.5 & 36 & J224808$-$011532 & 14.5 & 36 \\
J2305$+$3100 & J2307$+$3230 & 1.5 & 400 & J230655$+$305028 & 15.5 & 39 & J230655$+$305028 & 15.5 & 39 \\
J2317$+$1439 & J2327$+$1524 & 2.6 & 190 & J231619$+$143511 & 12.8 & 23 & J231619$+$143511 & 12.8 & 23 \\
 &  &  &  & J231715$+$145130 & 12.1 & 17 &  &  &  \\
J2317$+$2149 & J2318$+$2404 & 2.3 & 130 & J231657$+$220241 & 19.0 & 15 & J231657$+$220241 & 19.0 & 15 \\
 &  &  &  & J231643$+$220626 & 23.9 & 94 &  &  &  \\
J2325$+$6316 & J2302$+$6405 & 2.6 & 110 & J232445$+$633001 & 13.5 & 9 & J232519$+$631636 & 0.8 & 5 \\
 &  &  &  & J232519$+$631636 & 0.8 & 5 &  &  &  \\
J2346$-$0609 & J2348$-$0425 & 1.8 & 240 & J234636$-$060813 & 3.8 & 6 & J234636$-$060813 & 3.8 & 6 \\
 &  &  &  & J234728$-$060526 & 10.4 & 9 &  &  &  \\
J2354$+$6155 & J2339$+$6010 & 2.5 & 310 & J235440$+$613736 & 18.7 & 54 & J235440$+$613736 & 18.7 & 54 
\enddata
% Generated by ./makepsrpitable.py on bunker in /home/deller/svn_codebase/psrpi
% Note that mspsrpitable.py is on gygax
\tablenotetext{A}{Multiple entries indicate that data from two sources were combined to derive solutions.}
\label{tab:calibrators}
\end{deluxetable*}

This time series of measured positions and estimated uncertainties could then be processed using the \textit{pmpar}\footnote{\url{https://github.com/walterfb/pmpar}} package to perform least-squares minimization and fit for reference position, proper motion, and parallax.  Four of the \psrpi\ targets are a pulsar in a binary system, and where the orbital reflex motion is substantial additional steps were required in the fitting process.  PSR J1022$+$1001 and PSR~J2145$-$0750 are millisecond pulsars that have already been described in \citet{deller16a}, while PSR J0823$+$0159 is a slow pulsar in a long-period binary.  For these pulsars, the orbital period, longitude of periastron, eccentricity, and projected semi-major axis were all well-constrained by pulsar timing, and we therefore were only required to fit for inclination $i$ and longitude of ascending node $\Omega$.  For PSR J2317+1439, the orbital reflex motion is negligible compared to our positional uncertainties.  For the pulsars where fitting the orbital reflex motion was required, we included these two additional parameters in our least-squares minimization.

In almost all of the \psrpi\ pulsars, the reduced $\chi^2$ of the initial least-squares fit exceeded unity, often considerably.  This result is not surprising given that the initial input position uncertainties are purely based on the S/N of the pulsar (and position reference calibrator) images, and do not account for potential systematic position shifts.  In most cases, the dominant systematic contribution comes from the residual unmodeled ionosphere, but other possibilities such as source structure evolution in the source(s) defining the reference frame also exist.  In general, the distribution (both form and variance) of these systematic errors is extremely difficult to predict \textit{a priori}, as discussed in Section~\ref{sec:errorbudget}, which complicates efforts to accurately estimate the uncertainties on the fitted astrometric parameters.  The problem is exacerbated for datasets where the formal position uncertainties vary widely between epochs, as can be the case for pulsars that exhibit significant amplitude variability due to diffractive and/or refractive scintillation.  If no adjustment is made to the formal position errors, then the epochs with high-significance detections when the pulsar was ``scintillated up" will exhibit a disproportionate impact on the astrometric fit.

In Section~\ref{sec:errorbudget}, we investigate different methods for estimating a systematic error that can be added in quadrature to the formal position fit errors in order to mitigate this issue.  Unsurprisingly, we find that no method is perfect in all situations, but that the use of an estimator is better than neglecting systematic errors entirely.  Our final astrometric solutions therefore make use of the empiral systematic error estimator discussed in Section~\ref{sec:errorbudget}.

Once a position time series with final estimated uncertainties is available, best-fit values and uncertainties for the astrometric parameters must be produced.  Two options are available:

\begin{enumerate}
\item A simple least squares fit; or
\item A bootstrap fit.
\end{enumerate}

The least-squares fit has the advantage of simplicity, but is sensitively dependent on beginning with a good estimate of the input position errors.  If these are underestimated (which will generally result in a reduced $\chi^2$ that still significantly exceeds unity) then the errors on the astrometric observables will likewise be underestimated.  Conversely (but more rarely), overestimating the systematic errors will lead to inflated uncertainties on the astrometric observables. A bootstrap fit  \citep[e.g.,][]{efron91a} utilizes a large number of trials, where in each trial $N$ position measurements for the input dataset for each trial are selected randomly {\em with replacement} from the available $N$ astrometric position measurements for that pulsar.  In our case, $N$ is usually 8 or 9. For each trial dataset, a least-squares fit is made as usual, and the best-fit parameters are saved.  After many trials, a cumulative probability distribution for each of the fitted parameters is built, from which the most probable value and a desired confidence interval can be extracted.

A bootstrap fit has the advantage that the uncertainty on the fitted parameters is not determined solely by the uncertainty in the input position measurements, which as we have seen is hard to estimate accurately.  However, the bootstrap approach can exacerbate a problem already present for \psrpi\ and most VLBI astrometry programs: the small sample size.  With just 8 or 9 position measurements, a significant fraction of trials can end up with poor time coverage of one of the desired astrometric quantities, sampling a shorter time range or predominantly one side of the parallax signature.  This is especially problematic in cases where the pulsar scintillates and is detected only weakly (or not at all) in some epochs, further reducing the number of useful degrees of freedom.  An example is PSR J2317$+$1439, where non-detections due to unfavourable scintillation were concentrated in the December/January observations and resulted in a poor sampling of the parallax ellipse.

We favor a bootstrap approach for determining the final astrometric uncertainties, as it generally produces the most conservative error estimates (as can be seen in Section~\ref{sec:errorbudget}).  The results presented here are obtained from a bootstrap with 100,000 trials per pulsar.  We highlight the circumstances under which the bootstrap uncertainties may potentially be too conservative in the discussion.

\section{Results}
\label{sec:results}

\subsection{Astrometric fits for 60 pulsars}

The astrometric results for our 60 target pulsars are shown in Table~\ref{tab:allresults}.  Asymmetric error bars representing the 68\% confidence interval are listed along with the best-fit parameter values.  The median parallax uncertainty obtained was 46 $\mu$as, with 60\% of our targets meeting or exceeded the design goal of 50 $\mu$as parallax accuracy.  Almost all (53 of the 60) target pulsars have a significant ($>$95\% confidence) parallax measurement, while two-thirds of the sample provide a distance error of 20\% or less.

\begin{figure*}
\begin{tabular}{cc}
\includegraphics[width=0.48\textwidth]{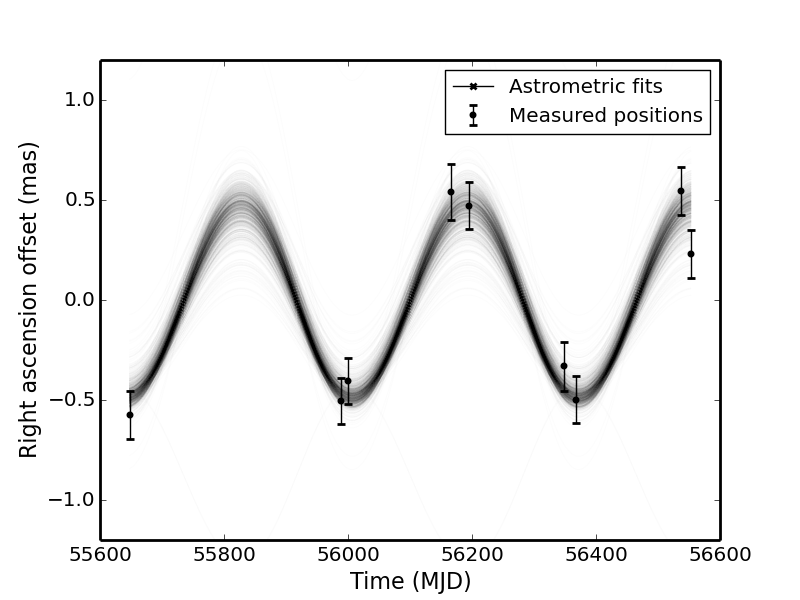} &
\includegraphics[width=0.48\textwidth]{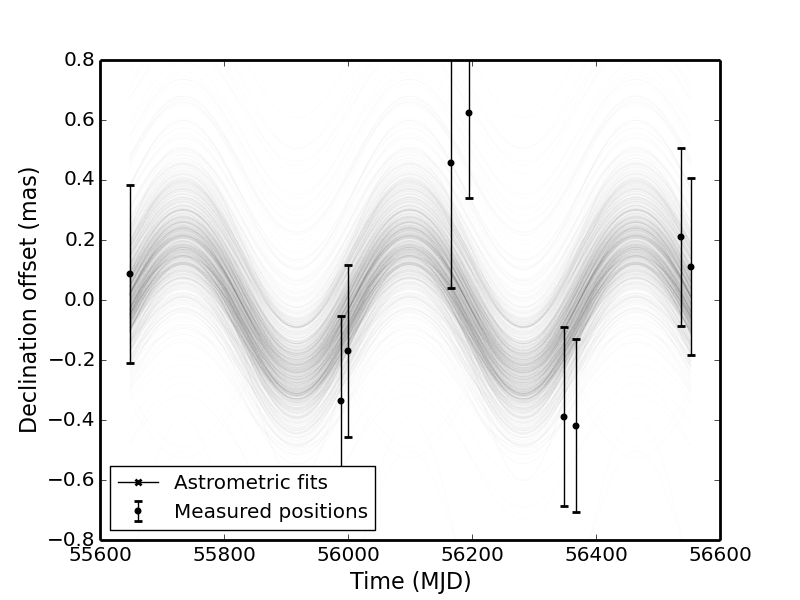} \\
\includegraphics[width=0.48\textwidth]{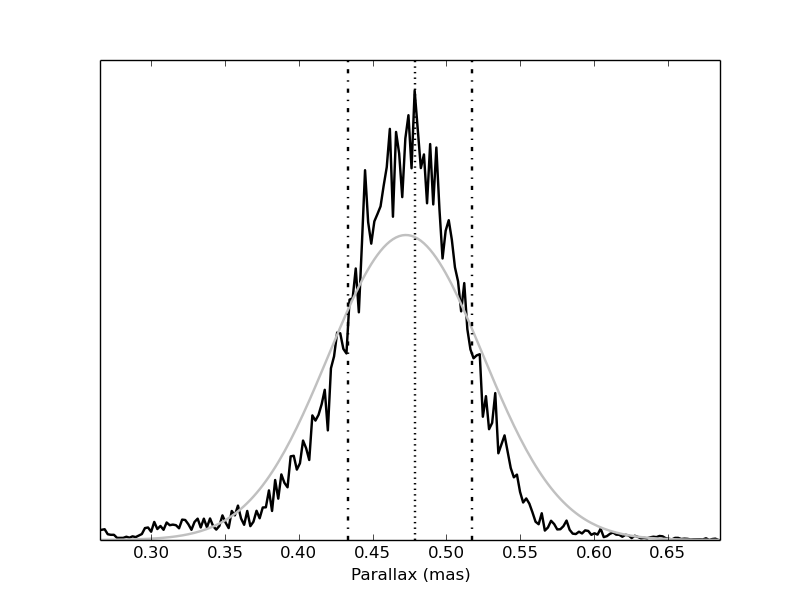} &
\includegraphics[width=0.48\textwidth]{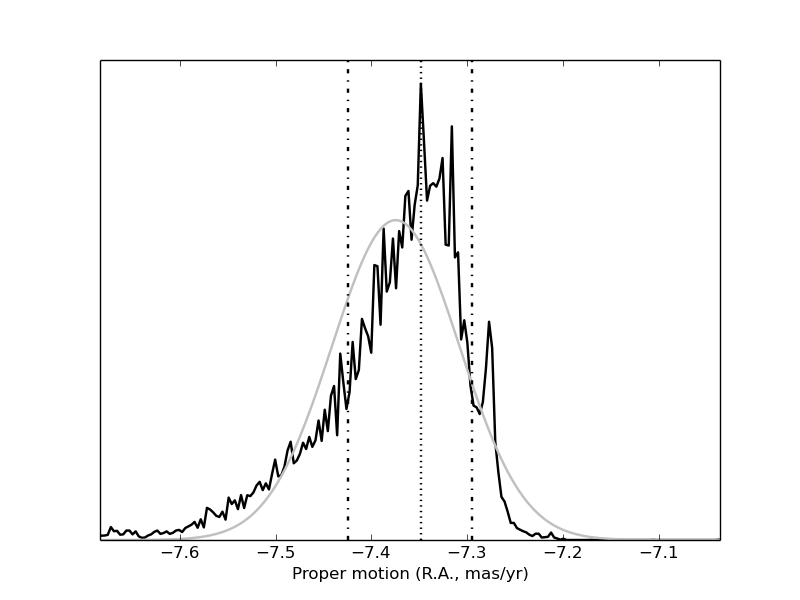}
\end{tabular}
\caption{\label{fig:examplefits}Top: Illustration of the bootstrap fits for PSR~J0601$-$0527, showing position offset in right ascension (left) and declination (right) after subtraction of the best-fit proper motion.  Each of the 100,000 fits is overplotted in a light greyscale, and so darker regions indicate the most likely parallax signature.  Bottom: Probability distribution functions for parallax (left) and proper motion in right ascension (right).  The black line shows the results of the bootstrap which are used in the text, while for comparison purposes, a light grey line shows the result of a least-squares fit after adding additional systematic error contributions to obtain a reduced $\chi^2$ of 1.0.}
\end{figure*}

Detailed results, including astrometric plots and calibrator images, can be found for each pulsar at \url{https://safe.nrao.edu/vlba/psrpi/release.html}.  As an example, the astrometric plots and bootstrap histograms for PSR~J0601$-$0527 are shown in Figure~\ref{fig:examplefits}.  This is a typical-to-challenging target---the signal-to-noise ratio on the target was $\sim$50, and the in-beam calibrator was slightly resolved, with a total flux density of  $\sim$25 mJy, and separated from the target by 16\arcmin. The reduced $\chi^2$ of the astrometric fit to the position time series using the empirical systematic error estimate discussed in Section~\ref{sec:errorbudget} is 1.3, and the attained parallax precision of $\sim$40\,\uas\ is very close to the median \psrpi\ value.

\begingroup
\setlength{\tabcolsep}{6pt}
\renewcommand\arraystretch{1.5}
\begin{longtable*}{lrrrrr}
\caption{\label{tab:allresults}Fitted astrometric parameters for all PSR$\pi$ targets. Position offsets are relative to the defined position for the chosen reference source, at epoch MJD 56000.0; the right ascension offset is calculated at the declination of the target pulsar.}\\
\hline
\tabletypesize{\scriptsize}
\tablewidth{1.00\textwidth}
\textbf{Pulsar} & \multicolumn{2}{r}{\textbf{Offset from reference}} & \multicolumn{2}{r}{\textbf{Proper motion}}  & \textbf{Parallax}\\[-2pt]
                 & \textbf{R.A.}  & \textbf{Decl.}                    & \textbf{R.A.}       & \textbf{Decl.}        & \\[-2pt]
                 & \textbf{(mas)} & \textbf{(mas)}                    & \textbf{(mas/yr)}   & \textbf{(mas/yr)}    & \textbf{(mas)}\\
\hline
\endfirsthead
\multicolumn{6}{c}%
{\tablename\ \thetable\ -- \textit{Continued from above}} \\
\hline
\textbf{Pulsar} & \multicolumn{2}{r}{\textbf{Offset from reference}} & \multicolumn{2}{r}{\textbf{Proper motion}}  & \textbf{Parallax}\\[-2pt]
                 & \textbf{R.A.}  & \textbf{Decl.}                    & \textbf{R.A.}       & \textbf{Decl.}        & \\[-2pt]
                 & \textbf{(mas)} & \textbf{(mas)}                    & \textbf{(mas/yr)}   & \textbf{(mas/yr)}    & \textbf{(mas)}\\
\hline
\endhead
\hline \multicolumn{6}{r}{\textit{Continued below}} \\
\endfoot
\hline
\endlastfoot
J0040+5716 & $-119735.120^{+0.037}_{-0.045}$ & $783183.975^{+0.064}_{-0.109}$ & $12.399^{+0.033}_{-0.057}$ & $-5.450^{+0.140}_{-0.082}$ & $0.102^{+0.051}_{-0.025}$ \\
J0055+5117 & $-332840.251^{+0.071}_{-0.023}$ & $-302297.263^{+0.085}_{-0.074}$ & $10.490^{+0.049}_{-0.085}$ & $-17.352^{+0.074}_{-0.204}$ & $0.349^{+0.055}_{-0.055}$ \\
J0102+6537 & $47321.472^{+0.074}_{-0.006}$ & $79472.077^{+0.041}_{-0.045}$ & $9.252^{+0.049}_{-0.081}$ & $1.828^{+0.093}_{-0.206}$ & $0.399^{+0.044}_{-0.045}$ \\
J0108+6608 & $-140564.314^{+0.039}_{-0.033}$ & $27077.203^{+0.050}_{-0.025}$ & $-32.754^{+0.036}_{-0.025}$ & $35.162^{+0.024}_{-0.051}$ & $0.468^{+0.035}_{-0.031}$ \\
J0147+5922 & $-744046.103^{+0.100}_{-0.038}$ & $-189510.613^{+0.093}_{-0.054}$ & $-6.380^{+0.083}_{-0.101}$ & $3.826^{+0.054}_{-0.097}$ & $0.495^{+0.042}_{-0.093}$ \\
J0151-0635 & $-582099.427^{+0.164}_{-0.014}$ & $-358146.639^{+0.066}_{-0.052}$ & $10.697^{+0.094}_{-0.145}$ & $-5.373^{+0.061}_{-0.078}$ & $0.217^{+0.098}_{-0.076}$ \\
J0152-1637 & $-1067819.606^{+0.158}_{-0.014}$ & $-399877.713^{+0.363}_{-0.330}$ & $0.804^{+0.234}_{-0.201}$ & $-31.372^{+0.424}_{-0.313}$ & $0.443^{+0.214}_{-0.181}$ \\
J0157+6212 & $814250.347^{+0.035}_{-0.007}$ & $325456.249^{+0.037}_{-0.010}$ & $1.521^{+0.105}_{-0.017}$ & $44.811^{+0.034}_{-0.048}$ & $0.554^{+0.039}_{-0.024}$ \\
J0323+3944 & $401715.413^{+0.046}_{-0.011}$ & $-189856.218^{+0.029}_{-0.017}$ & $26.484^{+0.059}_{-0.034}$ & $-30.780^{+0.029}_{-0.015}$ & $1.051^{+0.039}_{-0.040}$ \\
J0332+5434 & $-158517.350^{+0.035}_{-0.008}$ & $-328076.504^{+0.050}_{-0.029}$ & $16.969^{+0.027}_{-0.029}$ & $-10.379^{+0.058}_{-0.036}$ & $0.595^{+0.020}_{-0.025}$ \\
J0335+4555 & $939470.468^{+0.073}_{-0.010}$ & $-746024.523^{+0.066}_{-0.058}$ & $-3.638^{+0.023}_{-0.073}$ & $-0.097^{+0.134}_{-0.105}$ & $0.409^{+0.022}_{-0.027}$ \\
J0357+5236 & $-61073.169^{+0.128}_{-0.028}$ & $-744690.138^{+0.062}_{-0.089}$ & $13.908^{+0.062}_{-0.115}$ & $-10.633^{+0.098}_{-0.058}$ & $0.305^{+0.029}_{-0.077}$ \\
J0406+6138 & $-41240.544^{+0.050}_{-0.171}$ & $1377882.120^{+0.122}_{-0.093}$ & $12.400^{+0.151}_{-0.085}$ & $22.716^{+0.100}_{-0.060}$ & $0.218^{+0.051}_{-0.057}$ \\
J0601-0527 & $-765952.497^{+0.074}_{-0.014}$ & $606395.795^{+0.097}_{-0.144}$ & $-7.348^{+0.053}_{-0.077}$ & $-15.227^{+0.084}_{-0.105}$ & $0.478^{+0.039}_{-0.045}$ \\
J0614+2229 & $81743.228^{+0.050}_{-0.021}$ & $471984.943^{+0.062}_{-0.014}$ & $-0.233^{+0.036}_{-0.053}$ & $-1.224^{+0.011}_{-0.065}$ & $0.282^{+0.022}_{-0.031}$ \\
J0629+2415 & $-54102.959^{+0.179}_{-0.041}$ & $1069457.665^{+0.054}_{-0.047}$ & $3.629^{+0.050}_{-0.193}$ & $-4.607^{+0.013}_{-0.153}$ & $0.333^{+0.036}_{-0.054}$ \\
J0729-1836 & $862898.001^{+0.143}_{-0.043}$ & $-875533.906^{+0.400}_{-0.390}$ & $-13.072^{+0.125}_{-0.091}$ & $13.252^{+0.456}_{-0.418}$ & $0.489^{+0.098}_{-0.078}$ \\
J0823+0159 & $-525441.078^{+0.150}_{-0.134}$ & $-225189.223^{+0.080}_{-0.095}$ & $-3.797^{+0.073}_{-0.415}$ & $0.171^{+0.232}_{-0.281}$ & $0.376^{+0.129}_{-0.070}$ \\
J0826+2637 & $-566289.704^{+0.007}_{-0.007}$ & $5345.818^{+0.072}_{-0.012}$ & $62.994^{+0.021}_{-0.007}$ & $-96.733^{+0.045}_{-0.085}$ & $2.010^{+0.013}_{-0.009}$ \\
J1022+1001 & $-190717.466^{+0.043}_{-0.016}$ & $25883.813^{+0.039}_{-0.025}$ & $-14.921^{+0.050}_{-0.033}$ & $5.611^{+0.033}_{-0.035}$ & $1.387^{+0.041}_{-0.028}$ \\
J1136+1551 & $-87829.834^{+0.010}_{-0.016}$ & $-73945.169^{+0.047}_{-0.031}$ & $-73.785^{+0.031}_{-0.010}$ & $366.569^{+0.072}_{-0.055}$ & $2.687^{+0.018}_{-0.016}$ \\
J1257-1027 & $-135426.468^{+0.059}_{-0.059}$ & $-182136.895^{+0.087}_{-0.066}$ & $-7.164^{+0.140}_{-0.105}$ & $12.079^{+0.119}_{-0.110}$ & $0.141^{+0.064}_{-0.092}$ \\
J1321+8323 & $36.330^{+0.115}_{-0.070}$ & $446008.090^{+0.173}_{-0.130}$ & $-52.674^{+0.099}_{-0.076}$ & $32.373^{+0.204}_{-0.048}$ & $0.968^{+0.036}_{-0.140}$ \\
J1532+2745 & $-1068597.478^{+0.106}_{-0.029}$ & $646805.042^{+0.113}_{-0.118}$ & $1.542^{+0.082}_{-0.127}$ & $18.932^{+0.104}_{-0.118}$ & $0.624^{+0.031}_{-0.096}$ \\
J1543-0620 & $-690701.998^{+0.041}_{-0.008}$ & $1328301.506^{+0.120}_{-0.113}$ & $-16.774^{+0.026}_{-0.063}$ & $-0.312^{+0.147}_{-0.114}$ & $0.322^{+0.028}_{-0.045}$ \\
J1607-0032 & $1476474.720^{+0.060}_{-0.014}$ & $-94965.170^{+0.163}_{-0.266}$ & $-26.437^{+0.027}_{-0.099}$ & $-27.505^{+0.222}_{-0.200}$ & $0.934^{+0.026}_{-0.047}$ \\
J1623-0908 & $-1100560.359^{+0.175}_{-0.043}$ & $-352982.390^{+0.182}_{-0.097}$ & $-10.769^{+0.131}_{-0.120}$ & $23.509^{+0.166}_{-0.069}$ & $0.586^{+0.101}_{-0.099}$ \\
J1645-0317 & $770570.904^{+0.016}_{-0.016}$ & $-268070.730^{+0.107}_{-0.136}$ & $-1.011^{+0.003}_{-0.051}$ & $20.523^{+0.147}_{-0.205}$ & $0.252^{+0.028}_{-0.019}$ \\
J1650-1654 & $172228.769^{+0.061}_{-0.006}$ & $168064.225^{+0.097}_{-0.056}$ & $-15.024^{+0.002}_{-0.092}$ & $-6.556^{+0.148}_{-0.131}$ & $-0.089^{+0.031}_{-0.015}$ \\
J1703-1846 & $-717146.582^{+0.047}_{-0.025}$ & $712491.475^{+0.182}_{-0.194}$ & $-0.751^{+0.102}_{-0.056}$ & $16.962^{+0.146}_{-0.230}$ & $0.348^{+0.049}_{-0.047}$ \\
J1735-0724 & $61924.490^{+0.076}_{-0.043}$ & $508913.404^{+0.087}_{-0.093}$ & $0.791^{+0.087}_{-0.029}$ & $20.614^{+0.074}_{-0.046}$ & $0.150^{+0.041}_{-0.035}$ \\
J1741-0840 & $1190194.661^{+0.112}_{-0.069}$ & $-560681.159^{+0.066}_{-0.087}$ & $0.436^{+0.082}_{-0.126}$ & $6.876^{+0.109}_{-0.066}$ & $0.279^{+0.050}_{-0.058}$ \\
J1754+5201 & $-341736.402^{+0.074}_{-0.010}$ & $-1983.221^{+0.056}_{-0.083}$ & $-3.950^{+0.047}_{-0.046}$ & $1.101^{+0.072}_{-0.059}$ & $0.160^{+0.029}_{-0.022}$ \\
J1820-0427 & $-169381.766^{+0.076}_{-0.027}$ & $-63903.822^{+0.120}_{-0.109}$ & $-7.318^{+0.074}_{-0.055}$ & $15.883^{+0.088}_{-0.069}$ & $0.351^{+0.049}_{-0.055}$ \\
J1833-0338 & $269293.573^{+0.119}_{-0.060}$ & $-932810.642^{+0.182}_{-0.260}$ & $-17.409^{+0.158}_{-0.025}$ & $15.038^{+0.333}_{-0.337}$ & $0.408^{+0.050}_{-0.067}$ \\
J1840+5640 & $945467.477^{+0.053}_{-0.011}$ & $-260360.747^{+0.070}_{-0.029}$ & $-31.212^{+0.033}_{-0.022}$ & $-29.079^{+0.047}_{-0.082}$ & $0.657^{+0.065}_{-0.008}$ \\
J1901-0906 & $-880884.264^{+0.071}_{-0.014}$ & $-543910.907^{+0.113}_{-0.163}$ & $-7.531^{+0.034}_{-0.045}$ & $-18.211^{+0.143}_{-0.159}$ & $0.510^{+0.067}_{-0.042}$ \\
J1912+2104 & $-170443.421^{+0.121}_{-0.006}$ & $-180675.561^{+0.070}_{-0.085}$ & $-11.335^{+0.023}_{-0.097}$ & $-5.768^{+0.092}_{-0.122}$ & $0.024^{+0.171}_{-0.022}$ \\
J1913+1400 & $-6337.234^{+0.050}_{-0.018}$ & $-122121.054^{+0.047}_{-0.033}$ & $-5.265^{+0.040}_{-0.072}$ & $-8.927^{+0.038}_{-0.065}$ & $0.185^{+0.027}_{-0.023}$ \\
J1917+1353 & $316294.074^{+0.016}_{-0.016}$ & $-672694.400^{+0.039}_{-0.025}$ & $-1.253^{+0.022}_{-0.074}$ & $3.811^{+0.057}_{-0.064}$ & $0.142^{+0.068}_{-0.007}$ \\
J1919+0021 & $893781.316^{+0.091}_{-0.045}$ & $-7819.701^{+0.060}_{-0.083}$ & $10.167^{+0.029}_{-0.143}$ & $-4.713^{+0.102}_{-0.073}$ & $0.166^{+0.042}_{-0.042}$ \\
J1937+2544 & $-865965.163^{+0.039}_{-0.015}$ & $700670.139^{+0.029}_{-0.039}$ & $-10.049^{+0.042}_{-0.030}$ & $-13.055^{+0.034}_{-0.039}$ & $0.318^{+0.031}_{-0.029}$ \\
J2006-0807 & $-522282.331^{+0.076}_{-0.010}$ & $1163967.750^{+0.184}_{-0.126}$ & $-6.176^{+0.035}_{-0.070}$ & $-10.616^{+0.174}_{-0.123}$ & $0.424^{+0.010}_{-0.101}$ \\
J2010-1323 & $-225966.545^{+0.305}_{-0.205}$ & $1202928.433^{+0.188}_{-0.146}$ & $2.358^{+0.329}_{-0.210}$ & $-5.611^{+0.257}_{-0.303}$ & $0.484^{+0.166}_{-0.120}$ \\
J2046+1540 & $777607.703^{+0.099}_{-0.016}$ & $-413782.623^{+0.041}_{-0.072}$ & $-10.455^{+0.032}_{-0.090}$ & $0.681^{+0.039}_{-0.090}$ & $0.310^{+0.082}_{-0.076}$ \\
J2046-0421 & $357981.097^{+0.006}_{-0.037}$ & $848678.492^{+0.105}_{-0.268}$ & $10.760^{+0.038}_{-0.035}$ & $-4.404^{+0.373}_{-0.076}$ & $0.167^{+0.026}_{-0.042}$ \\
J2113+2754 & $-109967.706^{+0.033}_{-0.005}$ & $238352.997^{+0.047}_{-0.058}$ & $-27.981^{+0.052}_{-0.014}$ & $-54.432^{+0.040}_{-0.096}$ & $0.704^{+0.023}_{-0.022}$ \\
J2113+4644 & $-704691.944^{+0.078}_{-0.025}$ & $569536.867^{+0.066}_{-0.029}$ & $9.525^{+0.068}_{-0.148}$ & $8.846^{+0.076}_{-0.090}$ & $0.454^{+0.077}_{-0.065}$ \\
J2145-0750 & $-111529.383^{+0.076}_{-0.016}$ & $-149818.403^{+0.054}_{-0.080}$ & $-9.491^{+0.052}_{-0.042}$ & $-9.114^{+0.090}_{-0.076}$ & $1.603^{+0.063}_{-0.009}$ \\
J2149+6329 & $-809671.242^{+0.101}_{-0.027}$ & $-343634.854^{+0.184}_{-0.072}$ & $15.786^{+0.131}_{-0.082}$ & $11.255^{+0.092}_{-0.284}$ & $0.356^{+0.072}_{-0.061}$ \\
J2150+5247 & $1041961.339^{+0.176}_{-0.105}$ & $-373619.769^{+0.163}_{-0.179}$ & $8.377^{+0.226}_{-0.181}$ & $-4.427^{+0.247}_{-0.353}$ & $0.034^{+0.164}_{-0.081}$ \\
J2212+2933 & $207796.901^{+0.052}_{-0.052}$ & $-50710.117^{+0.050}_{-0.083}$ & $-5.513^{+0.067}_{-0.102}$ & $-11.322^{+0.102}_{-0.128}$ & $0.265^{+0.050}_{-0.120}$ \\
J2225+6535 & $782263.168^{+0.201}_{-0.065}$ & $-735020.424^{+0.050}_{-0.058}$ & $147.220^{+0.243}_{-0.223}$ & $126.532^{+0.076}_{-0.115}$ & $1.203^{+0.166}_{-0.204}$ \\
J2248-0101 & $279160.561^{+0.151}_{-0.010}$ & $824047.754^{+0.184}_{-0.083}$ & $-10.548^{+0.117}_{-0.027}$ & $-17.407^{+0.110}_{-0.267}$ & $0.256^{+0.049}_{-0.067}$ \\
J2305+3100 & $-733995.388^{+0.042}_{-0.009}$ & $572881.368^{+0.078}_{-0.068}$ & $-3.737^{+0.082}_{-0.006}$ & $-15.571^{+0.049}_{-0.163}$ & $0.223^{+0.033}_{-0.028}$ \\
J2317+1439 & $721838.143^{+0.100}_{-1.295}$ & $259459.178^{+0.827}_{-0.829}$ & $-1.476^{+0.465}_{-0.065}$ & $3.806^{+0.272}_{-0.704}$ & $0.603^{+1.533}_{-0.241}$ \\
J2317+2149 & $834190.412^{+0.040}_{-0.056}$ & $-773795.695^{+0.093}_{-0.101}$ & $8.522^{+0.035}_{-0.104}$ & $0.136^{+0.192}_{-0.084}$ & $0.510^{+0.057}_{-0.049}$ \\
J2325+6316 & $-45049.774^{+0.080}_{-0.012}$ & $16351.365^{+0.149}_{-0.083}$ & $-5.926^{+0.082}_{-0.073}$ & $-2.051^{+0.188}_{-0.192}$ & $-0.010^{+0.049}_{-0.043}$ \\
J2346-0609 & $203020.882^{+0.037}_{-0.006}$ & $-106887.144^{+0.060}_{-0.052}$ & $37.390^{+0.025}_{-0.042}$ & $-20.230^{+0.107}_{-0.070}$ & $0.275^{+0.021}_{-0.036}$ \\
J2354+6155 & $-253697.746^{+0.058}_{-0.005}$ & $1090745.164^{+0.035}_{-0.019}$ & $22.755^{+0.056}_{-0.040}$ & $4.888^{+0.033}_{-0.016}$ & $0.412^{+0.031}_{-0.043}$ \\
% Produced on bunker in /home/deller/data/vlbi/psrpi/final by ./makePulsarPages.py
\end{longtable*}
\endgroup

The fitted parallax and proper motion results can be used to derive distances, Galactic z-heights, and transverse velocities for the target pulsars, or lower limits in the case where the parallax was not measured to sufficient accuracy.  Likewise, the fitted offsets from the inbeam calibrators can be combined with an estimate of the in-beam calibrator position uncertainty (comprising contributions from core-shift, phase-referencing to the out-of-beam calibrator, and the out-of-beam calibrator absolute uncertainty added in quadrature as described in Section~\ref{sec:posextraction}), to produce an absolute pulsar position at the reference epoch and associated uncertainty.  All of these derived quantities are shown in Table~\ref{tab:derivedresults}.  We stress that the absolute positions are of a preliminary nature, since the calibrator positions and core-shifts have not been determined to high precision, and note in particular that the positional uncertainties for PSR J0614+2229, PSR J0629+2415, and PSR J1820-0427 could be substantially underestimated due to the fact that their out-of-beam calibrator source exhibits a compact double structure.

\begin{figure*}
\begin{tabular}{cc}
\includegraphics[width=0.48\textwidth]{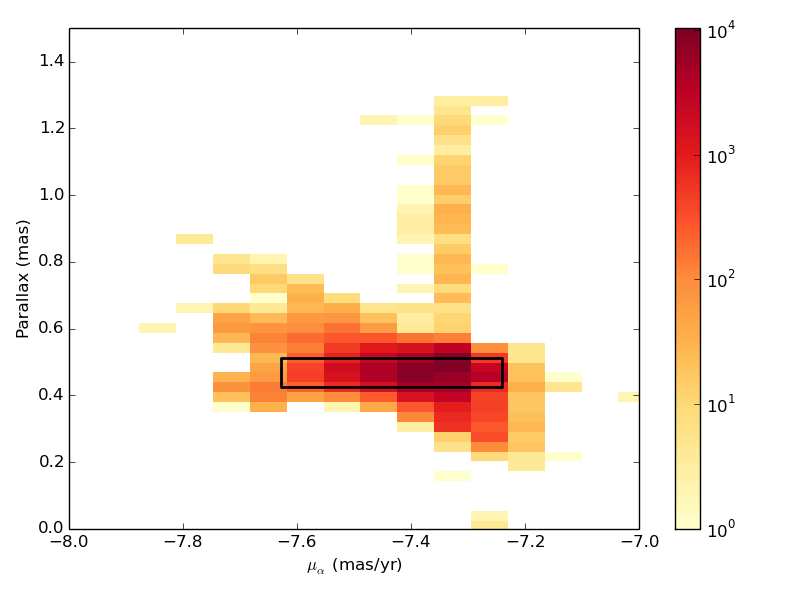} &
\includegraphics[width=0.48\textwidth]{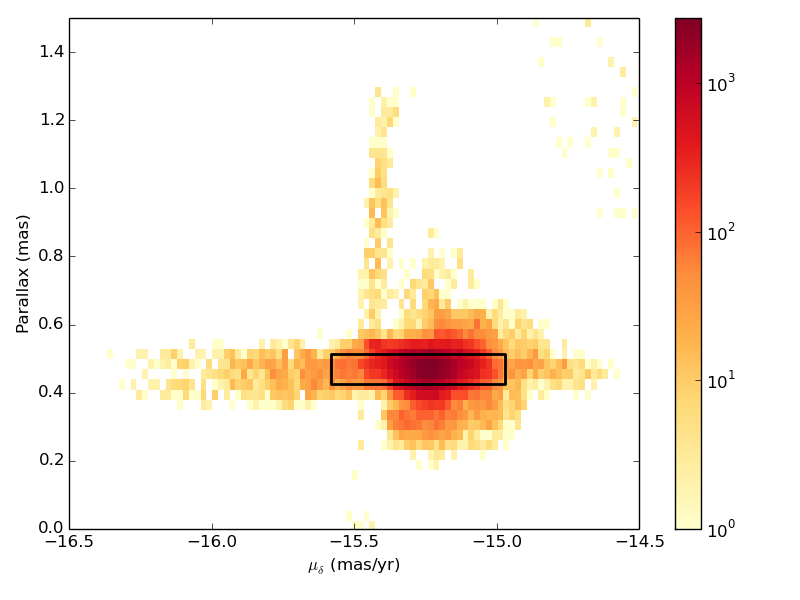}
\end{tabular}
\caption{\label{fig:vt-example}Probability density maps for parallax vs proper motion in right ascension (left) and parallax vs proper motion in declination (right) for PSR J0601$-$0527.  The dark rectangles on the plot show the projections of the cuboid selected for estimating the 68\% confidence interval for transverse velocity. The smallest cuboid contains the most compact 70\% interval along the parallax axis, the most compact 98.5\% interval along the proper motion (R.A.) axis, and the most compact 97\% interval along the proper motion (Dec.) axis, which collectively retains 68\% of the bootstrap trials.}
\end{figure*}

The most probable distance and the 68\% confidence interval were calculated directly from the fitted parallax and confidence interval, without applying any priors based on an assumed pulsar spatial distribution or luminosity distribution \citep[e.g.,][]{verbiest12a,igoshev16a}.  For high-significance parallax detections, the distance is relatively insensitive to the assumed priors, but we note that for low significance parallax detections (for instance, the 20 \psrpi\ pulsars with a parallax significance below 5 $\sigma$) the inferred distance can be substantially dependent on the assumed priors.  The most probable transverse velocity was estimated using the most probable distance and most probable proper motion, while the 68\% confidence interval was calculated by finding the smallest rectangular cuboid in (parallax, proper motion (R.A.), proper motion (Decl.)) space that encompassed 68\% of the bootstrap trial results, and taking the highest and lowest transverse velocity from these included trials.  Figure~\ref{fig:vt-example} shows an example of the transverse velocity estimator for PSR~J0601$-$0527.

\begingroup
\setlength{\tabcolsep}{6pt}
\renewcommand\arraystretch{1.5}
\begin{longtable*}{lllrrr}
\caption{\label{tab:derivedresults}Derived astrometric parameters for all PSR$\pi$ targets.  Positions are shown at the reference epoch of MJD 56000.0.  1$\sigma$ uncertainties are shown on measured values (parentheses show uncertainties on the last digit), and lower limits are shown with 95\% confidence.}\\
\hline
\tabletypesize{\scriptsize}
\tablewidth{1.00\textwidth}
\textbf{Pulsar} & \textbf{Right Ascension} & \textbf{Declination} & \textbf{Dist.} & \textbf{z-height}  & \textbf{Trans. vel.}\\[-2pt]
                 & \textbf{(J2000)}         & \textbf{(J2000)}     & \textbf{(kpc)}    & \textbf{(kpc)}  & \textbf{(km/s)}             \\
\hline
\endfirsthead
\multicolumn{4}{c}%
{\tablename\ \thetable\ -- \textit{Continued from above}} \\
\hline
\textbf{Pulsar} & \textbf{Right Ascension} & \textbf{Declination} & \textbf{Dist.} & \textbf{z-height}  & \textbf{Trans. vel.}\\[-2pt]
                 & \textbf{(J2000)}         & \textbf{(J2000)}     & \textbf{(kpc)}    & \textbf{(kpc)}     & \textbf{(km/s)}             \\
\hline
\endhead
\hline \multicolumn{4}{r}{\textit{Continued below}} \\
\endfoot
\hline
\endlastfoot
J0040+5716 & 00:40:32.3899(1) & +57:16:24.833(1) & $  9.77^{+3.13}_{-3.23}$ & $  0.95^{+0.30}_{-0.31}$ & $ 626.3^{+203.2}_{-221.5}$ \\
J0055+5117 & 00:55:45.3981(1) & +51:17:24.601(1) & $  2.87^{+0.54}_{-0.39}$ & $  0.58^{+0.11}_{-0.08}$ & $ 275.7^{+56.0}_{-42.5}$ \\
J0102+6537 & 01:02:32.9914(1) & +65:37:13.416(1) & $  2.51^{+0.32}_{-0.25}$ & $  0.12^{+0.02}_{-0.01}$ & $ 111.9^{+16.5}_{-12.9}$ \\
J0108+6608 & 01:08:22.5049(3) & +66:08:34.499(1) & $  2.14^{+0.15}_{-0.15}$ & $  0.12^{+0.01}_{-0.01}$ & $ 487.0^{+36.4}_{-35.2}$ \\
J0147+5922 & 01:47:44.6434(1) & +59:22:03.284(1) & $  2.02^{+0.46}_{-0.16}$ & $  0.10^{+0.02}_{-0.01}$ & $  71.1^{+17.4}_{-6.6}$ \\
J0151$-$0635 & 01:51:22.7179(2) & $-$06:35:02.987(2) & $  4.60^{+2.45}_{-1.43}$ & $  4.17^{+2.22}_{-1.30}$ & $ 261.0^{+116.8}_{-93.4}$ \\
J0152$-$1637 & 01:52:10.8539(1) & $-$16:37:53.641(2) & $  2.26^{+1.56}_{-0.73}$ & $  2.15^{+1.48}_{-0.70}$ & $ 335.8^{+219.4}_{-124.4}$ \\
J0157+6212 & 01:57:49.9434(1) & +62:12:26.648(1) & $  1.80^{+0.08}_{-0.12}$ & $  0.01^{+0.00}_{-0.00}$ & $ 383.7^{+17.8}_{-25.8}$ \\
J0323+3944 & 03:23:26.6619(1) & +39:44:52.403(1) & $  0.95^{+0.04}_{-0.03}$ & $  0.24^{+0.01}_{-0.01}$ & $ 183.2^{+7.6}_{-7.0}$ \\
J0332+5434 & 03:32:59.4096(1) & +54:34:43.329(1) & $  1.68^{+0.07}_{-0.06}$ & $  0.04^{+0.00}_{-0.00}$ & $ 158.6^{+7.5}_{-5.6}$ \\
J0335+4555 & 03:35:16.6416(1) & +45:55:53.452(1) & $  2.44^{+0.18}_{-0.12}$ & $  0.34^{+0.02}_{-0.02}$ & $  42.2^{+5.7}_{-2.4}$ \\
J0357+5236 & 03:57:44.8403(2) & +52:36:57.493(1) & $  3.27^{+1.10}_{-0.29}$ & $  0.03^{+0.01}_{-0.00}$ & $ 271.7^{+93.1}_{-30.4}$ \\
J0406+6138 & 04:06:30.0806(2) & +61:38:41.408(1) & $  4.58^{+1.63}_{-0.87}$ & $  0.56^{+0.20}_{-0.11}$ & $ 562.8^{+212.6}_{-107.7}$ \\
J0601$-$0527 & 06:01:58.9752(2) & $-$05:27:50.871(4) & $  2.09^{+0.22}_{-0.16}$ & $  0.49^{+0.05}_{-0.04}$ & $ 167.8^{+23.8}_{-13.4}$ \\
J0614+2229 & 06:14:17.0058(1) & +22:29:56.848(1) & $  3.55^{+0.44}_{-0.26}$ & $  0.15^{+0.02}_{-0.01}$ & $  21.0^{+4.3}_{-2.3}$ \\
J0629+2415 & 06:29:05.7273(1) & +24:15:41.546(1) & $  3.00^{+0.57}_{-0.29}$ & $  0.33^{+0.06}_{-0.03}$ & $  84.0^{+19.0}_{-11.6}$ \\
J0729$-$1836 & 07:29:32.3369(1) & $-$18:36:42.244(2) & $  2.04^{+0.39}_{-0.34}$ & $  0.01^{+0.00}_{-0.00}$ & $ 179.9^{+40.7}_{-36.5}$ \\
J0823+0159 & 08:23:09.7651(1) & +01:59:12.469(1) & $  2.66^{+0.60}_{-0.68}$ & $  0.96^{+0.22}_{-0.25}$ & $  47.8^{+18.9}_{-11.8}$ \\
J0826+2637 & 08:26:51.5068(1) & +26:37:21.297(1) & $  0.50^{+0.00}_{-0.00}$ & $  0.26^{+0.00}_{-0.00}$ & $ 272.3^{+1.3}_{-1.9}$ \\
J1022+1001 & 10:22:57.9957(1) & +10:01:52.765(2) & $  0.72^{+0.01}_{-0.02}$ & $  0.56^{+0.01}_{-0.02}$ & $  54.5^{+1.3}_{-2.0}$ \\
J1136+1551 & 11:36:03.1198(1) & +15:51:14.183(1) & $  0.37^{+0.00}_{-0.00}$ & $  0.35^{+0.00}_{-0.00}$ & $ 659.7^{+4.2}_{-4.5}$ \\
J1257$-$1027 & 12:57:04.7625(2) & $-$10:27:05.551(2) & $  7.09^{+13.18}_{-2.22}$ & $  5.62^{+10.44}_{-1.76}$ & $ 474.1^{+886.4}_{-160.0}$ \\
J1321+8323 & 13:21:45.6315(7) & +83:23:39.432(1) & $  1.03^{+0.17}_{-0.04}$ & $  0.57^{+0.10}_{-0.02}$ & $ 302.6^{+51.6}_{-12.0}$ \\
J1532+2745 & 15:32:10.3646(1) & +27:45:49.623(1) & $  1.60^{+0.29}_{-0.07}$ & $  1.31^{+0.24}_{-0.06}$ & $ 144.0^{+28.8}_{-7.6}$ \\
J1543$-$0620 & 15:43:30.1373(1) & $-$06:20:45.332(2) & $  3.11^{+0.51}_{-0.25}$ & $  1.85^{+0.30}_{-0.15}$ & $ 247.4^{+41.3}_{-22.2}$ \\
J1607$-$0032 & 16:07:12.0598(2) & $-$00:32:41.527(2) & $  1.07^{+0.06}_{-0.03}$ & $  0.62^{+0.03}_{-0.02}$ & $ 193.4^{+13.0}_{-7.0}$ \\
J1623$-$0908 & 16:23:17.6599(1) & $-$09:08:48.733(2) & $  1.71^{+0.34}_{-0.25}$ & $  0.78^{+0.16}_{-0.12}$ & $ 209.4^{+44.0}_{-34.3}$ \\
J1645$-$0317 & 16:45:02.0406(1) & $-$03:17:57.819(2) & $  3.97^{+0.33}_{-0.39}$ & $  1.74^{+0.14}_{-0.17}$ & $ 386.4^{+38.7}_{-43.2}$ \\
J1650$-$1654 & 16:50:27.1694(7) & $-$16:54:42.282(20) & $>$   3.1 & $>$   0.9 & $>$ 229.8 \\
J1703$-$1846 & 17:03:51.0915(2) & $-$18:46:14.845(6) & $  2.88^{+0.45}_{-0.36}$ & $  0.67^{+0.11}_{-0.08}$ & $ 231.5^{+40.4}_{-38.6}$ \\
J1735$-$0724 & 17:35:04.9730(1) & $-$07:24:52.130(1) & $  6.68^{+2.03}_{-1.43}$ & $  1.53^{+0.47}_{-0.33}$ & $ 653.1^{+205.8}_{-144.8}$ \\
J1741$-$0840 & 17:41:22.5629(1) & $-$08:40:31.711(1) & $  3.58^{+0.94}_{-0.55}$ & $  0.70^{+0.18}_{-0.11}$ & $ 116.7^{+35.7}_{-21.8}$ \\
J1754+5201 & 17:54:22.9068(1) & +52:01:12.244(1) & $  6.27^{+1.03}_{-0.98}$ & $  3.10^{+0.51}_{-0.48}$ & $ 122.0^{+22.4}_{-22.2}$ \\
J1820$-$0427 & 18:20:52.5934(1) & $-$04:27:37.712(2) & $  2.85^{+0.52}_{-0.35}$ & $  0.23^{+0.04}_{-0.03}$ & $ 236.0^{+47.0}_{-30.2}$ \\
J1833$-$0338 & 18:33:41.8945(1) & $-$03:39:04.258(1) & $  2.45^{+0.48}_{-0.27}$ & $  0.10^{+0.02}_{-0.01}$ & $ 266.7^{+56.8}_{-34.0}$ \\
J1840+5640 & 18:40:44.5372(1) & +56:40:54.852(1) & $  1.52^{+0.02}_{-0.14}$ & $  0.62^{+0.01}_{-0.06}$ & $ 307.9^{+5.2}_{-27.8}$ \\
J1901$-$0906 & 19:01:53.0087(3) & $-$09:06:11.146(10) & $  1.96^{+0.17}_{-0.23}$ & $  0.22^{+0.02}_{-0.03}$ & $ 183.1^{+20.4}_{-24.3}$ \\
J1912+2104 & 19:12:43.3391(1) & +21:04:33.926(1) & $ 41.02^{+377.75}_{-35.90}$ & $  3.57^{+32.86}_{-3.12}$ & $2482.4^{+6732.3}_{-2184.1}$ \\
J1913+1400 & 19:13:24.3527(1) & +14:00:52.559(1) & $  5.42^{+0.75}_{-0.70}$ & $  0.15^{+0.02}_{-0.02}$ & $ 266.3^{+39.5}_{-37.6}$ \\
J1917+1353 & 19:17:39.7864(1) & +13:53:57.077(1) & $  7.04^{+0.38}_{-2.29}$ & $  0.08^{+0.00}_{-0.02}$ & $ 134.5^{+9.7}_{-47.8}$ \\
J1919+0021 & 19:19:50.6715(1) & +00:21:39.722(2) & $  6.03^{+2.02}_{-1.21}$ & $  0.65^{+0.22}_{-0.13}$ & $ 319.6^{+107.2}_{-72.3}$ \\
J1937+2544 & 19:37:01.2544(1) & +25:44:13.436(1) & $  3.15^{+0.32}_{-0.28}$ & $  0.12^{+0.01}_{-0.01}$ & $ 245.9^{+25.1}_{-24.4}$ \\
J2006$-$0807 & 20:06:16.3650(1) & $-$08:07:02.167(3) & $  2.36^{+0.73}_{-0.06}$ & $  0.82^{+0.25}_{-0.02}$ & $ 137.1^{+45.4}_{-8.5}$ \\
J2010$-$1323 & 20:10:45.9211(1) & $-$13:23:56.083(4) & $  2.07^{+0.68}_{-0.53}$ & $  0.83^{+0.27}_{-0.21}$ & $  58.7^{+26.9}_{-16.0}$ \\
J2046+1540 & 20:46:39.3373(1) & +15:40:33.558(1) & $  3.22^{+1.04}_{-0.68}$ & $  0.93^{+0.30}_{-0.20}$ & $ 160.0^{+48.2}_{-35.4}$ \\
J2046$-$0421 & 20:46:00.1730(1) & $-$04:21:26.256(2) & $  5.98^{+2.00}_{-0.81}$ & $  2.75^{+0.92}_{-0.37}$ & $ 329.3^{+112.3}_{-50.6}$ \\
J2113+2754 & 21:13:04.3506(1) & +27:54:01.160(1) & $  1.42^{+0.04}_{-0.04}$ & $  0.34^{+0.01}_{-0.01}$ & $ 412.1^{+14.2}_{-13.8}$ \\
J2113+4644 & 21:13:24.3295(1) & +46:44:08.844(1) & $  2.20^{+0.36}_{-0.32}$ & $  0.05^{+0.01}_{-0.01}$ & $ 135.8^{+18.4}_{-24.8}$ \\
J2145$-$0750 & 21:45:50.4588(1) & $-$07:50:18.514(4) & $  0.62^{+0.00}_{-0.02}$ & $  0.42^{+0.00}_{-0.02}$ & $  38.9^{+0.5}_{-1.9}$ \\
J2149+6329 & 21:49:58.7033(2) & +63:29:44.277(2) & $  2.81^{+0.58}_{-0.47}$ & $  0.36^{+0.07}_{-0.06}$ & $ 258.3^{+52.6}_{-49.4}$ \\
J2150+5247 & 21:50:37.7499(1) & +52:47:49.556(1) & $>$   2.4 & $>$   0.0 & $>$  89.1 \\
J2212+2933 & 22:12:23.3444(1) & +29:33:05.411(1) & $  3.77^{+3.14}_{-0.60}$ & $  1.40^{+1.16}_{-0.22}$ & $ 225.7^{+193.8}_{-38.2}$ \\
J2225+6535 & 22:25:52.8627(3) & +65:35:36.371(1) & $  0.83^{+0.17}_{-0.10}$ & $  0.10^{+0.02}_{-0.01}$ & $ 765.2^{+157.6}_{-94.5}$ \\
J2248$-$0101 & 22:48:26.8859(1) & $-$01:01:48.085(1) & $  3.90^{+1.40}_{-0.63}$ & $  3.02^{+1.08}_{-0.49}$ & $ 377.1^{+149.2}_{-66.7}$ \\
J2305+3100 & 23:05:58.3212(1) & +31:00:01.281(1) & $  4.47^{+0.65}_{-0.58}$ & $  2.01^{+0.29}_{-0.26}$ & $ 341.1^{+56.4}_{-47.2}$ \\
J2317+1439 & 23:17:09.2364(1) & +14:39:31.265(1) & $  1.66^{+1.10}_{-1.19}$ & $  1.12^{+0.74}_{-0.80}$ & $  31.4^{+30.0}_{-22.9}$ \\
J2317+2149 & 23:17:57.8419(1) & +21:49:48.019(1) & $  1.96^{+0.21}_{-0.20}$ & $  1.16^{+0.12}_{-0.12}$ & $  79.3^{+8.9}_{-9.6}$ \\
J2325+6316 & 23:25:13.3196(2) & +63:16:52.362(1) & $>$  12.1 & $>$   0.4 & $>$ 327.8 \\
J2346$-$0609 & 23:46:50.4978(1) & $-$06:09:59.899(2) & $  3.64^{+0.55}_{-0.26}$ & $  3.27^{+0.50}_{-0.23}$ & $ 732.5^{+114.5}_{-53.1}$ \\
J2354+6155 & 23:54:04.7830(1) & +61:55:46.845(1) & $  2.42^{+0.28}_{-0.17}$ & $  0.01^{+0.00}_{-0.00}$ & $ 268.0^{+32.7}_{-20.3}$ \\
% Produced on bunker in /home/deller/data/vlbi/psrpi/final by ./makePulsarPages.py
\end{longtable*}
\endgroup

\subsection{Analysing the astrometric error budget}

As shown above, correctly estimating the total uncertainty of the position measurements used for the astrometric fit is challenging.  Below, we summarise the primary contributions to the error budget:

\label{sec:errorbudget}

\begin{enumerate}
\item {\em Thermal noise in the target image:} This is the most readily quantified, as it can be easily extracted from the image-plane fitting.  It is inversely proportional to the instrumental resolution and inversely proportional to the signal-to-noise in the pulsar image.  This term generally dominates for faint sources.
\item {\em Systematic offsets introduced by  differential propagation effects between the target and calibrator}: This is usually the dominant term for bright sources, where the signal-to-noise on the target is not the limiting factor.  At 1600~MHz, the ionosphere dominates these path length differences, which vary on a sub-epoch timescale (minutes to hours).  The solutions on the calibrator must be extrapolated spatially to the target, and are  averaged over a time interval during which the ionosphere can change.  Generally, the spatial extrapolation introduces the largest error, meaning this term is most dependent on the calibrator-target separation, along with factors influencing the mean path length through the ionosphere such as the solar activity level, time of day, and antenna elevation.  The amount of temporal averaging required depends on the calibrator flux density and structure -- for a typical 20 mJy calibrator, a solution interval of order 1.25 minutes was typical.

Refraction in the interstellar medium also produces a differential offset between a pulsar and calibrator.  For the \psrpi\ sample, the predicted angular wandering \citep[using the predictions of the NE2001 model and assuming a Kolmogorov density spectrum;][]{cordes02a} at the observing frequency (1660 or 2270 MHz) for our target pulsars due to refraction has a median value for a given observation of $\sim 0.05$~mas.  For most pulsars, this is a negligible component of the astrometric error.  However, the predicted scattering disk diameter exceeds 1 mas and the predicted refractive wander exceeds 0.1 mas for four pulsars in our sample: PSR~J0601$-$0527, PSR~J1833$-$0338, PSR~J2212+2933, and PSR J2325+6313.  In cases such as these, and others where the astrometric precision is extremely high, refractive wander of the pulsar may be a significant component of the error budget.

As well as the pulsar, refractive wander also affects the in-beam calibrator sources, which is another potential source of error in the target--calibrator separation.  Generally, the refractive wander is larger for calibrators than for the pulsars, as the radiation from the calibrators passes by all of the Galactic electrons along the line of sight, leading to a larger scattering disk.  For PSR~J1833$-$0338, for instance, the NE2001 model predicts refractive wander with an rms deviation of 0.4 mas for the position reference calibrator source, nearly three times larger than that of the pulsar.  However, the refractive wander timescale for the calibrator sources is typically much longer (years), meaning that reference position will be affected more than proper motion, which will itself be affected more than parallax.
\item {\em Systematic variations in the image reference frame:} An imperfect model of the calibrator source will lead to an offset in the obtained pulsar position.  If this is constant in time, it does not impact the measurement of parallax or proper motion, but time variability is an important source of error.  Time variability could be the result of evolution intrinsic to the source itself (which is present, at least at a low level, in all compact sources), or from changes in the observing setup (different frequency or $uv$ coverage between observing epochs).  This can be the dominant term if the calibrator source is bright and close to the target (minimizing the ionospheric terms) but is a blazar-like source which displays large and rapid variations in the jet structure.  Over the 1~yr--2~yr timescale typical for pulsar astrometry programs such as \psrpi, it is often possible to fit a significant component of this reference source offset with a linear function with time, meaning it can corrupt the proper motion measured for the pulsar.  However, for most reference sources the likely effect is small compared to our measurement error \citep[the median apparent proper motion seen by ][ was 19\,\uas\ yr$^{-1}$, versus 106\,\uas\ yr$^{-1}$ for our relative astrometric uncertainty]{moor11a}, and parallax (which has a sinusoidal signature with time) is much less affected.  As well as effects intrinsic to the source, time-variable position shifts due to the changing Galactic gravitational potential field can be expected \citep{r:lar17}, albeit only at the level of up to $\sim$10 \uas\ over our timescales, and hence smaller than the reference source structure effects.
\item {\em Stochastic noise in the image reference frame:} The phase solutions on the calibrator source will have some noise dependent on the signal-to-noise in each solution interval, which is determined by the source flux density, instrumental sensitivity, and calibration interval.  Fainter calibrators will generally lead to an increase in this term; although this could be compensated by increasing the solution interval, that would be reduce the ability to compensate for time-variable ionospheric effects.  Changing the solution interval can thus shift error between this term and term 2 above; we seek to choose a value on a per-source basis that minimises their sum.
\end{enumerate}

Accordingly, there are five main factors we would expect to influence the total uncertainty of a position measurement in a given epoch:
\begin{enumerate}
\item Pulsar flux density
\item Calibrator flux density
\item Pulsar-calibrator angular separation
\item Ionospheric conditions and average observing elevation
\item Calibrator stability
\end{enumerate}

We can straightforwardly measure all but the last of these factors, and as noted the calibrator stability is not usually expected to be a significant contributor to the total error budget.  We undertook a number of approaches to try and determine the contributions of each of the remaining factors to our error budget.  To illustrate the results, we again use the typical-to-challenging source PSR~J0601$-$0527, where the thermal noise errors alone substantially underestimate the total error budget, as can be seen from the reduced $\chi^2 \sim 15$ for a simple least-squares fit to the unmodified position data.  

First, we examined the apparent shifts in pulsar position within an observation, by subdividing the data into two halves and imaging each one separately.  This approach is particularly useful for identifying observations where short-term ionospheric conditions were unstable---if the two positions differ by considerably more than their formal error bars, it is likely that the mean position over the whole observation also has an underestimated position uncertainty.  However, while reliable, this approach is likely not complete, as it would fail to pick up observations with large but relatively stable residual ionospheric ``wedges" that lead to a fairly constant position offset over the whole observation duration. Also, when the target source is faint, making a significant measurement of the offset between the two halves of the observation may not be possible.

\begin{figure*}
\begin{tabular}{cc}
% produced on bunker with pylab_pmparplot.py -f local.jmfit.pmpar.in --plottype=pdf --target=J0601-0527 ; mv ratime_nopm.J0601-0527.pdf J0601-0527.full.ratime.pdf ; scp bunker:data/vlbi/psrpi/final/J0601-0527/pulsar/full/J0601-0527.full.ratime.pdf .
\includegraphics[width=0.48\textwidth]{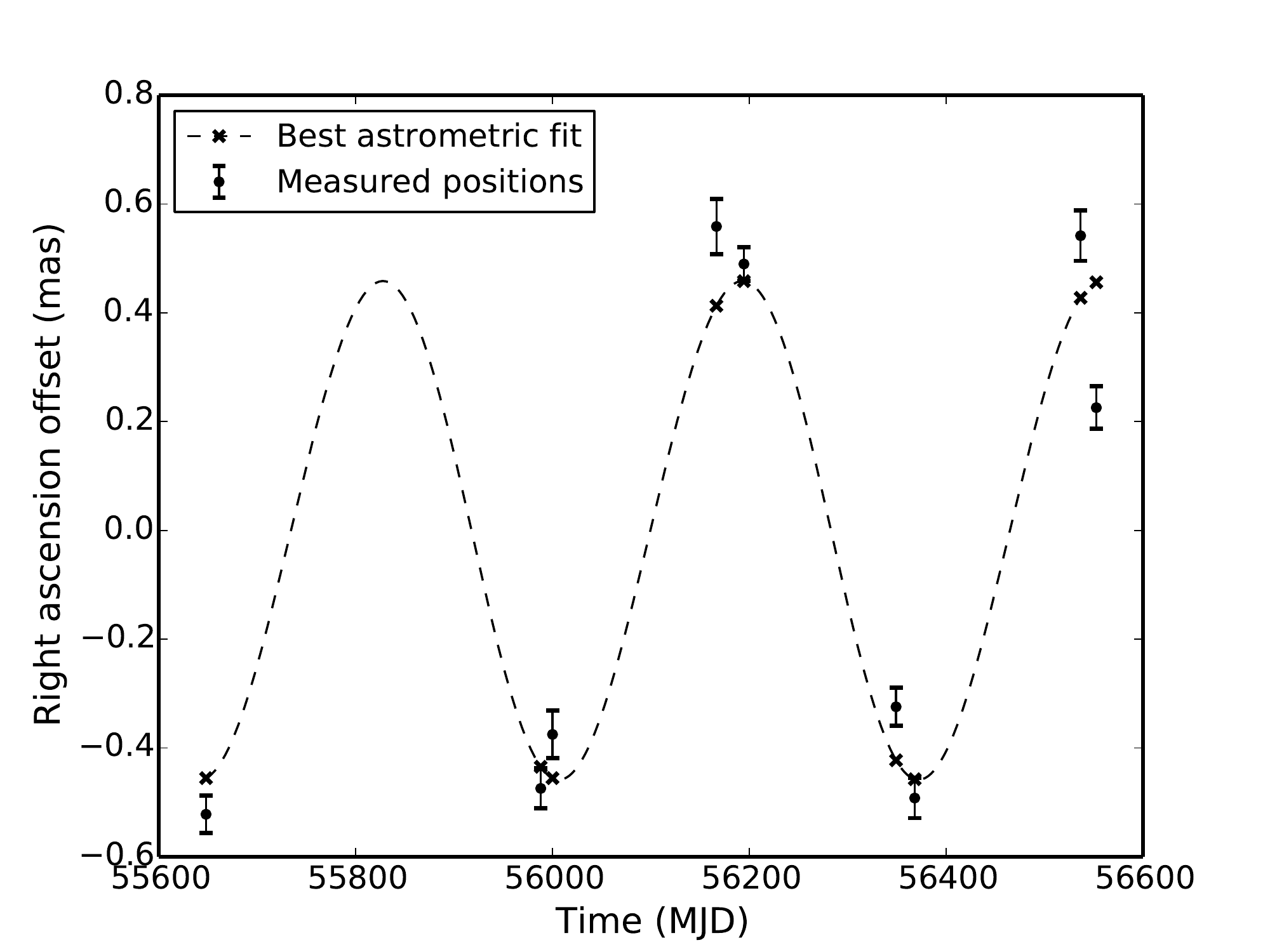} &
% produced on bunker with pylab_pmparplot.py -f local.jmfit.pmpar.in --plottype=pdf --target=J0601-0527 --legendlocation='upper left' --ralim2=-0.8,1.2 --file2=halves.jmfit.pmpar.in; mv ratime_nopm.J0601-0527.pdf J0601-0527.halves.ratime.pdf ; scp bunker:data/vlbi/psrpi/final/J0601-0527/pulsar/full/with-half-systematic/J0601-0527.halves.ratime.pdf .
\includegraphics[width=0.48\textwidth]{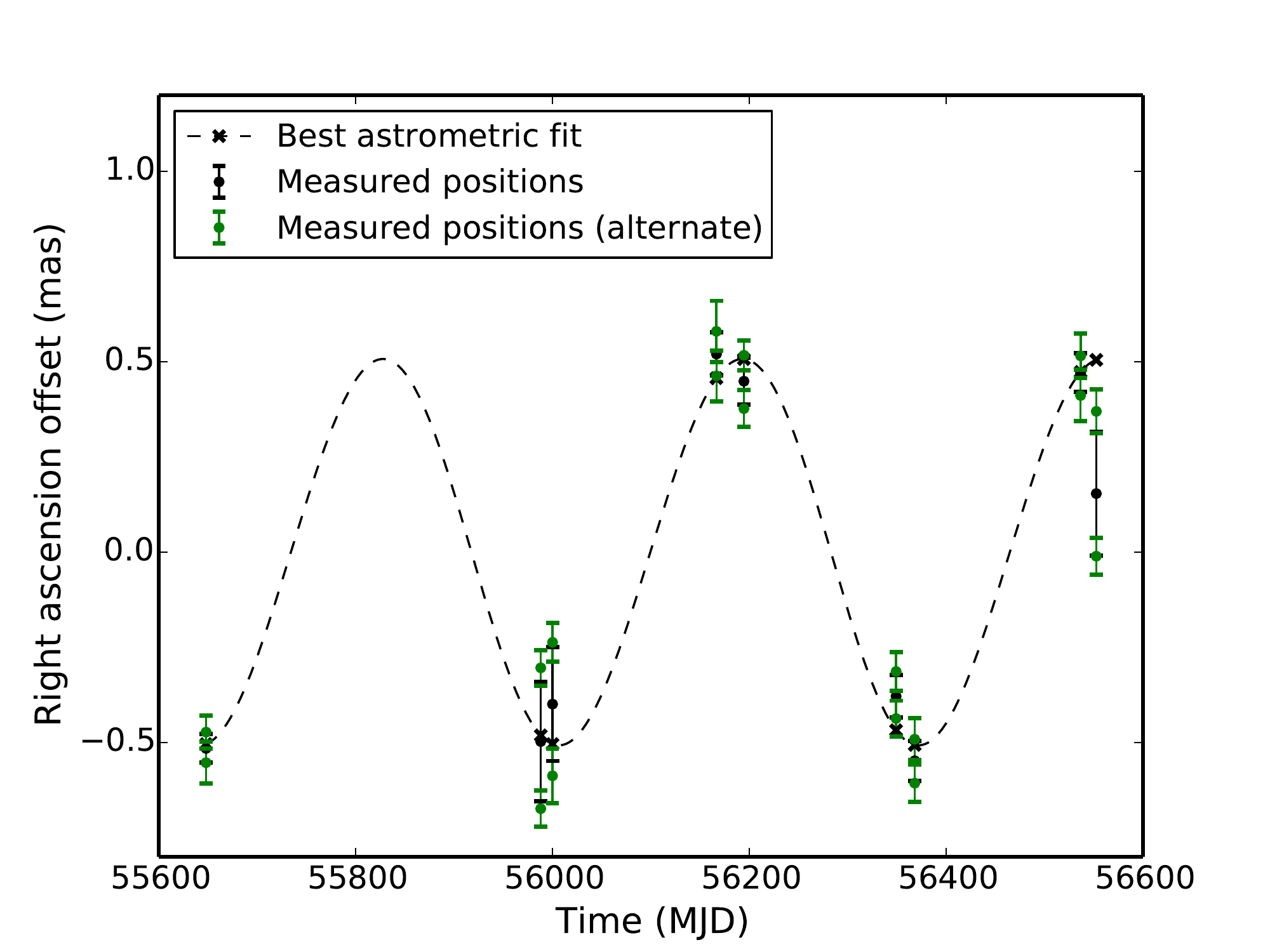} \\
% scp bunker:data/vlbi/psrpi/final/J0601-0527/pulsar/full/bootstrap_parallax.png J0601-0527.full.bootstrapparallax.png
\includegraphics[width=0.48\textwidth]{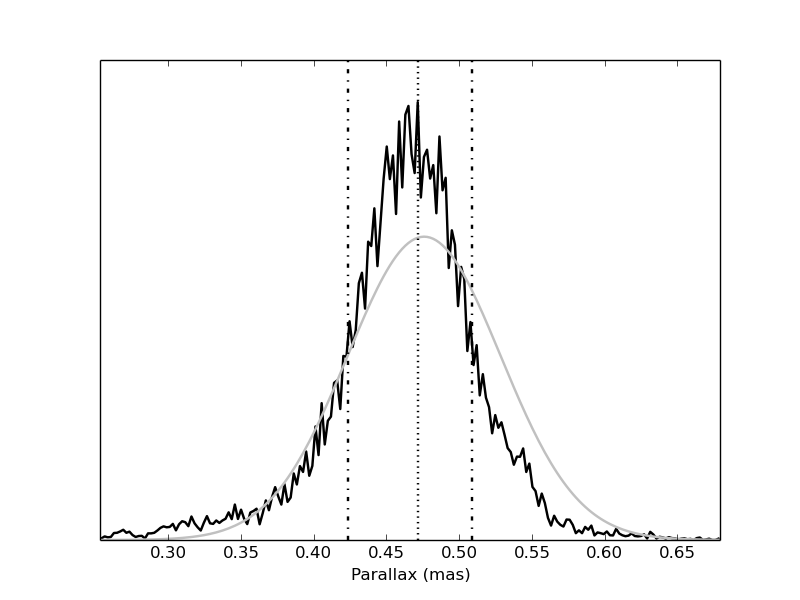} &
% scp bunker:data/vlbi/psrpi/final/J0601-0527/pulsar/full/with-half-systematic/bootstrap_parallax.png J0601-0527.halvessystematic.bootstrapparallax.png
\includegraphics[width=0.48\textwidth]{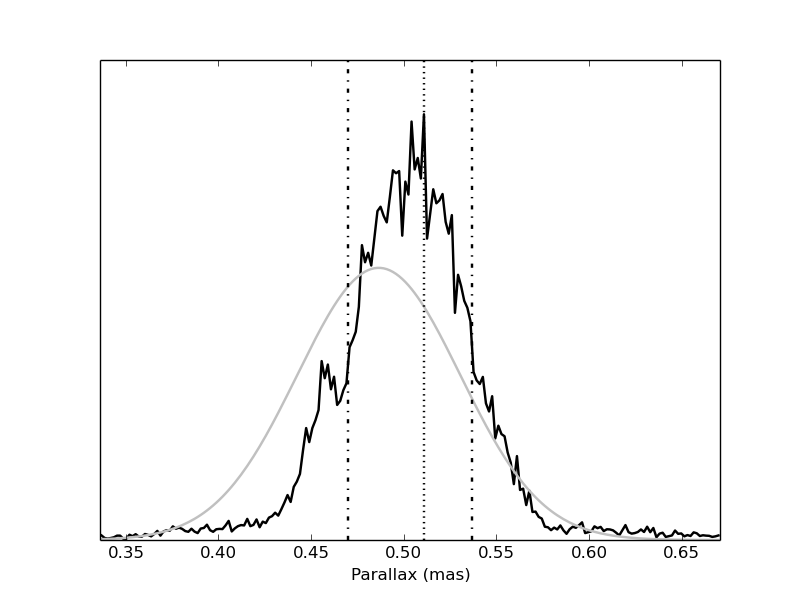}
\end{tabular}
\caption{The effect of including an estimate of the systematic error based on apparent positional wander during an observation for PSR~J0601$-$0527.  The plots in the first row show the offset in right ascension as a function of time, after subtracting the best-fit proper motion.  Top left: least-squares fit to dataset without systematic errors.  The reduced $\chi^2$ is 15.  Top right: least squares fit to dataset including systematic error estimate.  The measurements with error bars in green show the positions obtained from each half-observation---these are not fit, but show the size of the potential systematic offserts.  The reduced $\chi^2$ is 10. Bottom left: the bootstrap fit results for parallax, from the dataset without systematic error estimates.  The parallax obtained is $0.47 \pm 0.04$ mas.  Bottom right: the bootstrap fit results for parallax, from the dataset with systematic error estimates.  The parallax obtained is $0.51^{+0.03}_{-0.04}$ $\mu$as.}
\label{fig:twohalves}
\end{figure*}

We evaluated an approach in which we recorded the minimum systematic offset between the two observation halves (accounting for the uncertainty in the position measurements) and set the systematic error contribution to the whole epoch to be half of this value.  The right ascension and declination axes are treated separately.  The effect, as expected, was to lower the chi-squared of the resultant astrometric fit, although the position errors remained underestimated (as determined by a $\chi^2$ value well in excess of 1.0) in many cases.  Figure~\ref{fig:twohalves} shows the result for PSR~J0601$-$0527.  As expected, this approach yields a systematic error estimate that is too low; the reduced $\chi^2$ remains at 10.  Using this refined position set as the input for bootstrap fits results in a change in the best-fit value at the 1$\sigma$ level, and gives a small reduction in the estimated parameter uncertainty.  For most targets, the impact on both best-fit value and uncertainty was smaller than in this example.

Second, we processed the datasets multiple times making use of different ionospheric models, and examined the resultant position scatter for each epoch.  If different ionospheric models of comparable quality give widely divergent positions, then the residual error from our chosen ionospheric model is likely high, since we have no way of determining which of the ionospheric models is correct.  As with the previous approach comparing the two halves of an observation in time, this method is likely reliable but not necessarily complete, since it will not pick up cases where all models suffer from the same deficiencies.

We investigated all of the products covering our complete observing timespan from \url{ftp://cddis.gsfc.nasa.gov/gps/products/ionex/}, and found that the {\tt jplg}, {\tt codg}, {\tt igsg}, and {\tt esag} models consistently gave the best results, with the lowest residuals on the astrometric fit.  The {\tt igsg} model was chosen for the final data reduction.  We therefore trialed an appoach in which, for every epoch, we computed the rms scatter in the positions provided by processing using these 4 models, and used this as the estimate for systematic error for that epoch.  As with the epoch-splitting approach above, the right ascension and declination axes were treated separately.  We found that for many sources, a given ionospheric model yielded a statistically significant mean offset across all epochs in addition to random scatter, and so for all sources we subtracted (per model) the mean positional offset from all epochs before computing the rms.

\begin{figure*}
\begin{tabular}{cc}
% produced on bunker with pylab_pmparplot.py -f local.jmfit.pmpar.in --plottype=pdf --target=J0601-0527 --legendlocation='upper left'; mv ratime_nopm.J0601-0527.pdf J0601-0527.ionosystematic.ratime.pdf ; scp bunker:data/vlbi/psrpi/final/J0601-0527/pulsar/full/withionosystematic/J0601-0527.ionosystematic.ratime.pdf .
\includegraphics[width=0.48\textwidth]{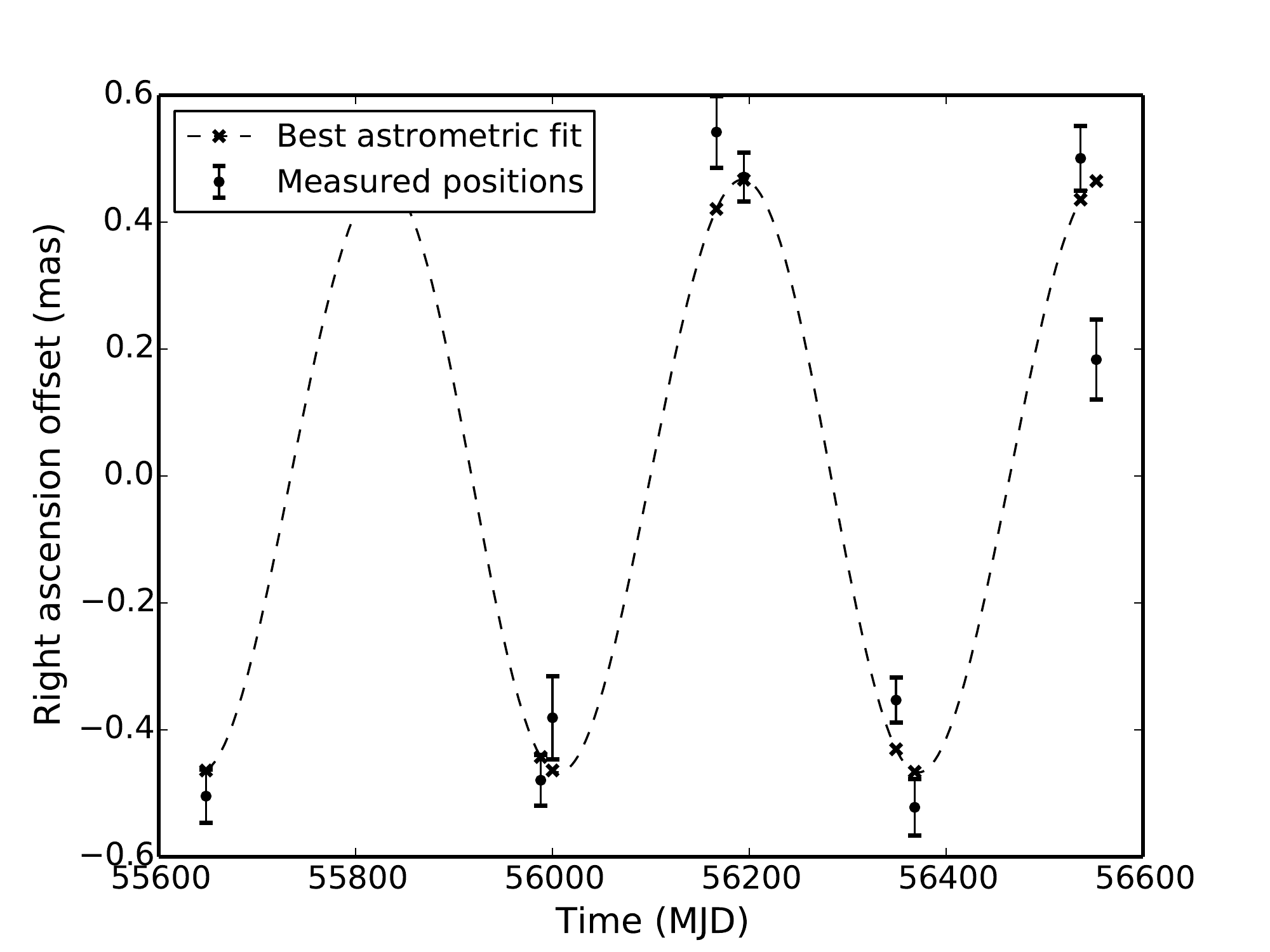} &
% scp bunker:data/vlbi/psrpi/final/J0601-0527/pulsar/full/withionosystematic/bootstrap_parallax.png J0601-0527.ionosystematic.bootstrapparallax.png
\includegraphics[width=0.48\textwidth]{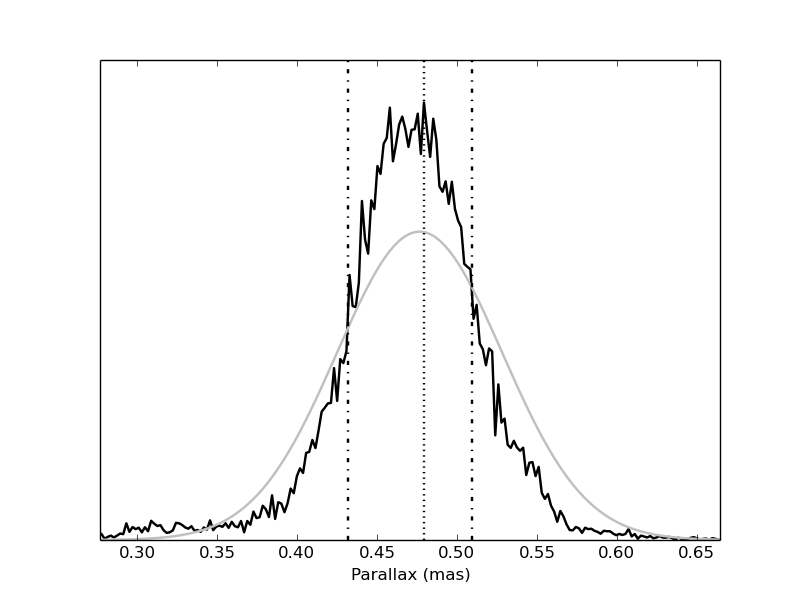}
\end{tabular}
\caption{The effect of including an estimate of the systematic error based on the scatter between ionospheric models for PSR~J0601$-$0527.  Left: least-squares fit to dataset with systematic error estimates included: offset in right ascension is shown as a function of time, after subtracting the best-fit proper motion.  The reduced $\chi^2$ is 8.  Right:  the bootstrap fit results for parallax, from the dataset with systematic error estimates. The parallax obtained is $0.47 \pm 0.04$ $\mu$as.}
\label{fig:ionosystematic}
\end{figure*}

As expected and as with the previous approach splitting the observation in halves, and as expected, we typically capture some but not all of the systematic error with this technique.  Figure~\ref{fig:ionosystematic} shows the result for PSR~J0601$-$0527; the reduced $\chi^2$ is still 8.  The best-fit value for parallax changes by approximately 0.2$\sigma$, and the estimated parallax uncertainty from the bootstrap fit is reduced by $\sim$10\%.

Our third approach to estimate systematic error made use of the residual position errors across our entire data set of 60 pulsars.  As shown above, we expect the dominant error sources to depend on the calibrator-target separation, the observing elevation, and the calibrator brightness.  As a simplification, we consider that the systematic error should be proportional to two quantities: the mean ``deprojected" calibrator-target separation (calculated as the angular separation multiplied by the cosecant of the observing elevation, averaged over all antennas and all scans in the observation) and the signal-to-noise ratio achieved on the inbeam calibrator source(s).  For each epoch, we added a systematic error estimate given by:

\begin{equation}
\label{eq:apriorsyserror}
\Delta_{\mathrm{sys}} = A \times \frac{s \times \sum_{a=1}^{N} \sum_{o=1}^{M}  \mathrm{cosec} \left( \mathrm{el_{a,o}} \right) }{M\times N} + B / S
\end{equation}
where $\Delta_{\mathrm{sys}}$ is the systematic error estimate in fractions of a synthesized beam, $s$ is the calibrator-target separation in arcminutes, $\mathrm{el}_{a,o}$ is the observing elevation for antenna $a$ in scan $o$ on the target pulsar, $N$ and $M$ are the number of antennas and target scans respectively, and $S$ is the signal-to-noise ratio on the calibrator source (added in quadrature if multiple sources were used).  We conducted a brute-force grid search for the optimal values of the coefficients $A$ and $B$, seeking the values that gave the tightest grouping of reduced $\chi^2$ values around 1.0 for our ensemble of 60 target pulsars.

\begin{figure*}
\begin{tabular}{cc}
% produced on bunker with pylab_pmparplot.py -f local.jmfit.pmpar.in --plottype=pdf --target=J0601-0527 --legendlocation='upper left'; mv ratime_nopm.J0601-0527.pdf J0601-0527.apriorisystematic.ratime.pdf ; scp bunker:data/vlbi/psrpi/final/J0601-0527/pulsar/full/withsystematic/J0601-0527.apriorisystematic.ratime.pdf .
\includegraphics[width=0.48\textwidth]{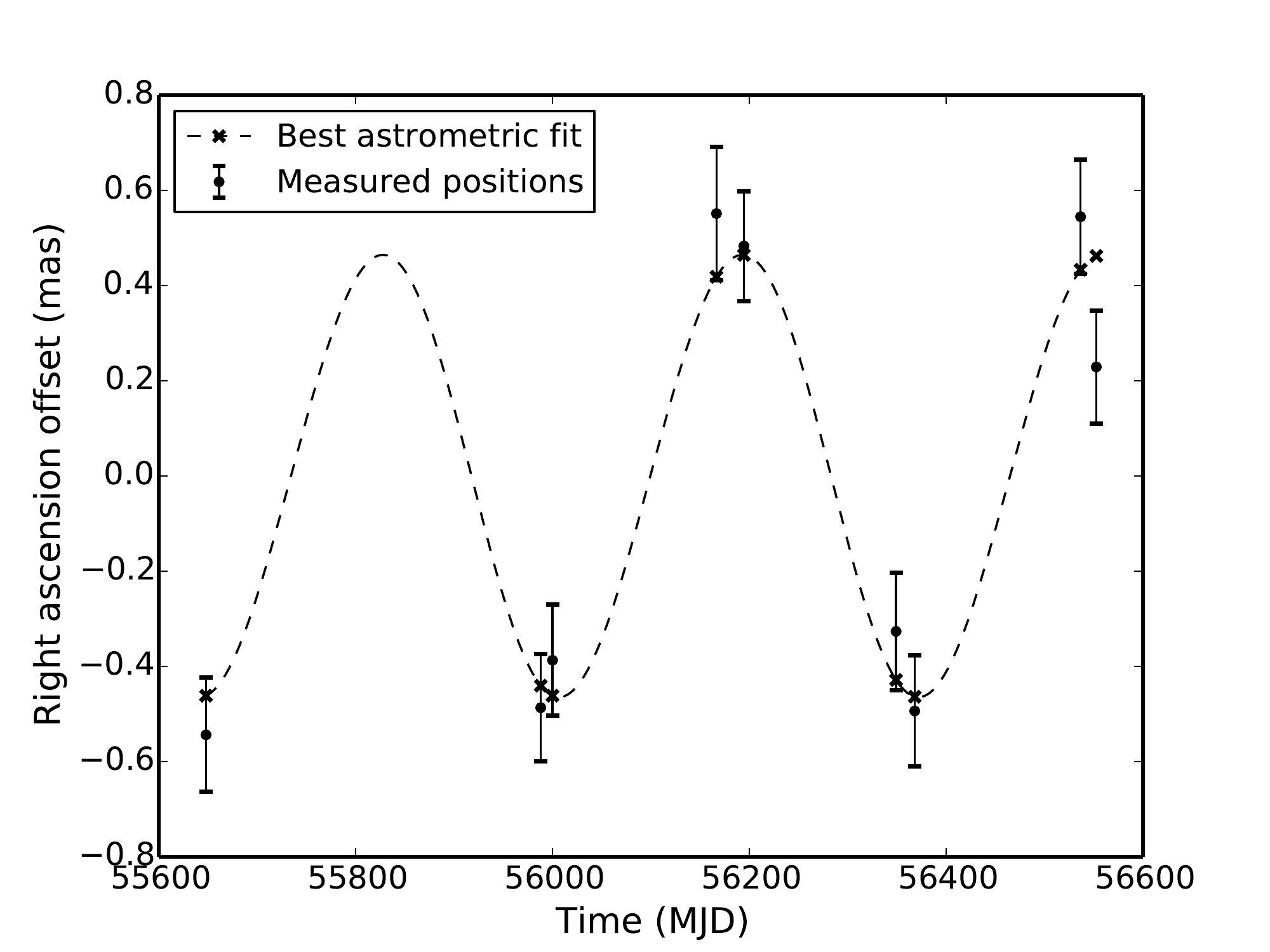} &
% scp bunker:data/vlbi/psrpi/final/J0601-0527/pulsar/full/withsystematic2/bootstrap_parallax.png J0601-0527.apriorisystematic.bootstrapparallax.png
\includegraphics[width=0.48\textwidth]{J0601-0527.apriorisystematic.bootstrapparallax.png}
\end{tabular}
\caption{The effect of including an empirical estimate of the systematic error based on the calibrator-target separation and calibrator flux density.  Left: least-squares fit to dataset with systematic error estimates included: offset in right ascension is shown as a function of time, after subtracting the best-fit proper motion.  The reduced $\chi^2$ is 1.4.  Right:  the bootstrap fit results for parallax, from the dataset with systematic error estimates. The parallax obtained is $0.48 \pm 0.04$ $\mu$as.}
\label{fig:apriorisystematic}
\end{figure*}

The optimal values were found to be $A = 0.001$, $B = 0.6$.  When this estimate of systematic error is added for all pulsars, the 25\%, median, and 75\% values of reduced $\chi^2$ across all pulsars becomes 0.64,1.08, and 1.69, compared to 1.78, 4.81, and 11.96 when no estimate of systematic error is added.  The spread in reduced $\chi^2$ values is comparable to that expected given the typical number of degrees of freedom ($\sim$11) in the astrometric fits.  The results for our example pulsar PSR~J0601$-$0527 are shown in Figure~\ref{fig:apriorisystematic}; the typical systematic error contribution is 110 \uas\ in right ascension and 280 \uas\ in declination at each epoch, and the revised reduced $\chi^2$ is 1.35.

We summarise the results of our different estimates of systematic error for PSR~J0601$-$0527 in Table~\ref{tab:systematicomparison}.  Three things are immediately apparent:

\begin{enumerate}
\item The inclusion of systematic error (however estimated) pushes the estimated parallax in one direction.  When the position errors are severely underestimated (as they are initially), individual discrepant epochs can have overly large effects on the fit.  This diminishes once a more realistic error is applied.
\item The fitted parameters and their uncertainties remain relatively unchanged regardless of the systematic error estimate used when estimated using a bootstrap.
\item The fitted parameters and their uncertainties exhibit good agreement between a bootstrap and a simple least-squares fit when the systematic errors are reasonably well estimated (as appears to be the case for the empirically estimated values).
\end{enumerate}

For our quoted results, we choose to use the bootstrap fit to the dataset including empirically estimated systematic errors, which is generally the most conservative (and we believe) correct error estimate we can make with our available information.  We do however note that in some cases, generally when the parallax has been poorly sampled due to non-detections, this bootstrap error estimate may be overly conservative (since many trials have effectively no sensitivity to parallax).  In these cases, better constraints could be obtained by interpreting, with caution, the least-squares fit to the dataset incorporating empirically estimated systematic errors.  We highlight this for individual pulsars in the discussion that follows in Section~\ref{sec:individualpulsars}.

\begin{deluxetable}{lccc}
\tabletypesize{\small}
%\tablewidth{0pt}
\tablecaption{Comparison between systematic error estimators}
\tablehead{
\colhead{Estimator} 	& \colhead{Reduced $\chi^2$} 	& \colhead{Change in fitted } &\colhead{Relative parallax}  \\
\colhead{} 		& \colhead{} 				& \colhead{parallax (mas)\tablenotemark{A}} &\colhead{uncertainty\tablenotemark{A}}
}
\startdata
None		& 15.2 	& 0.000	& 1.00 	 \\
Time division	& 10.2 	& 0.029	& 0.79 	 \\
Ionosphere	& 7.8 	& 0.010	& 0.91 	 \\
Empirical 		& 1.4 	& 0.013	& 0.99
\enddata
\tablenotetext{A}{Compared to the reference case of no systematic error estimate, using the results for PSR~J0601$-$0527.  The relative parallax uncertainty is obtained by dividing the size of the 68\% confidence interval by that of the reference case.}
\label{tab:systematicomparison}
\end{deluxetable}

Finally, since the unmodeled ionosphere dominates the error budget in many cases, we should expect that the level of solar activity should significantly impact the results obtained.  While the solar cycle peaking in 2013 was not particularly active by historical standards, our observations were nevertheless made near the solar maximum, and accordingly we would expect that the same observations repeated a half-decade later would yield better results.  Likewise, the precise results seen in, e.g., \citet{deller12b} and \citet{deller13a} might have been more difficult to obtain at the time of the observations presented here.  Importantly, our empirical estimates of systematic error should be used with caution when applied to observations in different observing conditions.  An even larger observing program might consider including a measure of ionospheric activity as a parameter in the empirical fit to account for this.

\section{Discussion}
\label{sec:discussion}

\subsection{Notes on individual pulsars}
\label{sec:individualpulsars}
From our sample of 60 pulsars, three sources display discrepancies which indicate a potentially biased parallax estimation.  We consider the results for these sources in detail and estimate the probability that any of the remaining 57 sources have comparable but undetected errors.

\subsubsection{PSR J1650--1654}
PSR J1650--1654 has a significant negative parallax of $-0.089^{+0.030}_{-0.019}$ mas (where the uncertainty denotes the 68\% confidence interval from the bootstrap fit).  Since a negative parallax is unphysical, this indicates that the obtained value is incorrect by at least 3$\sigma$, but possibly more as the NE2001 distance obtained using the pulsar's dispersion measure is just 1.5 kpc.

\begin{figure*}
\begin{tabular}{cc}
\includegraphics[width=0.43\textwidth]{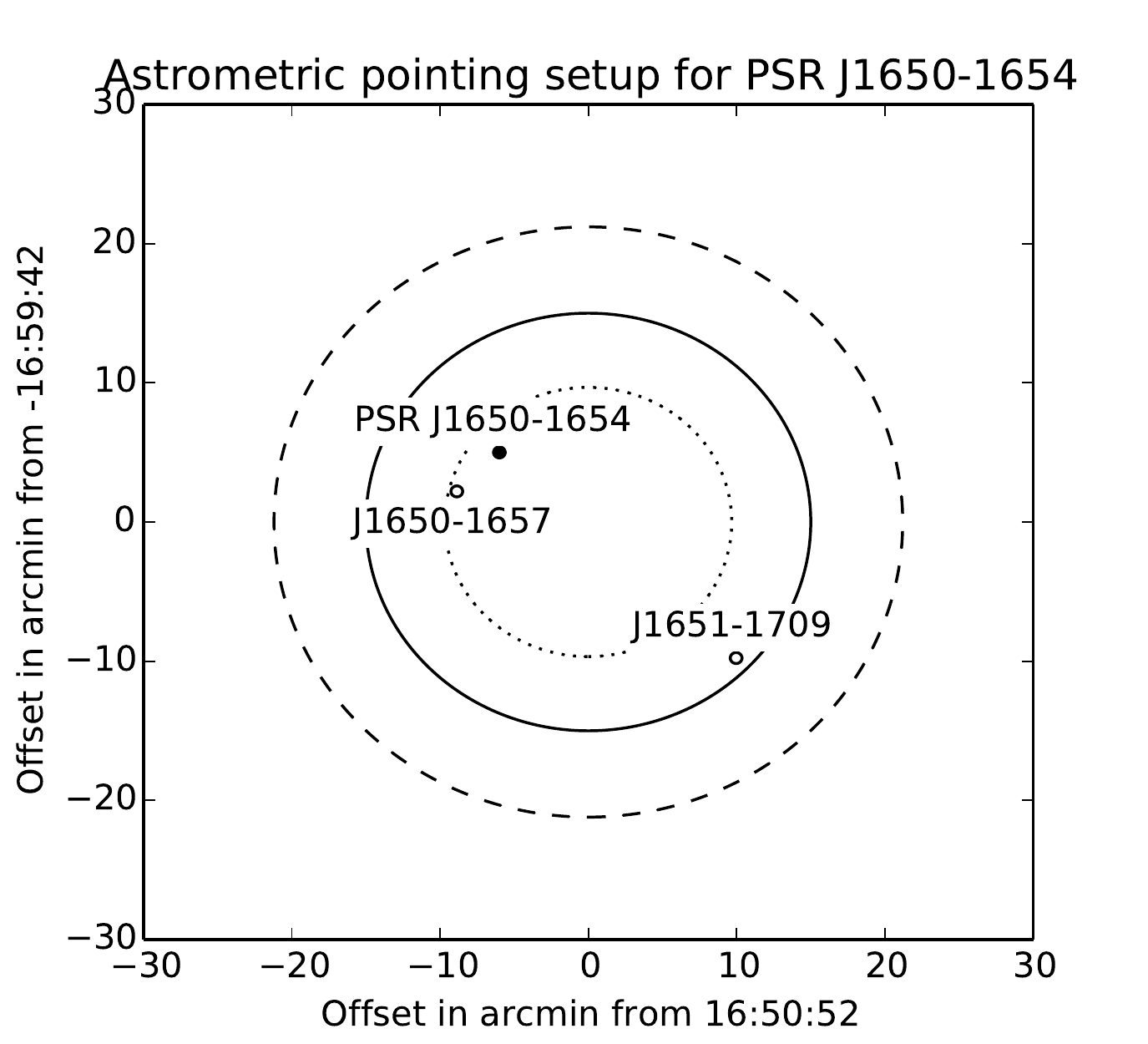} & 
% Made on bunker in /home/deller/data/vlbi/psrpi/final/pointingplots/temp/J1650/, using make_oneoff_beamplot.py
\includegraphics[width=0.53\textwidth]{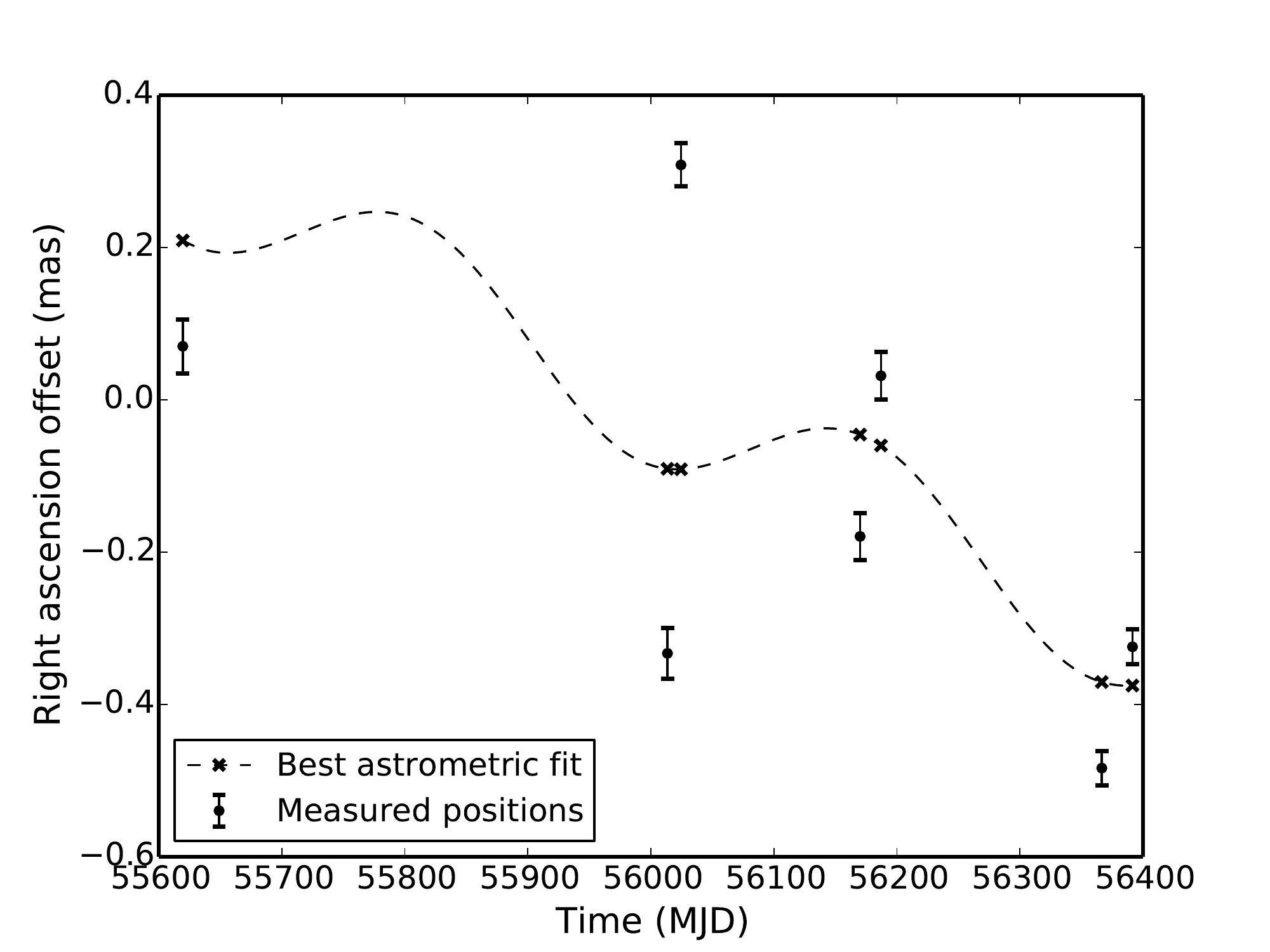} \\
% Made on bunker in /home/deller/data/vlbi/psrpi/final/J1650-1654/IBC-6-0104-divided/full/, using pylab_pmparplot.py -f local.jmfit.pmpar.in --plottype=pdf --target=J165015-165730 --nopmsubtract
\includegraphics[width=0.48\textwidth,trim={1cm 6mm 0mm 0},clip]{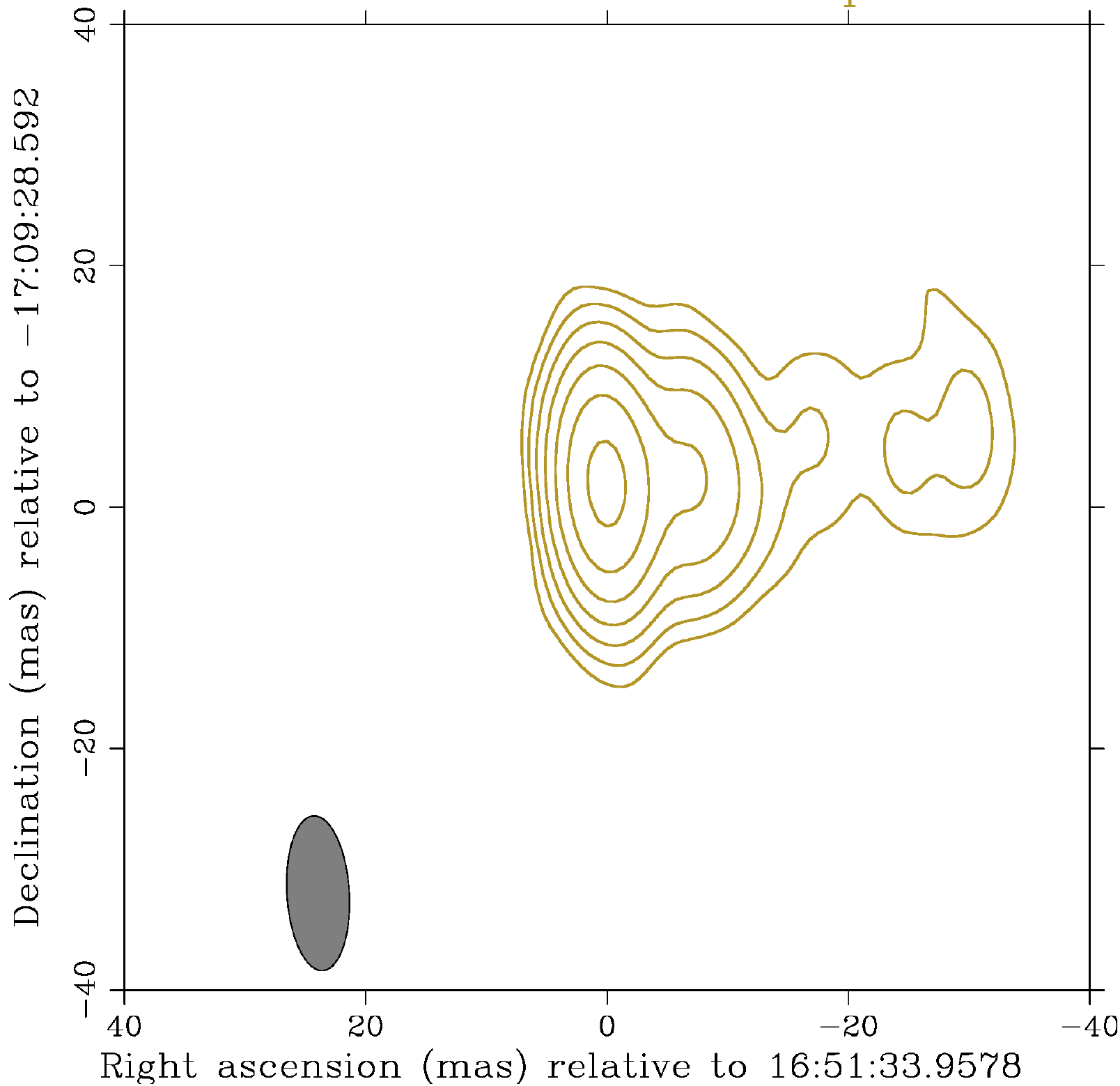} & 
\includegraphics[width=0.48\textwidth]{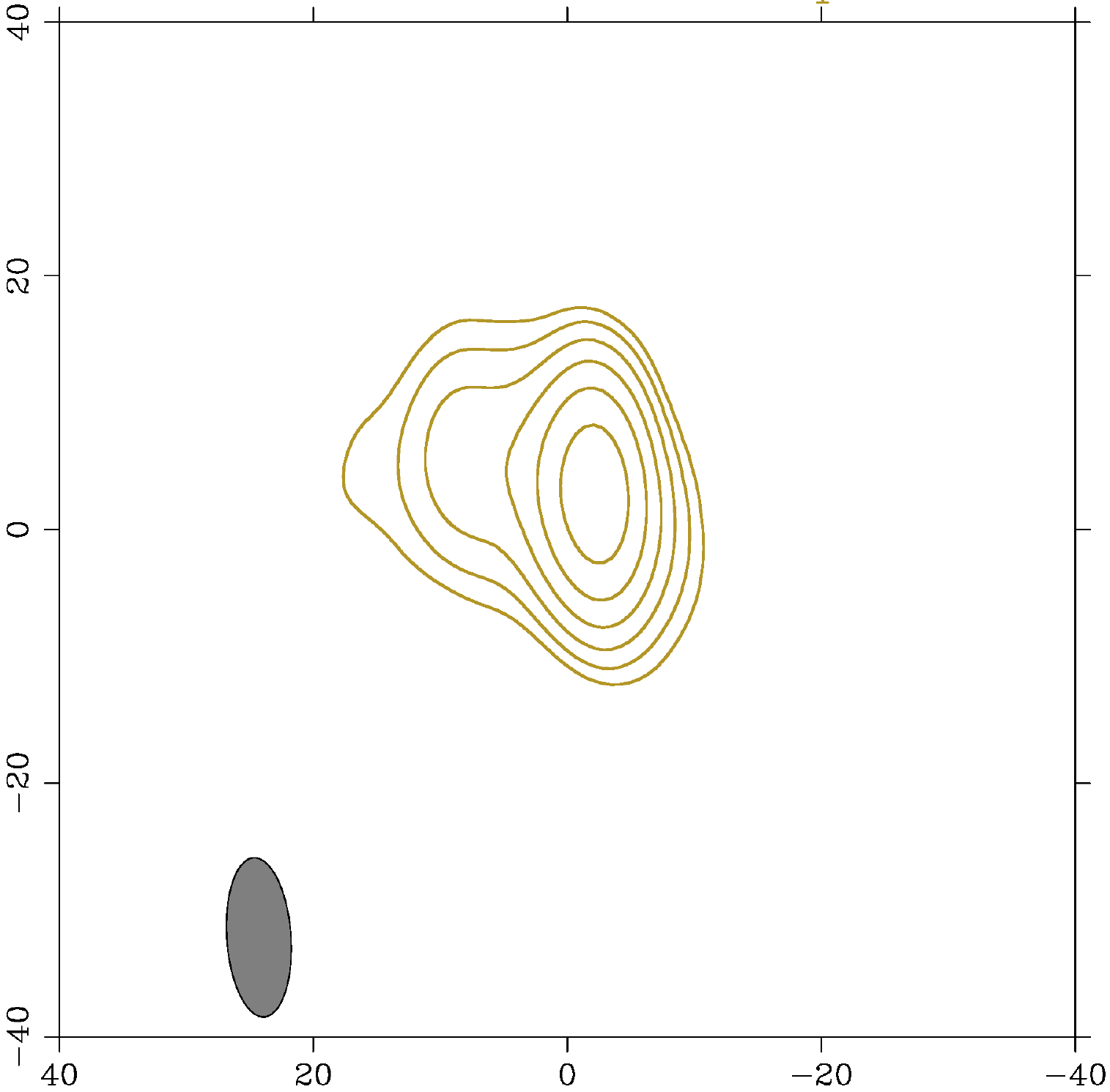}
\end{tabular}
\caption{\label{fig:1650}Top left: The pointing layout of the target pulsar and in-beam calibrator sources for PSR J1650--1654.  The dotted, solid and dashed lines show the 75\%, 50\% and 25\% response point of the primary beam at the center frequency of 1650 MHz. Top right: relative offset between the two calibrator sources -- error bars show measurements and the dashed line shows the best fit.  Bottom left: In-beam phase calibrator source J165133--170928, where the axis scale is milliarcseconds from the reference position.  Bottom right: Position reference source J165015--165730.}
\end{figure*}

Two in-beam calibration sources are present, allowing us to cross-check results.  The relative position separation of the in--beam phase calibrator source J165133--170928 from the second in-beam source J165015--165730 displays quite a large scatter (0.2 mas rms in right ascension, 0.25 mas rms in declination, as seen in Figure~\ref{fig:1650}), and an astrometric fit to this relative separation time series gives a best-fit parallax of 0.05 mas with an uncertainty of $\sim$0.1 mas.  This is not unexpected given the differential ionosphere across the $\sim20$\arcmin\ separation.  However, as shown in Figure~\ref{fig:1650}, the position reference J165015--165730 is separated by just 4\arcmin\ from the pulsar, meaning that differential ionospheric effects should be smaller than between the two calibrators.  An alternative is that the position reference source J165015--165730 exhibits substantial structure variations that lead to position offsets up to a few tenths of a milliarcsecond between epochs, and a substantial amount of the power this introduces into the pulsar's position time series can be fit by the parallax term.  The structure of the two calibrator sources can be seen in Figure~\ref{fig:1650} -- both exhibit prominent milliarcsecond-scale jets, meaning structure evolution is likely at some level.  

At present, we have no definitive explanation for the discrepant parallax for PSR J1650--1654, but the results indicate that the parallax uncertainty is underestimated by a factor of several for this pulsar.

\subsubsection{PSR J1820--0427}

PSR J1820--0427  has two in-beam calibrator sources (J182043--042412 and J182103--042633), which are close to each other and to the pulsar on the sky (angular separations 3\arcmin\ -- 5\arcmin).  PSR J1820--0427 was observed at 2.3 GHz due to the strong scattering along this low Galactic latitude line of sight.  

\begin{figure*}
\begin{tabular}{cc}
\includegraphics[width=0.43\textwidth]{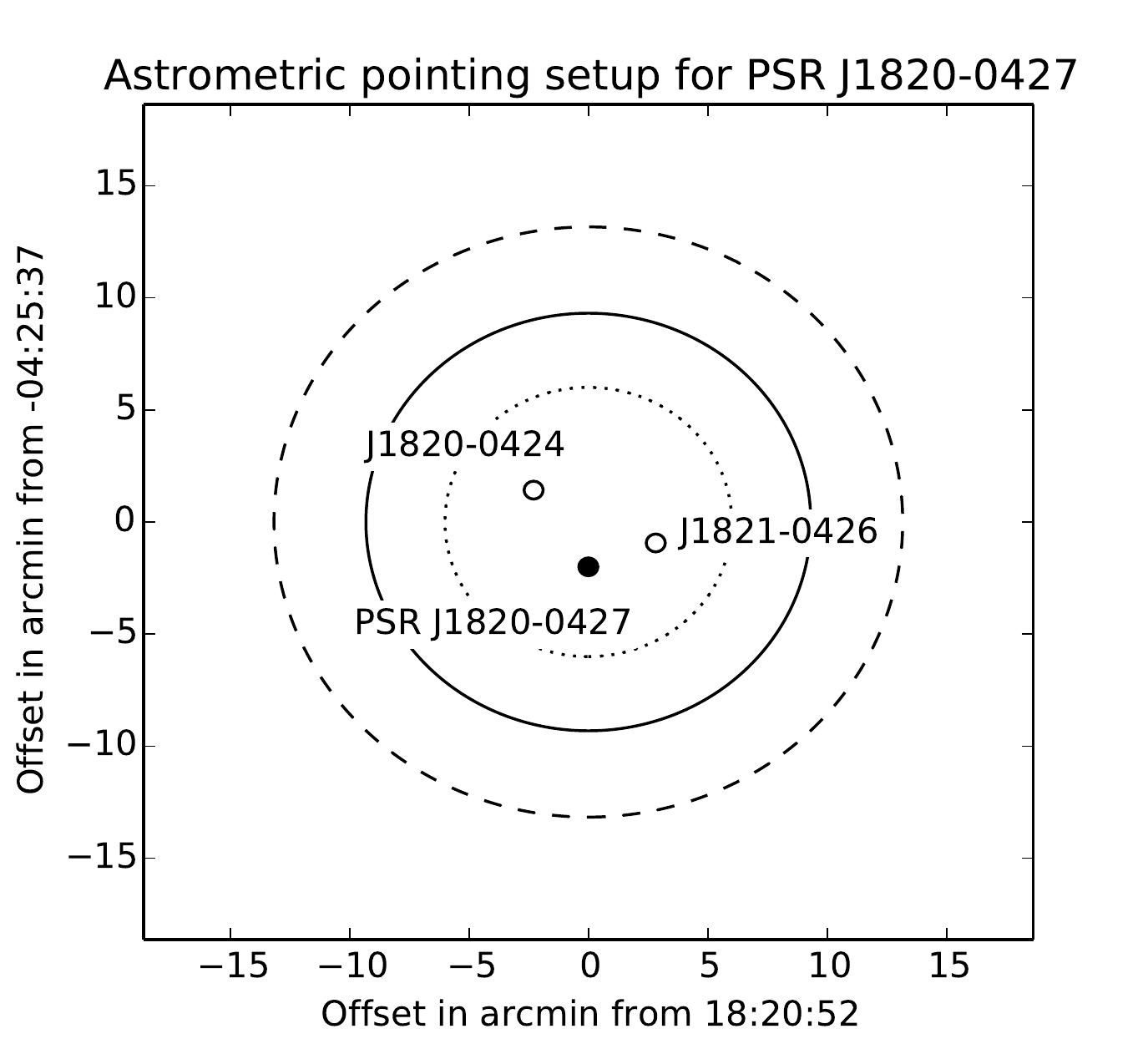} & 
% Made on bunker in /home/deller/data/vlbi/psrpi/final/pointingplots/temp/J1820/, using make_oneoff_beamplot.py
\includegraphics[width=0.53\textwidth]{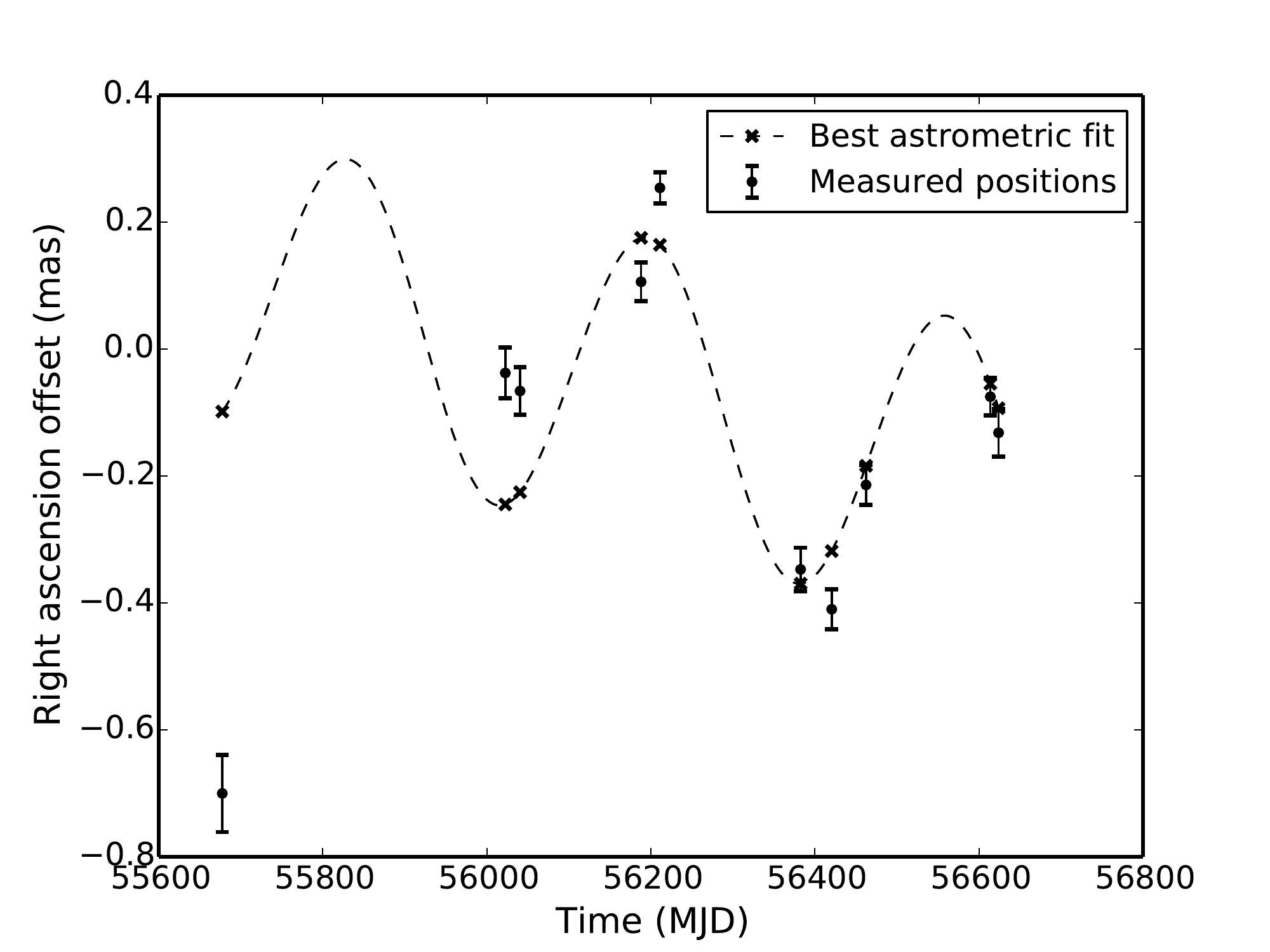} \\
% Made on bunker in /home/deller/data/vlbi/psrpi/final/J1820-0427/IBC-20-0152-divided/full/, using pylab_pmparplot.py -f local.jmfit.pmpar.in --plottype=pdf --target=J182103--042633 --nopmsubtract
\includegraphics[width=0.48\textwidth,trim={1cm 6mm 0mm 0},clip]{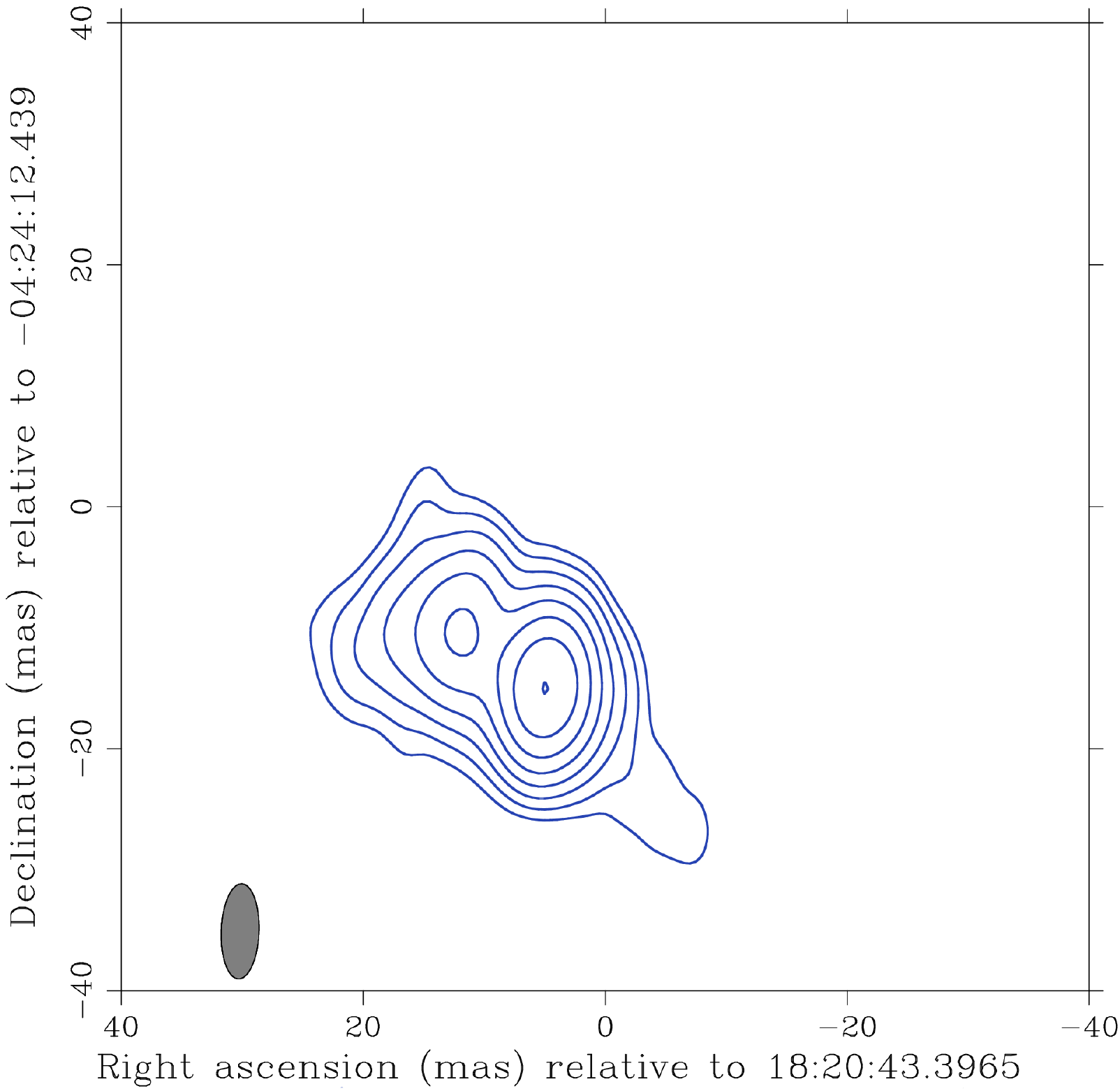} & 
\includegraphics[width=0.48\textwidth,trim={1cm 0mm 0mm 0},clip]{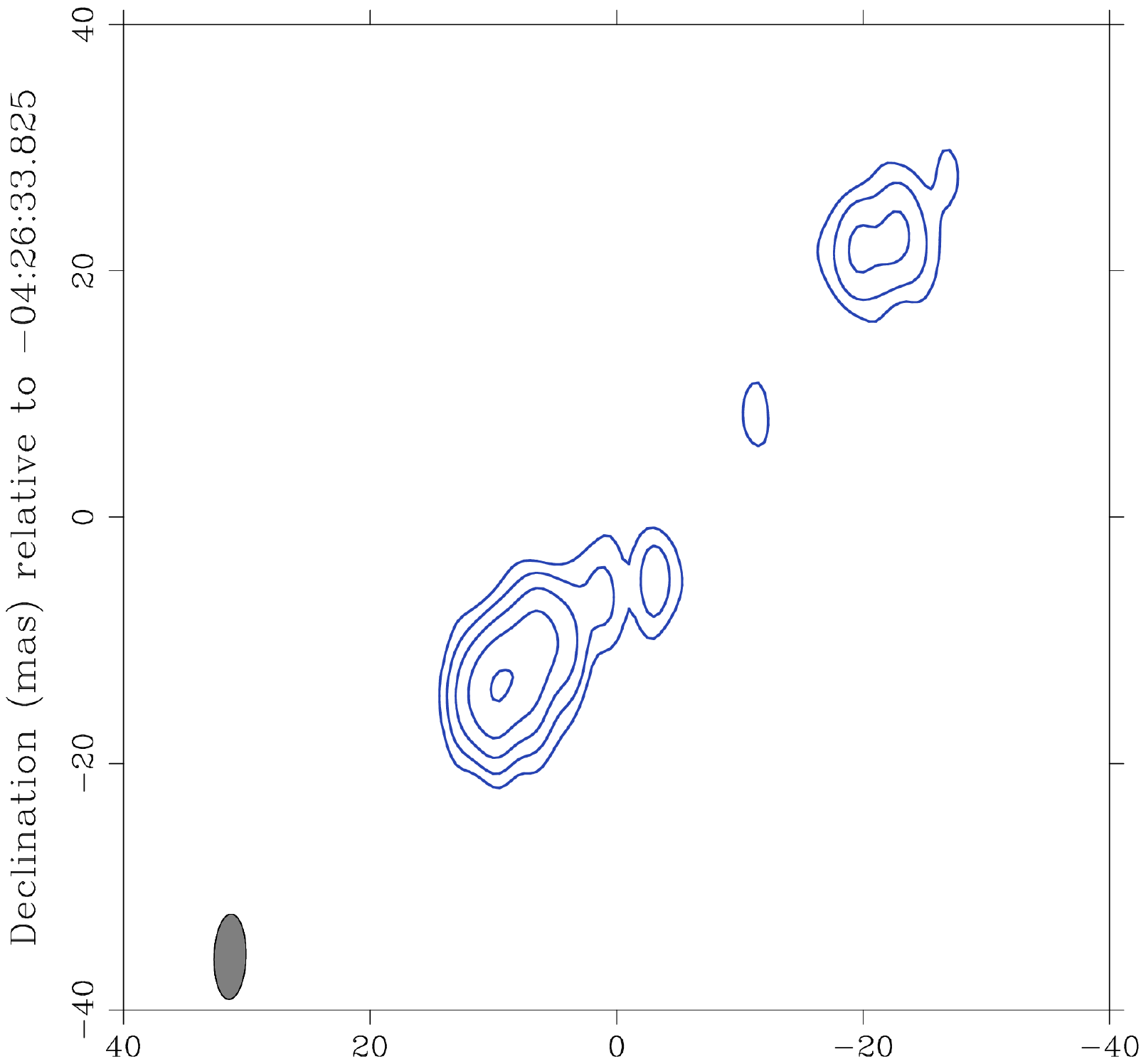}
\end{tabular}
\caption{\label{fig:1820}Top left: The pointing layout of the target pulsar and in-beam calibrator sources for PSR J1820--0427.  The dotted, solid and dashed lines show the 75\%, 50\% and 25\% response point of the primary beam at the center frequency of 2360 MHz. Top right: relative offset between the two calibrator sources  -- error bars show measurements and the dashed line shows the best fit.  Bottom left: In-beam phase calibrator source J182043--042412, where the axis scale is milliarcseconds from the reference position.  Bottom right: In-beam phase calibrator and position reference source J182103--042633.}
\end{figure*}

Figure~\ref{fig:1820} shows the clear structure in the residuals between the two in--beam calibrators.  The large offset in the first observation is likely due to the different observing frequency (1650 MHz vs 2360 MHz), which would result in a different observed source structure.  Neglecting this, however, it is clear that a significant parallax is measured between the two calibrators.  The higher observing frequency, along with the small angular separation, should have given relatively small ionospheric errors (which in any case should not yield a parallax-like signature).  Both in-beam calibrators do, however, exhibit significant milliarcsecond-scale jet structure, and as with the PSR J1650--1654 calibrators, evolution in this structure over time is a potential source of error.  PSR J1820--0427 is also at low Galactic latitude and the NE2001 predictions for scattering disk size and refractive wander of the in-beam calibrators are among the largest in the \psrpi\ sample, meaning refractive effects may also be a contributing factor.

The possibility that one of the two in--beam calibrators is actually Galactic cannot be definitively ruled out, but both are relatively bright (tens of mJy) and stable, arguing strongly against possibilities such as a microquasar or magnetically active protostar.

Since the discrepancy between the two calibrator sources is larger than parallax uncertainty calculated for the pulsar, there is a substantial possibliity that the distance uncertainty of PSR J1820--0427 is underestimated. If J182043--042412 had been used as a position reference instead of J182103--042633, the best--fit distance would change from $\sim$3 to $\sim$9 kpc.  Arguing in favour of using J182103--042633 is the fact that it is the closer source to the pulsar on the sky; this is the criterium applied throughout the \psrpi\ sample.  And since the pulsar--calibrator separation is smaller than the calibrator--calibrator separation, if the offsets are due to a differential term such as the ionosphere (as opposed to an offset created by one or the other calibrator source that is independent of angular separation), then the effect on the pulsar should be smaller than those seen here between the two calibrators.

\subsubsection{PSR J2325+6316}

PSR J2325+6316 has two in-beam calibrator sources, J232445+633001 and J232519+631636, with the latter being located extremely close to PSR J2325+6316 on the sky (angular separation 48\arcsec). Neither calibrator source was particularly bright and so they were summed in the calibration solution to improve S/N, but J232519+631636 was used as the position reference to minimise the differential calibration effects on the target pulsar's position time series.  The results for PSR~J2325+6316 under these conditions are in mild tension with the NE2001 model distance prediction of 8 kpc, with a parallax measurement of $-0.009^{+0.048}_{-0.044}$ mas.

J232519+631636 displays a parallax with respect to J232445+633001 of $0.26^{+0.10}_{-0.09}$ mas, indicating that had J232519+631636 not been available to serve as a position reference, a significantly different ($\sim$2.5$\sigma$) parallax would have been obtained for the target pulsar.  One (exceedingly unlikely) explanation for the discrepancy would be if J232519+631636 was a Galactic object located at a similar distance to PSR~J2325+6316, but sufficiently bright Galactic radio sources of milliarcsecond size are much rarer than radio AGN.  Systematic errors due to the differential ionosphere should be minimal due to the extremely small angular separation to the target; however, structure evolution in this nearby calibrator cannot be ruled out.  PSR J2325+6316 has the second-highest DM of the \psrpi\ targets and has the third largest predicted refractive wander based on the NE2001 model (rms 0.12 mas), and so refractive effects in the ISM are a potential explanation, particularly if the NE2001 refractive wander prediction is an underestimate along this line of sight.

\subsubsection{Implications for the remainder of the \psrpi\ sample}
In our sample, 39 pulsars have two or more in-beam reference sources. Of these, three show questionable astrometric results: PSR J1820--0427 and PSR J2325+6316 exhibit discrepancies in the relative positions between the two in-beam sources that exceed expectations, while PSR J1650--1654 shows an unphysical parallax result (while retaining low-precision consistency between the calibrators).  Using these values, we can estimate the likelihood that other sources in the \psrpi\ sample have under-estimated uncertainties on the fitted parameters.  The rate of discrepancies between in--beam sources is 3/39 or around 8\%, so from our remaining 21 pulsars, we expect that 1 or 2 more sources will have underestimated uncertainties.  

These findings highlight the the fact that the parallax fits based on small numbers of epochs, especially when the time baseline is short, should be treated with some caution when only a single calibrator source is available and hence independently estimating the systematic errors is not possible.

\subsubsection{PSR J2317+1439}
\label{sec:j2317}
Diffractive scintillation led to PSR J2317+1439 only being detected in 5 out of the 8 astrometric epochs.  The 3 non-detections all occurred on the same side of the parallax ellipse, meaning that the single detection at this parallax extremum carries a disproportionate weight in determining the parallax.  Because of the small number of measurements available, the bootstrap technique used to estimate the astrometric parameters and their uncertainties returns only weak constraints ($\pi = 0.6^{+1.5}_{-0.2}$): any bootstrap trial in which the crucial epoch is not selected has little ability to discriminate the parallax.

In this case, where the bootstrap sampling technique is overly pessimistic, we can with care make use of the simple least squares fit.  After accounting for systematic uncertainties to the position measurements for PSR J2317+1439 in the way described by Equation~\ref{eq:apriorsyserror} in Section~\ref{sec:errorbudget}, the reduced $\chi^2$ of a least squares fit is  1.35, indicating a reasonable fit, with a much smaller uncertainty than the bootstrap and a consistent best-fit value ($\pi = 0.65 \pm 0.07$).  This may be a fair reflection of the true parallax uncertainty, or it may underestimate the true uncertainty somewhat, but even if the input position uncertainties were doubled (a pessimistic case that would give a reduced $\chi^2$ well under 1) the parallax uncertainty would still be well under that estimated by the bootstrap.

\subsubsection{Pulsars with previous VLBI astrometry}
\label{sec:previousvlbi}
Three pulsars from the \psrpi\ sample have previously been the subject of VLBI astrometry: PSR J0332+5434 and PSR J1136+1551 \citep{brisken02a} and PSR J0826+2637 \citep{gwinn86a}.  Table~\ref{tab:previousvlbi} shows the \psrpi\ results for these pulsars compared against the previous results, for proper motion in right ascension and declination ($\mu_{\alpha}$, $\mu_{\delta}$) measured in mas yr$^{-1}$ and parallax in mas.  While agreement is good in most cases, with five of the nine measured parameters agreeing to better than 1$\sigma$, three values have a discrepancy exceeding 2$\sigma$, which is not expected statistically.

For PSR J0332+5434, the parallax and proper motion in declination measured by \citet{brisken02a} differ from the more precise \psrpi\ values by 2.5--3$\sigma$.  In \citet{brisken02a}, only 4 position measurements were made for this pulsar, meaning the resultant astrometric fit had only three degrees of freedom and would be susceptible to larger errors induced by poor fits in one or two epochs.  The final and most dioscrepant case is proper motion in declination for PSR J1136+1551, whiere the \citet{brisken02a} value (based on five observations) differs from the more precise \psrpi\ value by around 5$\sigma$.  While the shorter timespan (12 months) and small number of observations in the \citet{brisken02a} program would make it more susceptible to potential biases to proper motion such as calibrator source structure evolution, such a large discrepancy remains difficult to explain.

% Brisken+02 correlation positions:
% BB118E (Dec99): 11:36:03.18374 +15:51:09.709
% BB118I (Mar00): 11:36:03.18374 +15:51:09.709
% BB118O (Jun00): 11:36:03.180371 +15:51:09.899
% BB126D (Sep00): 11:36:03.17926 +15:51:09.983
% BB126I (Nov00): 11:36:03.177952 +15:51:10.060

\begin{deluxetable*}{lccc}
\tabletypesize{\small}
%%\tablewidth{0pt}
\tablecaption{Comparison against previous VLBI astrometry}
\tablehead{
\colhead{Astrometric quantity} & \colhead{PSR J0332+5434} & \colhead{PSR J0826+2637} & \colhead{PSR J1136+1551}}
\startdata
$\mu_{\alpha}$ (\psrpi) 	& $16.969^{+0.027}_{-0.029}$  	& $62.994^{+0.021}_{-0.007}$ 	& $-73.785^{+0.031}_{-0.010}$  \\
$\mu_{\alpha}$ (previous)	& 17.00 $\pm$ 0.27 	 		& 62.6 $\pm$ 2.4			&  $-$73.95 $\pm$ 0.38 \\
$\mu_{\delta}$ (\psrpi)	& $-10.379^{+0.058}_{-0.036}$	& $-96.733^{+0.045}_{-0.085}$ & $366.569^{+0.072}_{-0.055}$  \\
$\mu_{\delta}$ (previous)	& $-$9.48 $\pm$ 0.37 			& $-$95.3 $\pm$ 2.4 			& 368.05 $\pm$ 0.28 \\
Parallax (\psrpi) 		& $0.595^{+0.020}_{-0.025}$ 	& $2.010^{+0.013}_{-0.009}$ 	& $2.687^{+0.018}_{-0.016}$ \\
Parallax (previous)		& 0.94 $\pm$ 0.11			&  2.8 $\pm$ 0.6 			&  2.8 $\pm$ 0.16		 
\enddata
\label{tab:previousvlbi}
\end{deluxetable*}

\subsection{Galactic electron densities and models}
\label{sec:electrondensitymodels}
% electron density models
\newcommand{\nelec}{n_{\rm e}}
\newcommand{\DM}{{\rm DM}}

The \psrpi\ parallax sample allows estimation of the line-of-sight (LoS)  average electron density and will provide important input to the next generation Galactic electron density model\footnote{In particular, some of the authors are explicitly developing a follow-on model to the NE2001 model.}.

The mean electron density (cm$^{-3}$) for a given LoS is
$\nelec = \DM({\rm pc~cm^{-3}})\times{\rm parallax(mas)}/1000.$
The mean across the sample is $\langle \nelec \rangle = 0.020$~cm$^{-3}$ and the rms value is
$\sigma_{\nelec} = 0.016$~cm$^{-3}$.    The median distance $\sim 2.5$~kpc  implies that the electron densities
are representative of the solar region in the Galaxy. However, the sample includes a few objects that are well above
any realistic scale height ($\lesssim 2$~kpc) for the electrons, so the sample-mean density is biased lower than the mid-plane value.    Restricting the sample to six objects within 1~kpc of the Sun, we obtain 
a larger value, $\langle \nelec \rangle = 0.026$~cm$^{-3}$.

\begin{figure}
\includegraphics[width=0.48\textwidth]{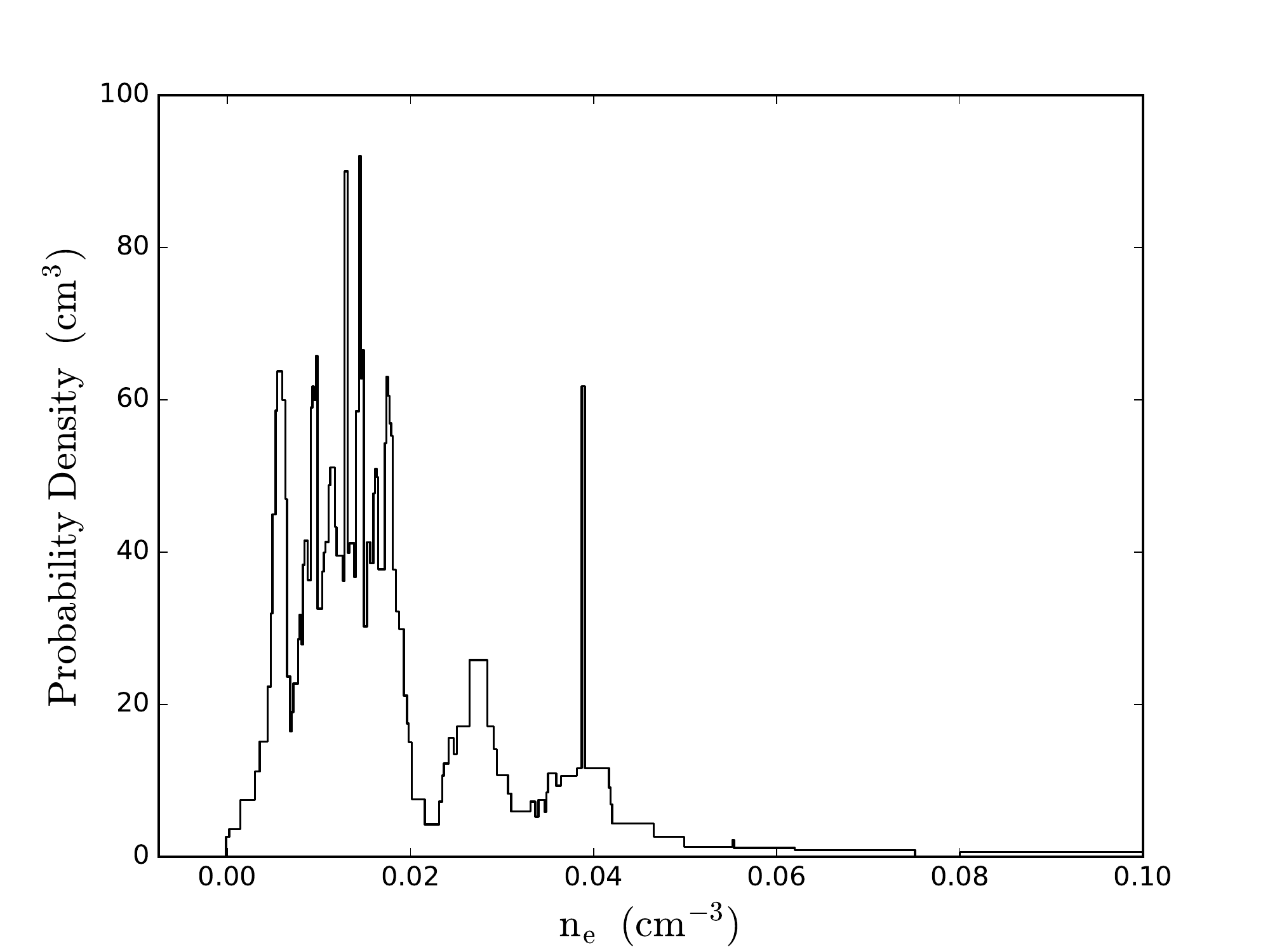}
\caption{
Probability density function for the mean electron density calculated as
$\nelec = \DM \times {\rm parallax ( mas})/1000$.    Errors on parallaxes are included by
representing the contribution from each pulsar as a rectangle function centered on the nominal value and  width equal to the sum of the positive and negative-going errors.  The three objects with lower limits on their distances are also included.  The PDF is normalized to unit area. 
\label{fig:pdfne}
}
\end{figure}

The distribution of LoS electron densities is positively skewed. 
Figure~\ref{fig:pdfne} shows the probability density function (PDF) for the  electron density.  It is calculated as the sum of rectangle functions, each centered on the nominal value of $\nelec$  and having a width equal to the
68\% confidence interval given in Table~\ref{tab:allresults}.   For the three pulsars with lower limits on the distance, the rectangle extends from $\nelec = 0$ to its upper bound. The PDF is normalized to unit area. Its width reflects the wide variation of electron densities between LoSs caused by Galactic structure on both small and large scales. 

\begin{figure}
\includegraphics[width=0.48\textwidth]{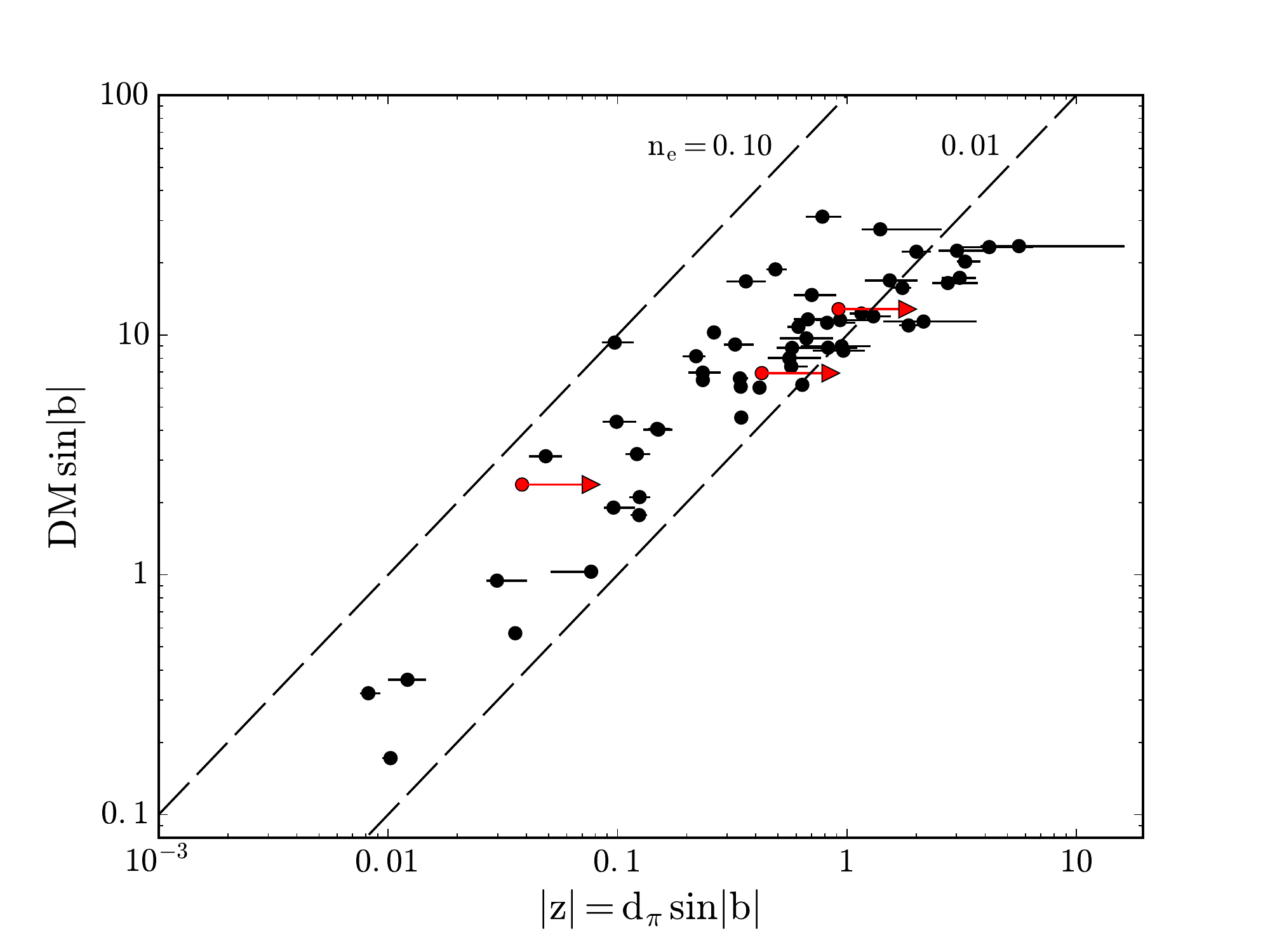} 
\caption{Plot of the $z$ component of DM vs. distance $z$ from the Galactic plane for the
\psrpi data.  The dashed lines show the variation expected for constant values of electron density,
$0.01$~cm$^{-3}$ and $0.1$~cm$^{-3}$.    The arrows denote objects with lower bounds on their distances, J1650$-$1654,  J2150+5247, and J2325+6316. 
\label{fig:dmz_vs_z}
}
\end{figure}

\begin{figure}
\includegraphics[width=0.48\textwidth]{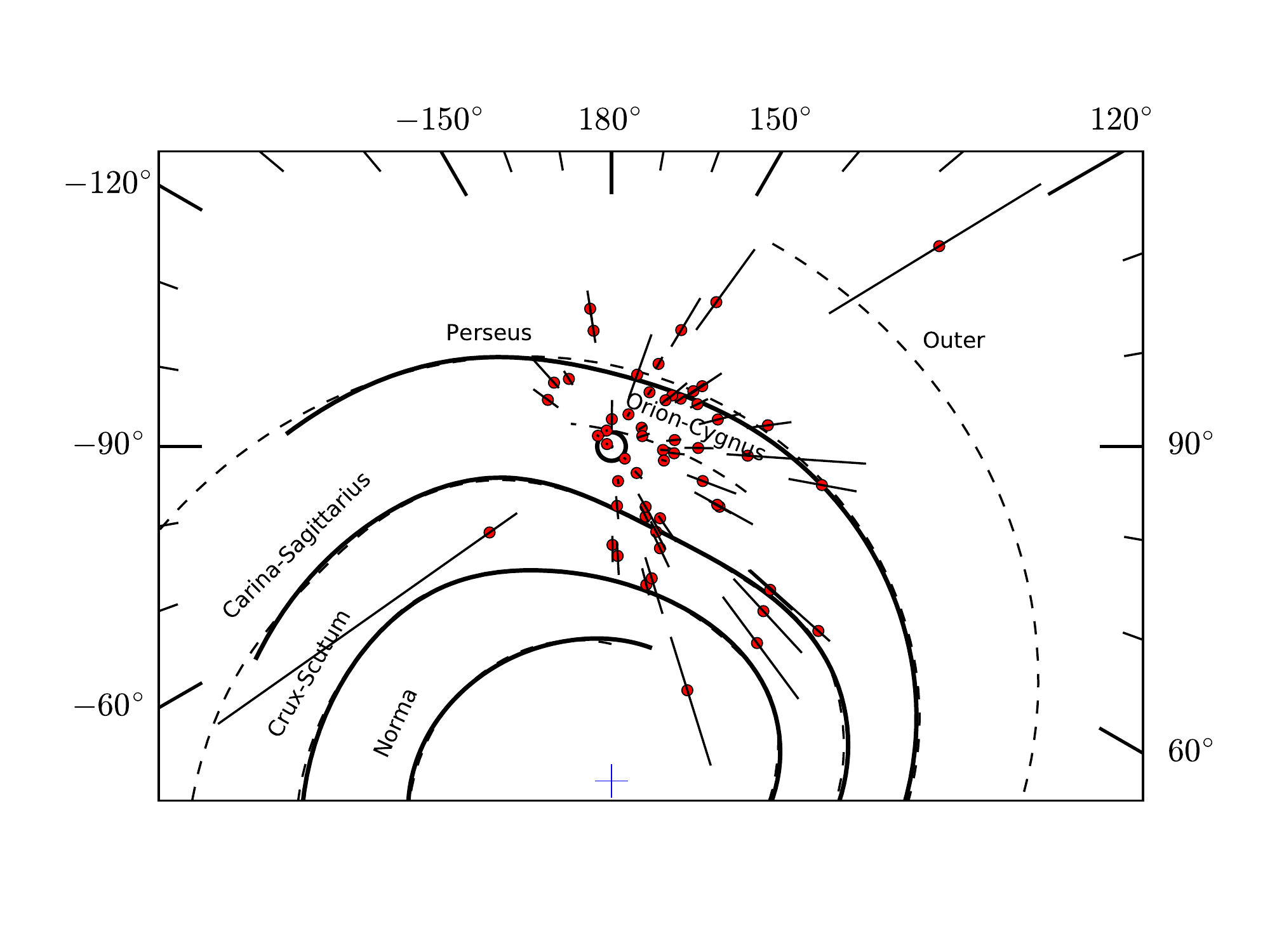}
\caption{\psrpi\ pulsars projected onto the Galactic plane using  parallax distances and their errors.  Objects with lower bounds on their distances and objects with greater than a 5:1 distance ratio (maximum to minimum parallax distance) are excluded.
The spiral arms and their labels are identical to those used in the NE2001 model. 
\label{fig:gplane}
}
\end{figure}

Figure~\ref{fig:dmz_vs_z} shows the perpendicular component $\DM_z = \DM \sin\vert b\vert$ plotted against the $z$ distance of each pulsar from the Galactic plane.   Dashed lines show the expected values for constant densities 
of $0.01$ and $0.1$~cm$^{-3}$.   Most of the points are between these two lines except at large values of $z$, where there appears to be a maximum in $\DM_z$ even though about eight of the \psrpi\ pulsars extend from 
two to five kpc above the plane. The \psrpi\ sample by itself therefore verifies that the Galaxy must have an electron density component similar to the thick disk component in the NE2001 \citep[][]{cordes02a} and YMW16 \citep[][]{YMW17} models. 

Distances of \psrpi\ pulsars projected onto the Galactic plane (Fig.~\ref{fig:gplane})  show that the sample extends  to large distances parallel to the plane and sample several spiral arms (as defined in the NE2001 model).   

Galactic electron density models are based on a wide variety of measurements but ultimately require independently obtained distances for as many pulsars as possible.   The NE2001 model  used distance constraints on 112 pulsars of which only 14 were parallax measurements; the remainder were mostly from
HI-absorption constrained distances and from pulsars in globular clusters\footnote{Multiple pulsars in a globular cluster were counted as only one distinct line of sight.  }.   For the YMW16 model, 73 parallax measurements were used, many from pulsar timing, but 29\% had a constrained distance range (maximum to minimum ratio) of 1.5:1 and 13\% had more than a 2:1 range. 

The performance of the NE2001 and YMW16 models can be compared against the \psrpi\ distances as a `blind' test because the \psrpi\ sample was not used in the construction of either model.   The comparison is particularly useful for more distant pulsars; the median distance of the \psrpi\ sample is 2.5~kpc, while the median distance to pulsars with  previously published VLBI parallaxes is 1.1~kpc.

\begin{figure}
\begin{tabular}{c}
\includegraphics[width=0.48\textwidth]{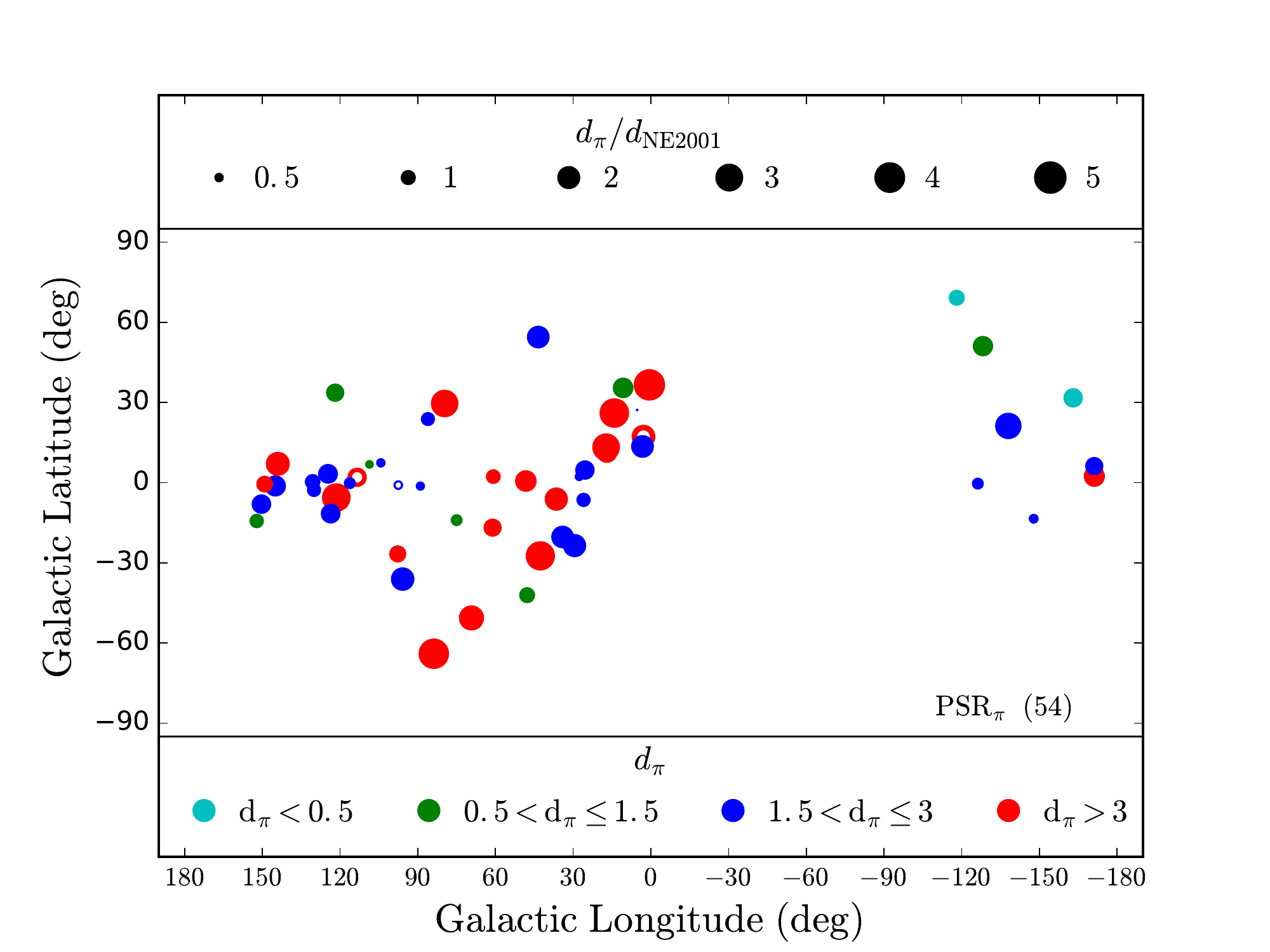} \\
\includegraphics[width=0.48\textwidth]{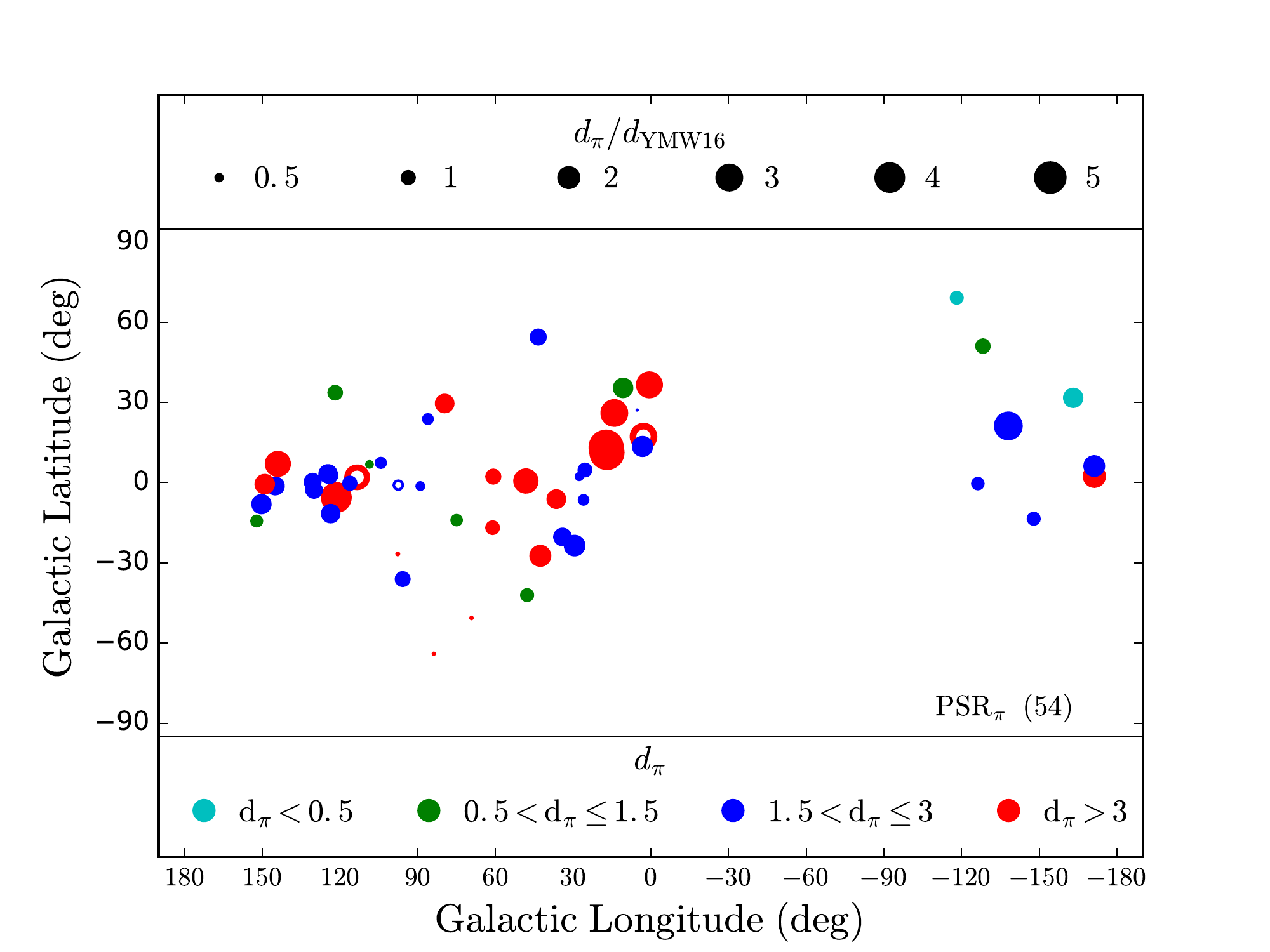}
\end{tabular}
\caption{Ratio of VLBI parallax distance to the NE2001 model distance (top panel) and the YMW16 model
distance (bottom panel).  The results are shown as a function of Galactic 
longitude and latitude.   Circle sizes indicate the value of the ratio, as shown in the upper legend in each panel.
Circle colors denote parallax distance in kpc as in the lower legend in each panel. 
Filled circles denote distance measurements while the three open circles indicate lower bounds on 
pulsar distances. 
\label{fig:dpi_dne2001ymw16_ratio}
}
\end{figure}

Figure~\ref{fig:dpi_dne2001ymw16_ratio} shows the ratio $d_\pi / d_{\rm model}$ for the NE2001 and YMW16 models in the top and bottom panels respectively.   Circle sizes indicate values of this ratio while colors indicate approximate distances.   Comparison of the two figures indicates that both models show large errors for some objects, with the NE2001 model doing better on some objects, and the YMW16 model on others.   
The YMW16 model performs somewhat better on a few high-latitude pulsars than the NE2001 model.  The distances of two pulsars with the most  negative latitudes (the red circles between 60$^{\circ}$ and 90$^{\circ}$
longitude)  are overestimated in the YMW16 model and underestimated by the NE2001 model. 
The median distance ratio exceeds unity for both models (1.5 and 1.1 for NE2001 and YMW16, respectively) and the RMS values of this ratio are 1.1 and 0.8, respectively.   
These results demonstrate that the new \psrpi\ sample will be extremely valuable for the next generation Galactic electron density distribution model. 

\citet{YMW17} claim that in 95\% of cases, the YMW16 predicted distance $d_{\mathrm{YMW}}$ will fall within the range $\left( 0.1\times d_{\mathrm{actual}}, 1.9\times d_{\mathrm{actual}} \right)$.  This claim can be examined using the \psrpi\ dataset.  We restrict ourselves to pulsars whose VLBI parallax significance is at least 5$\sigma$, of which there are 42.  Of these 42 pulsars, six have a predicted YMW16 distance that falls outside the range $\left( 0.1\times d_{\pi_\mathrm{min}}, 1.9\times d_{\pi_\mathrm{max}} \right)$, where $d_{\pi_\mathrm{min}}$ and $d_{\pi_\mathrm{max}}$ are the values given by inverting the 95\% confidence interval for parallax.  This would be inconsistent with 95\% of the YMW16 distances falling within the range $\left( 0.1\times d_{\mathrm{actual}}, 1.9\times d_{\mathrm{actual}} \right)$, if the \psrpi\ sample was representative of the entire pulsar population.  However, the \psrpi\ sample is explicitly not an unbiased sample of pulsars, and in particular pulsars at high Galactic latitudes are intentionally over-represented in order to help constrain the Galactic scale height.  Of the six discrepant YMW16 predictions, five are at moderate to high Galactic latitudes 
($|b| > 20$\degrees\ 
and the YMW16 model places them beyond the edge of the Galaxy, while the median Galactic latitude of the 42 pulsars with a significant parallax distance is 14\degrees.  The sixth source, at 
$b = 11.3$\degrees,
is underpredicted by an order of magnitude.  Accordingly, based on the \psrpi\ sample we advise that all DM--based distance estimates be used with caution, especially for high Galactic latitude pulsars.

Finally, as noted in Section~\ref{sec:errorbudget}, refractive wander is potentially a significant contributor to the differential astrometry error budget for some lines of sight, particularly at low Galactic latitude.  While beyond the scope of this work, future analysis using the \psrpi\ sample could refine the scattering disk and refractive wander predictions of the NE2001 model, leading to improved estimates of systematic error contributions for future studies.

\subsection{Transverse velocities}
\label{sec:transversevelocities}
Galactic pulsars have a larger scale height than their progenitor massive stars, leading to the early inference that they have high velocities \citep{go70}.  Individual high velocity objects such as PSR~B1508+55 \citep[$v_\perp \sim 1000$~km~s$^{-1}$;][]{cvb+05} and PSR~J2225+6535, the Guitar Nebula pulsar \citep[$v_\perp \sim 800$~km~s$^{-1}$; this work;][]{crl93} establish stringent constraints on natal kicks and the minimum asymmetry requirements in simulations of supernova core collapse \citep[e.g.,][]{f04}, and the
overall population velocities inform models for neutron star birth, supernova explosions, and the evolution of close binary systems. The pulsar velocity distribution has therefore been a topic of continued interest \citep[e.g.][]{ll94,hp97,cc98,acc02,hllk05,verbunt17a}.

The parallax and proper motion measurements presented here provide model-independent estimates of pulsar distances and transverse velocities (Table~\ref{tab:derivedresults}) and thus
mitigate a key uncertainty in deriving the pulsar velocity distribution. We note, however, that the astrophysically relevant quantity is the 3-dimensional birth velocity for the entire pulsar population. For individual pulsars, their uncertain age limits the precision of any extrapolation in the Galactic gravitational potential, and their radial velocity is unknown, rendering a full 3-dimensional birth velocity unmeasurable. While such uncertainties can be addressed statistically, the inference of population parameters is further affected by biases in the sample of objects with precise astrometry  \citep[see, e.g.,][]{acc02}. The targets in this work were selected based on flux density and calibrator availability (Section~2.1), and inferring the properties of the population would require the addressing of selection effects in the original detection surveys. Such selection effects are not trivial: for example, many pulsar surveys focus on the Galactic plane where the stellar progenitors of pulsars are concentrated\footnote{Pulsar surveys typically follow the ``Willie Sutton rule" and focus on areas where the expected discovery rate is highest.}, but on average, the higher velocity neutron stars will spend less time near the Galactic plane compared to the lower velocity objects. Thus the high velocity tail of the pulsar population will be suppressed in a typical survey yield.

\begin{figure}[hb]
\centerline{\includegraphics[width=0.5\textwidth]{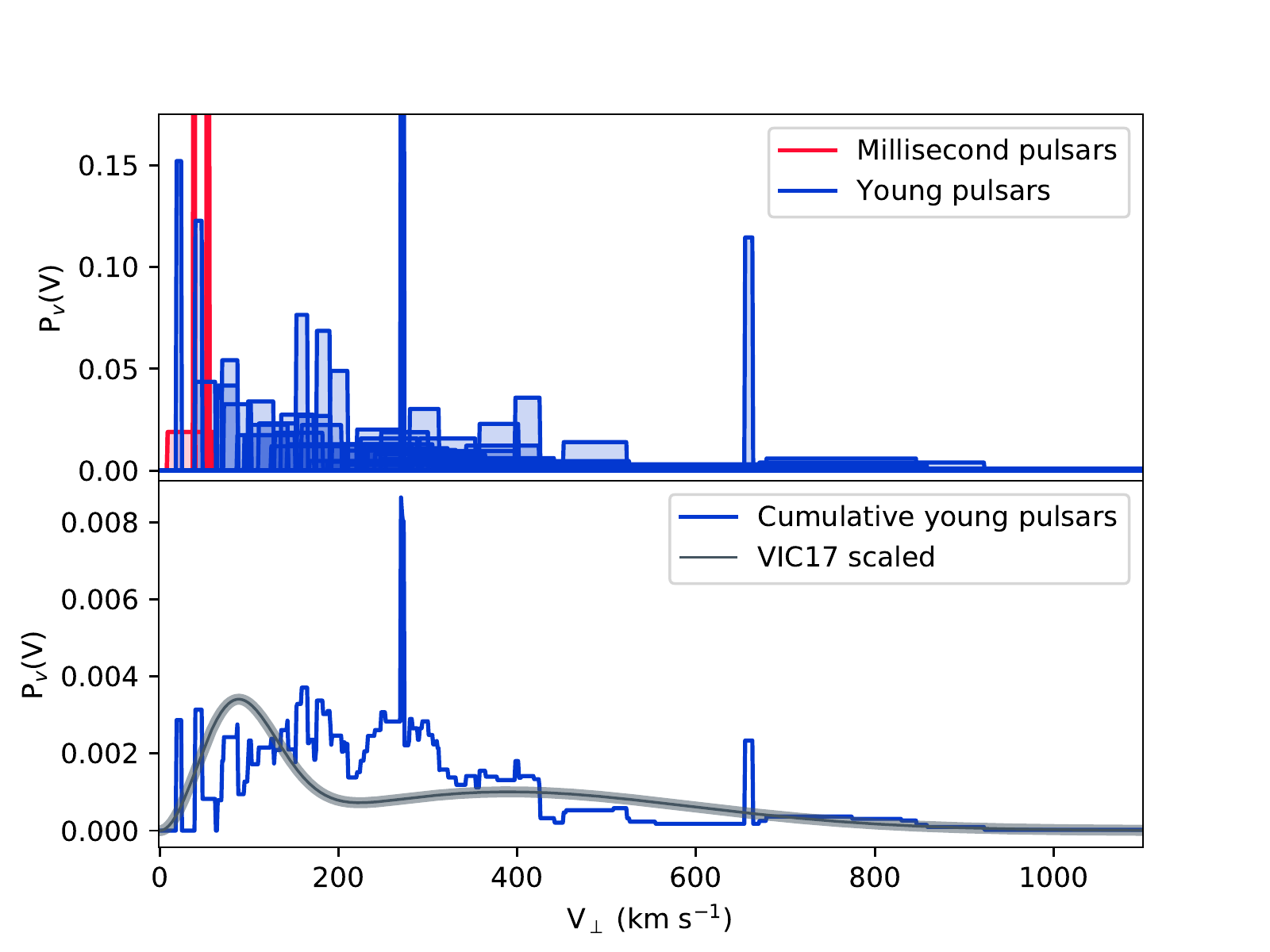}}
\caption{The transverse velocity distribution of the PSR$\pi$ sample of pulsars. Top: Histograms of the measured velocities ($\pm1\sigma$) of the young (blue) and millisecond (recycled; red) pulsars, as listed in Table~\ref{tab:derivedresults}. Bottom: The transverse velocity probability distribution, summed across all young pulsars in our sample, with a model for the pulsar velocity distribution from \citet{verbunt17a} scaled to two dimensions and overlaid for comparison. \label{fig:vdist}}
\end{figure}

We defer the treatment of the full range of selection effects to future work, but here we present histograms of the measured transverse velocities for each pulsar in our sample, along with a normalised probability distribution summing across all pulsars (Figure~\ref{fig:vdist}). Note that the four millisecond (recycled) pulsars in our sample, PSRs~J1022$+$1001, J2010$-$1323, J2145$-$0750, and J2317$+$1439, are excluded from the cumulative distribution, since recycled pulsars are an older population with a lower characteristic velocity distribution \citep[e.g.,][]{cc97}. As a comparison for our sample of transverse velocities, we plot a recent velocity distribution model for young pulsars \citep{verbunt17a}, scaled to two dimensions. The distribution of transverse velocities of the sample of young pulsars presented here is broadly compatible with previous published models, and a detailed treatment of selection effects is required before the models can be usefully discriminated between.

As an aside, we note that the present work provides the first model-independent distance and velocity estimate for PSR~J2225+6535, the Guitar Nebula pulsar ($v_\perp \sim 800$~km~s$^{-1}$; Table~\ref{tab:derivedresults}). That is the highest well-measured velocity in the current sample, although somewhat lower than previous estimates \citep[e.g.,][]{chatterjee04b}; further analysis and comparison to long-term optical monitoring of the time-evolution of the H$\alpha$\ bow shock nebula is underway.

\subsection{Comparison to timing astrometry}
\label{sec:timingcomparison}
As highlighted in Section~\ref{sec:introduction}, having independent measurements of
pulsar distances and astrometric parameters is extremely valuable
for several pulsar science cases.  While pulsar timing can provide
the pulsar DM, extracting a distance from this measurement 
is dependent on having an accurate model of the Galactic electron
density (see Section~\ref{sec:electrondensitymodels}). Moreover,
multi-frequency (or wide-band) observations are required to obtain a good
DM measurement.

Astrometric terms in pulsar timing models are covariant with red noise in timing data that arises from fluctuations in the pulsar spin-down and/or propagation delays through the ISM that are not completely captured in the pulsar ephemeris.   As shown in \citet{deller16a} and \citet{madison13a},  red noise can lead to substantial errors in the values and underestimates of the uncertainty for timing-derived parameters. These errors are especially large for unrecycled pulsars with surface field strengths $\sim 10^{11}$ - $10^{13}$~G, but are present at some level for all pulsars.

The highest precision measurements from pulsar timing are obtained with millisecond
pulsars (MSPs) due to their frequent, short pulses and stable rotation. In our sample we have four
MSPs: PSR\,J1022$+$1001, PSR\,J2010$-$1323, PSR\,J2145$-$0750, and
PSR\,J2317$+$1439. 
The VLBI results for PSR\,J1022$+$1001 and PSR\,J2145$-$0750 were
already presented and discussed in \citet{deller16a} and compared with
the most recent pulsar timing measurements. For completeness, we
summarize the results here for all MSPs in the \psrpi\ sample.
Each of these pulsars is observed by at least one pulsar timing array (PTA)
in the search for low-frequency gravitational waves \citep[e.g.,][]{verbiest16a}.

PSR\,J1022$+$1001 is located extemely close to the ecliptic plane
(ecliptic latitude $\beta$=-0.06\degrees) , which in pulsar timing leads to suboptimal measurements
of the position and proper motion, due to the components being
covariant in equatorial coordinates. Moreover, low-ecliptic
latitude pulsars have their line-of-sight passing close to the Sun
every year which leads to annual increases of DM which may not be
modelled optimally in pulsar timing \citep[e.g.][]{tiburzi18a}.

\begin{figure*}
\begin{tabular}{ccc}
\includegraphics[height=0.37\textwidth]{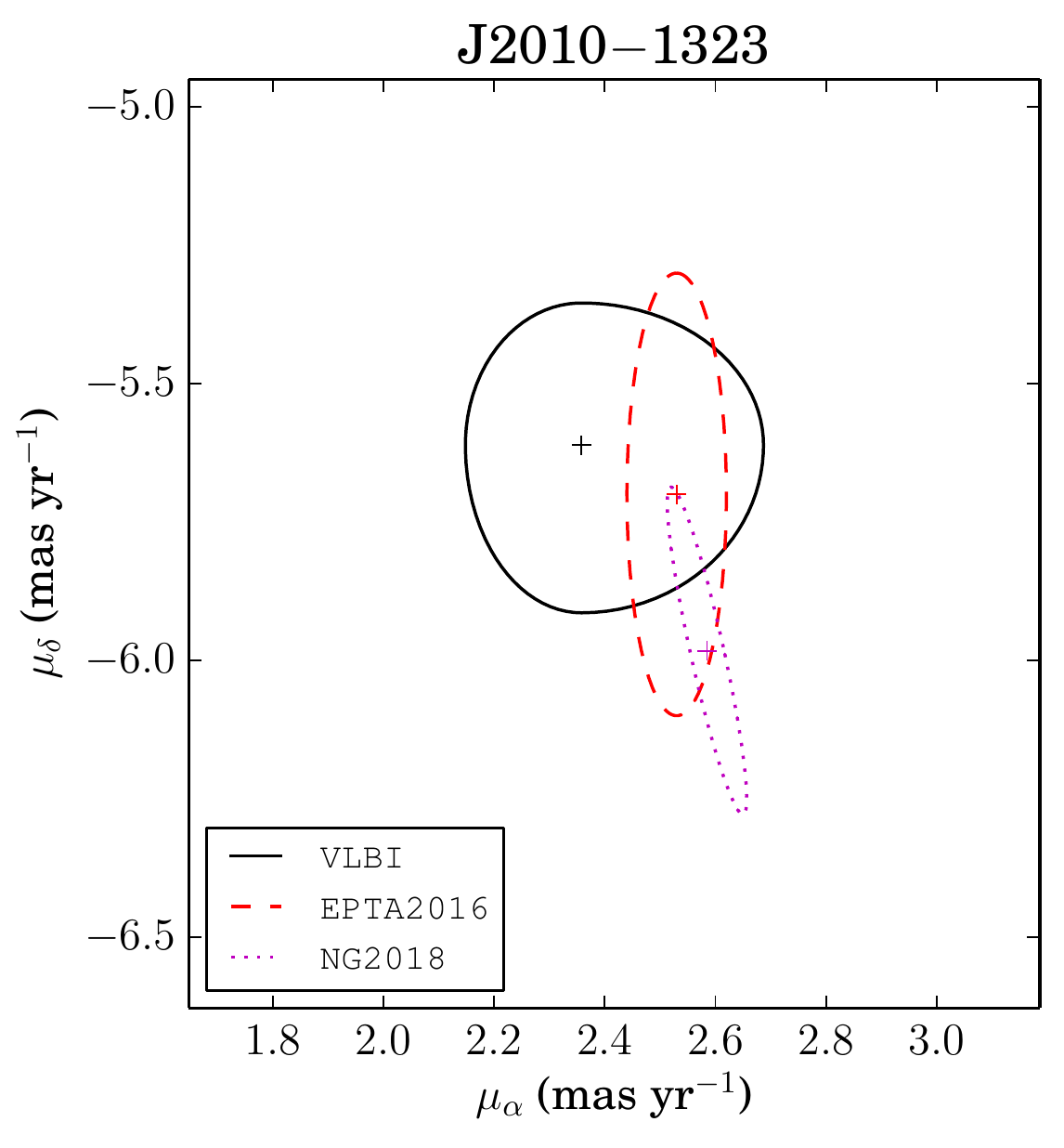} &
\includegraphics[height=0.37\textwidth]{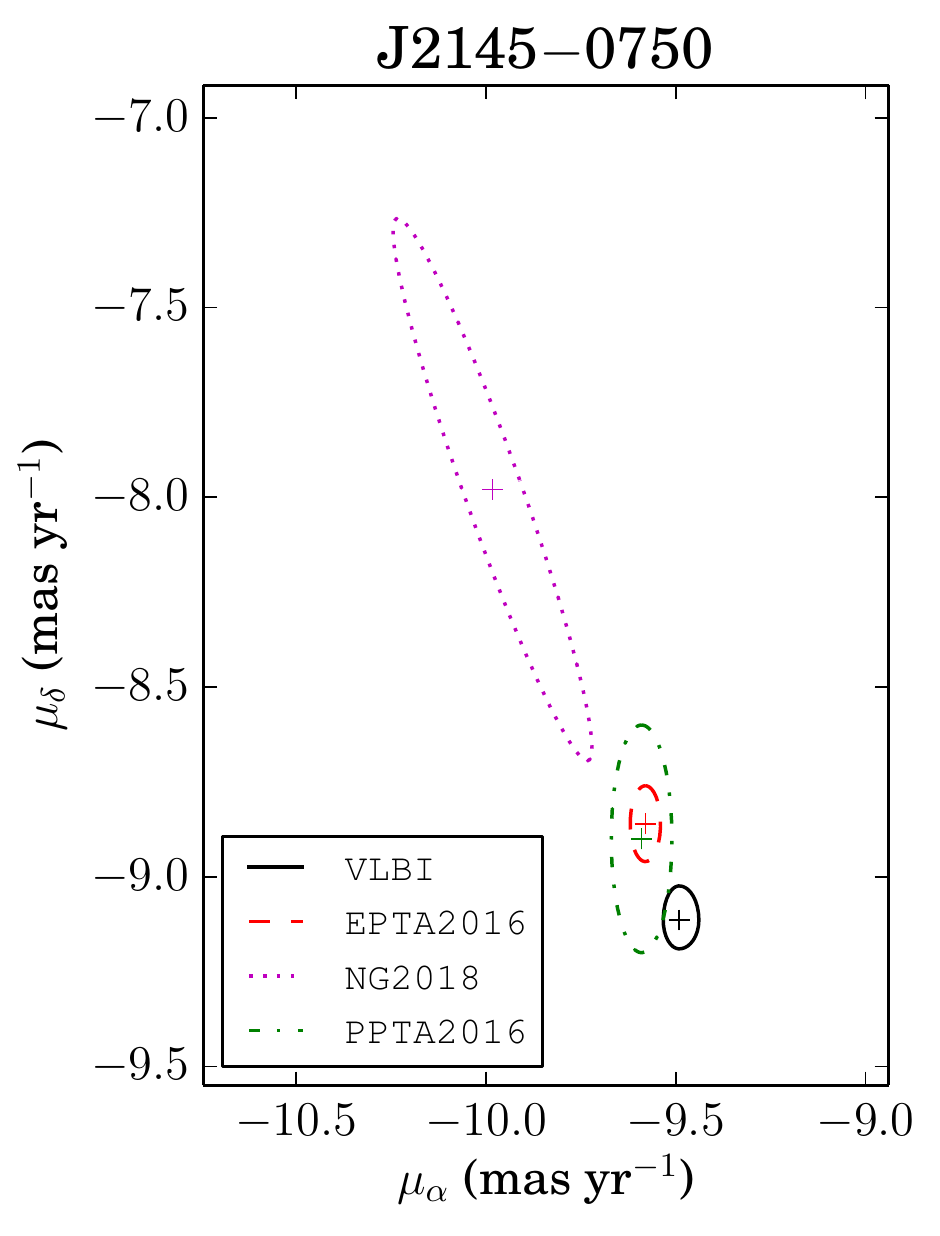} &
\includegraphics[height=0.37\textwidth]{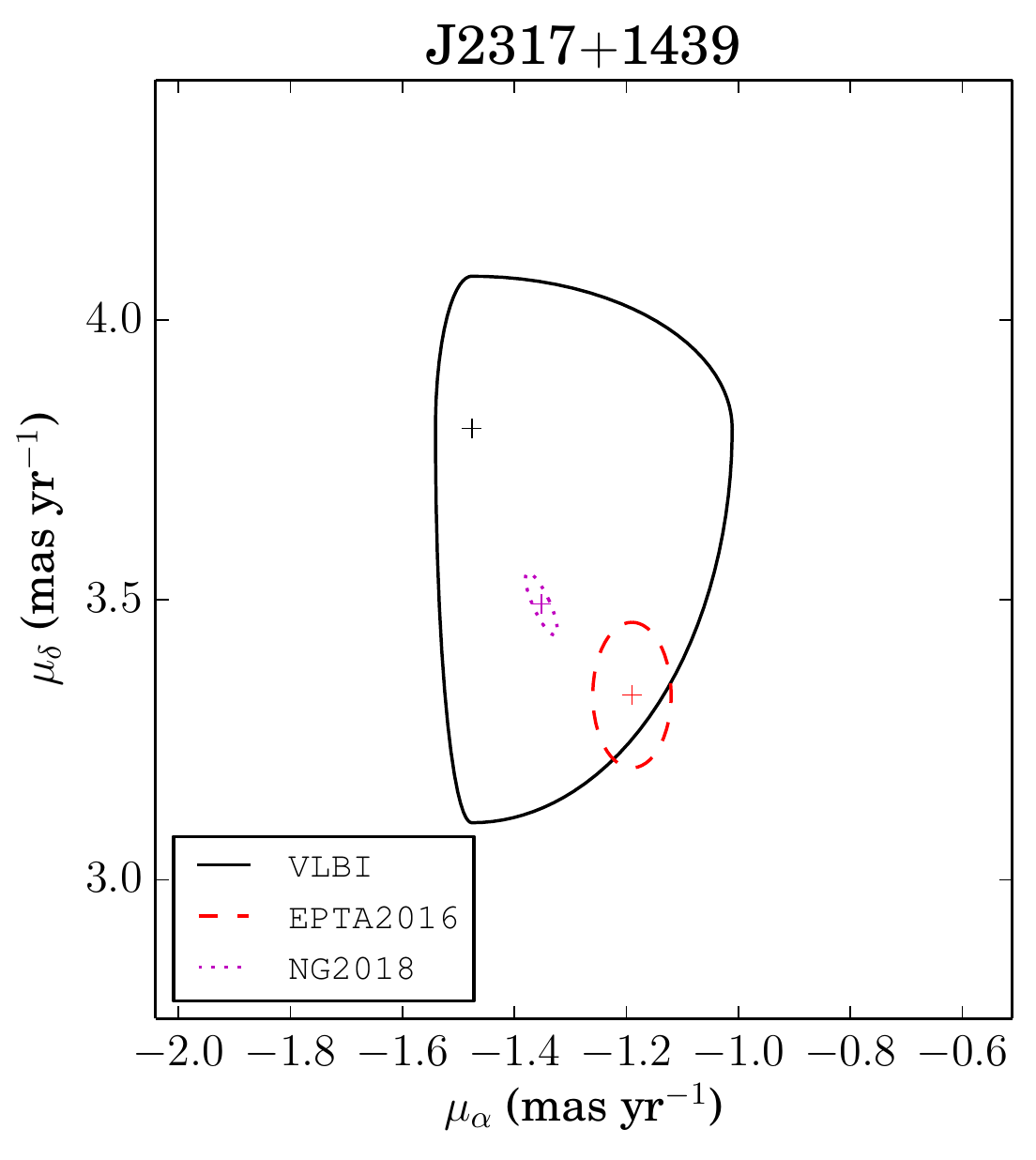} 
\end{tabular}
\caption{\label{fig:propermotioncomp}Comparison of proper motion measurements from VLBI and different PTA programmes. References: EPTA: \citet{dcl+16}; NANOGrav: \citet{abb+18}; PPTA: \citet{rhc+16}.}
\end{figure*}

Although PSR\,J2010$-$1323 is also located relatively close to
the ecliptic plane ($\beta$=6.49\degrees), pulsar timing has been able to measure
the proper motion of this pulsar with relatively high accuracy, as shown in
Table\,\ref{tab:pmcomparison}.  This pulsar is observed by the 
European Pulsar Timing Array (EPTA) and the North American Nanohertz
Observatory for Gravitational Waves (NANOGrav), but not the Parkes
Pulsar Timing Array (PPTA).
Figure\,\ref{fig:propermotioncomp} shows that the
uncertainties from their timing programmes are comparable to VLBI, and no 
significant discrepancies are seen. It is noteworthy that the 
11-yr NANOGrav dataset values \citep{abb+18} changed significantly from 
the 9-yr values \citep{matthews16a}, which were inconsistent at the $\gtrsim$2$\sigma$
level with the VLBI and EPTA values \citep{dcl+16}.

PSR\,J2145$-$0750 is observed by all three PTAs, and is a good
example where VLBI measurements give up to an order-of-magnitude
improved accuracy compared to some timing measurements. The ecliptic
latitude of this pulsar is 5.31 degrees, and this pulsar is known to
show DM variations that affect the timing observations \citep[e.g.][]{abb+18}.  As was the case for
PSR~J2010-1323, the proper motion obtained for PSR\,J2145$-$0750 from the NANOGrav 11-yr dataset is considerably
less discrepant with other measurements (our VLBI results, and also the PPTA and EPTA timing 
results) than the proper motion from the NANOGrav 9-yr dataset was.

As described in Section\,\ref{sec:individualpulsars}, the VLBI
astrometry of PSR\,J2317$+$1439 resulted in relatively poor constraints
due to failed observations, which all fell on the same side of the parallax
extrema. This also resulted in very conservative and skewed
uncertainties for the proper motion parameters when using the
bootstrapping method (green error ellipse in
Figure\,\ref{fig:propermotioncomp}), and all timing-derived measurement
are consistent with the VLBI values. When using the least-squares
fitting method, the uncertainties are reduced (black curve in
Figure\,\ref{fig:propermotioncomp}), and the EPTA timing value is
offset by about $1\sigma$ from the VLBI value.  Additional observations
for PSR~J2317$+$1439 could greatly reduce the VLBI proper motion
uncertainty and provide a much more stringest comparison against timing.

Overall, Figure\,\ref{fig:propermotioncomp} shows that although for MSPs
in some cases the timing measurements of proper motion parameters are
comparable to the accuracy of VLBI measurements, the actual values can
differ significantly between PTAs. As discussed in \citet{deller16a},
there could be multiple explanations, such as contamination by annual
DM variations, systematic instrumental noise, the use of different
versions of Solar System ephemeris (SSE), or including different levels of
noise modelling in the timing solutions.  \citet{abb+18} find the
effect of using a different SSE on the proper motion to be on the order
of 10 \uas\ yr$^{-1}$ or less, which is insignificant compared to the current uncertainty levels.

Besides further comparison between timing models and e.g. the effect
of including different types of noise modelling and DM correction, an
extended set of independent and improved VLBI measurements of
MSP proper motions will be extremely useful to find the underlying
causes of any discrepancies between measured values. 

Table\,\ref{tab:pxcomparison} shows that in order to derive an
independent distance measurement, VLBI observations can play an essential role 
to improve pulsar timing. In general, measuring a parallax
signature in pulsar timing data is dependent on having a long baseline
of observations, and as seen in Table\,\ref{tab:pxcomparison} the
uncertainties are between a factor of 2 and an order of magnitude
larger compared to what is achieved with VLBI within 2 years.

Finally, a comparison was made between the proper motions of the non-MSPs 
in our sample and the proper motions derived from timing observations. However, none
of these pulsars had sufficiently significant detections from timing \citep{hobbs04a,zhw+05,lwy+16}
to make a useful comparison with our VLBI-derived values. When
compared to previously derived interferometric measurements (see Table \ref{tab:pmslowcomparison}) we
find that al most all are consistent within the given (generally low precision) error bounds; the exceptions are 
PSR B0329+54 and PSR B1133+16, for which the previous VLBI proper motion measurements of \citet{brisken02a} are 
discrepant at the $\gtrsim$2$\sigma$ level in the declination coordinate.

\begin{deluxetable*}{lcccc}
\tabletypesize{\small}
%%\tablewidth{0pt}
\tablecaption{Comparison between proper motion measurements}
\tablehead{
\colhead{Pulsar} & \colhead{VLBI}&\colhead{EPTA}&\colhead{NANOGrav}&\colhead{PPTA}}
\startdata
J1022$-$1001 $\mu_{\alpha}$	& -14.92$^{+0.05}_{-0.03}$ & -18.2(64)  & --        & -17.09(3) \\
J1022$-$1001 $\mu_{\delta}$	&   5.61$^{+0.03}_{-0.04}$ &  -3(16)    & --        & -- \\
J2010$-$1323 $\mu_{\alpha}$	&   2.36$^{+0.33}_{-0.21}$ &   2.53(9)  &   2.59(5) & --	 \\
J2010$-$1323 $\mu_{\delta}$	&  -5.61$^{+0.26}_{-0.30}$ &  -5.7(4)   &  -6.0(2)  & --	 \\
J2145$-$0750 $\mu_{\alpha}$       &  -9.49$^{+0.05}_{-0.04}$ &  -9.58(4)  & -10.0(2)  & -9.59(8) \\
J2145$-$0750 $\mu_{\delta}$      &  -9.11$^{+0.09}_{-0.08}$ &  -8.86(10) &  -8.0(5)  & -8.9(3) \\
J2317$+$1439 $\mu_{\alpha}$ &  -1.43$^{+0.08}_{-0.08}$ &  -1.19(7)  &  -1.36(2) & --	 \\
J2317$+$1439 $\mu_{\delta}$ &   3.74$^{+0.18}_{-0.18}$ &   3.33(13) &   3.49(4) & --	 
\enddata
\tablenotetext{}{All values are given in mas/yr. For PSR\,J2317+1439, the VLBI value presents the results of the least-squares fit.  Timing references as mentioned in the caption of Figure\,\ref{fig:propermotioncomp}. Uncertainties on the timing parameters refer to the last digit(s) quoted.}
\label{tab:pmcomparison}
\end{deluxetable*}

\begin{deluxetable}{lcccc}
\tabletypesize{\small}
%%\tablewidth{0pt}
\tablecaption{Comparison between parallax measurements}
\tablehead{
\colhead{Pulsar} & \colhead{VLBI}&\colhead{EPTA}&\colhead{NANOGrav}&\colhead{PPTA}}
\startdata
J1022$-$1001 & 1.39$^{+0.04}_{-0.03}$ & 0.72(20) & -- & 1.1(3) \\
J2010$-$1323 & 0.48$^{+0.17}_{-0.12}$ & -- & 0.3(1) & -- \\
J2145$-$0750 & 1.60$^{+0.06}_{-0.01}$ & 1.53(11) & 1.6(4) & 1.84(17) \\
J2317$+$1439 & 0.65$^{+0.07}_{-0.07}$ & 0.7(3) & 0.50(8) & --
\enddata
\tablenotetext{}{All values are given in mas. For PSR\,J2317+1439, the VLBI value presents the results of the least-squares fit. References for the timing values are identical to those given in the caption of Figure\,\ref{fig:propermotioncomp}. Uncertainties on the timing parameters refer to the last digit(s) quoted.}
\label{tab:pxcomparison}
\end{deluxetable}

\begin{deluxetable*}{llccc}
\tabletypesize{\small}
%%\tablewidth{0pt}
\tablecaption{Previous interferometric proper motion measurements of normal pulsars from the literature}
\tablehead{
\colhead{Pulsar name (B1950)} & \colhead{Pulsar name (J2000)}&\colhead{PMRA (mas yr$^{-1}$)}&\colhead{PMDEC (mas yr$^{-1}$)}&\colhead{Reference}}
\startdata 
B0148-06 & J0151-0635 & 15(47) & -30(34) & \cite{harrison93a}
      \\  
B0149-16 & J0152-1637 & 3.1(12) & -27(2) & \cite{bfg+03}\\
B0320+39 & J0323+3944 & 16(6) & -30(5) & \cite{harrison93a}\\ 
B0329+54 & J0332+5434 & 17.0(3) & -9.5(4) & \cite{brisken02a}\\
B0559-05 & J0601-0527 & 18(8) & -16(7) & \cite{harrison93a}\\ 
B0611+22 & J0614+2229 & -4(5) & -3(7) & \cite{harrison93a}\\ 
B0626+24 & J0629+2415 & -7(12) & 2(12) & \cite{harrison93a}\\ 
B0823+26 & J0826+2637 & 62.6(24) & 95.3(24) & \cite{gwinn86a}\\
B1133+16 & J1136+1551 & -74.0(4) & 368.1(3) & \cite{brisken02a}\\ 
B1322+83 & J1321+8323 & -53(20) & 13(7) & \cite{harrison93a}\\ 
B1540-06 & J1543-0620 & -17(2) & -4(3) & \cite{bfg+03}\\ 
B1604-00 & J1607-0032 & -1(14) & -7(9) & \cite{las82}\\
B1642-03 & J1645-0317 & -3.7(15) & 30.0(16) & \cite{bfg+03}\\ 
B1732-07 & J1735-0724 & -2.4(17) & 28(3) & \cite{bfg+03}\\ 
B1839+56 & J1840+5640 & -30(4) & -21(2) & \cite{harrison93a}\\ 
B1917+00 & J1919+0021 & -2(30) & -1(10) & \cite{harrison93a}\\ 
B2044+15 & J2046+1540 & -13(6) & 3(4) & \cite{harrison93a}\\
B2110+27 & J2113+2754 & -23(2) & -54(3) & \cite{harrison93a}\\
B2148+63 & J2149+6329 & 14(3) & 10(4) & \cite{harrison93a}\\
B2224+65 & J2225+6535 & 144(3) & 112(3) & \cite{harrison93a}\\
B2303+30 & J2305+3100 & 2(2) & -20(2) & \cite{bfg+03}\\
B2351+61 & J2354+6155 & 22(3) & 6(2) & \cite{harrison93a}\\
B0820+02 & J0823+0159 & 5(11) & -1(8) & \cite{harrison93a}\\
B0823+26& J0826+2637 & 61(3) & -90(2) & \cite{las82}\\
B2043-04 & J2046-0421 & 9(16) & -7(8) & \cite{harrison93a}
\enddata
\tablenotetext{}{Uncertainties on the proper motion parameters given in the
  parentheses refer to the last digit(s) quoted.}
\label{tab:pmslowcomparison}
\end{deluxetable*}

\subsection{Absolute positional accuracy}
Three factors contribute to the accuracy of pulsar absolute positions
that we can obtain from phase-referenced VLBI observations: 
\begin{enumerate}
\item the accuracy of the off-beam calibrator absolute position; 
\item a contribution from the frequency-dependent core-shift of the off-beam calibrator; and
\item the accuracy of the determination of position offsets with respect to the off-beam 
calibrators.
\end{enumerate}
We now consider each of these effects in turn for the \psrpi\ sample.

Among 60 sources used as off-beam calibrators, the absolute position accuracy 
(as recorded in the Radio Fundamental Catalog; \url{http://astrogeo.org/rfc/}) ranged
from 0.10 to 0.37~mas, with a median of 0.17~mas.

We have not measured the core-shift of any of the off-beam calibrators in our observations. 
Therefore, we can present only a rough
estimate of its unaccounted contribution. \citet{Sokol_cs2011} presented 
results of multifrequency observations of core-shift. The core-shift of 
17~AGNs at 1.6~GHz varied from 0.4 to 2.2~mas with the median 1.1~mas. 
We can take this estimate and assume it is typical for the sources used by the \psrpi\ sample.

According to Table~\ref{tab:allresults}, uncertainties in 
the position offsets from the pulsars to their position reference source range from 
0.04 to 1.1~mas, with a median of 0.09~mas. The offset from the position reference source to the
out-of-beam calibrator is typically be an order of magnitude greater, given the typical angular separations
(14\arcmin\ median separation from position reference to target, vs 1.9\degrees\ for off-beam calibrator to 
target).  

Assuming these sources of errors to be independent, 
the overall uncertainties in the \psrpi\ pulsar absolute positions 
range from 0.4 to $\sim$10~mas, with a median of 1.4~mas (Table~\ref{tab:derivedresults}). The 
unaccounted core-shift in the off-beam calibrator and the position 
offset from the off-beam calibrator to the position reference source
contribute roughly equally in most cases, and generally dominate over the absolute position
uncertainty of the off-beam calibrator.

The absolute pulsar position does not play a significant role in the context 
of our study, but this does not mean it is not valuable at all.  Comparison
of the VLBI pulsar positions with positions determined with pulsar timing
provides important information. First, a determination of the net rotation of
pulsar positions determined with timing against the positions
determined with VLBI can be used to improve three parameters that describe
the orientation of the Earth orbit in the inertial space, i.e., the position
of the ecliptic pole and the point of vernal equinox. Second, analysis
of the residual position differences between timing and VLBI after the removal of net rotation
gives us a measure of possible systematic errors in VLBI and/or timing.
Timing and VLBI position estimates are to a great extent independent,
and therefore, their intercomparson provides us a unique opportunity
to make an assessment of their accuracy.

Dedicated VLBA observations are able to bring the absolute accuracy of 
calibrator positions down to at least 0.1~mas. As \citet{r:icrf2} demonstrated via 
decimation tests, a level of 0.05~mas for random position errors can even 
be reached if sources are observed long enough. Reaching that level of
accuracy requires additional observations in the mode of absolute
astrometry, similar to the regular geodesy ``RDV" program conducted on the
VLBA \citep{r:rdv}.

The most important step for improvement of accuracy of pulsar VLBI
absolute astrometry is determination of the core-shift. This requires
multi-frequency dedicated observations and analysis, which we have 
not utilised for \psrpi. \citet{Sokol_cs2011} describes the technique of
such observations. As it was shown in this and following works, 
frequency dependence of observed core-shifts obeys the power law 
$\nu^{-1/r}$, with $r$ close to 1. It was shown theoretically by
\citet{lobanov_cs1998} that in a case if i) the plasma is in the state of 
equipartition with the magnetic field, ii) the dominating absorption 
mechanism is synchrotron self-absorption, and iii) the jet has
a conical shape, then $r=1$. 

Determination of the core-shift requires significant observational
resources and it is not practical to do it for every pulsar.
The \psrpi\ program is presently being continued with a second sample 
consisting exclusively of millisecond pulsars: \mspsrpi, with observations of 18 millisecond
pulsars taking place during the period 2015 -- 2018 and first 
results reported in \citet{vigeland18a}.  To facilitate a comparison between high-precision
pulsar timing and VLBI absolute positions, we have decided to focus on a list of
17 millisecond pulsars drawn from \psrpi\ and \mspsrpi\ that have timing positions accurate
to 1~mas or better.  In March 2018 we commenced a VLBA program targeting 
their off-beam calibrators, aiming to improve the absolute positions of these
sources to the 0.1~mas level, as well as measuring the core-shift at 1.6 GHz. As of August 2018, 
28\% of planned observations have been conducted.

For pulsars in binary systems, an optically-visible companion offers the possibility
of comparing VLBI positions against \Gaia\ positions for truly point-like sources,
avoiding the problems of systematic VLBI--\Gaia\ offsets that are
located preferentially along the jet direction of AGN \citep{r:vlbi_gaia2}.  Unfortunately, pulsars with
bright optical companions are rare; of the four binary pulsars in \psrpi, none have a companion 
above the \Gaia\ magnitude limit. Given that over 9,000 common sources can be identified in 
VLBI and \Gaia\ catalogs \citep{vlbi_gaia4}, the sheer weight of numbers means that an ensemble
comparison of radio/optical AGN will provide a more accurate alignment of VLBI and \Gaia\ positions
than will be possible with pulsar companions.

\subsection{Achievable accuracy of differential VLBI pulsar astrometry}
\label{accuracydiscussion}
\begin{figure}
\includegraphics[width=0.48\textwidth]{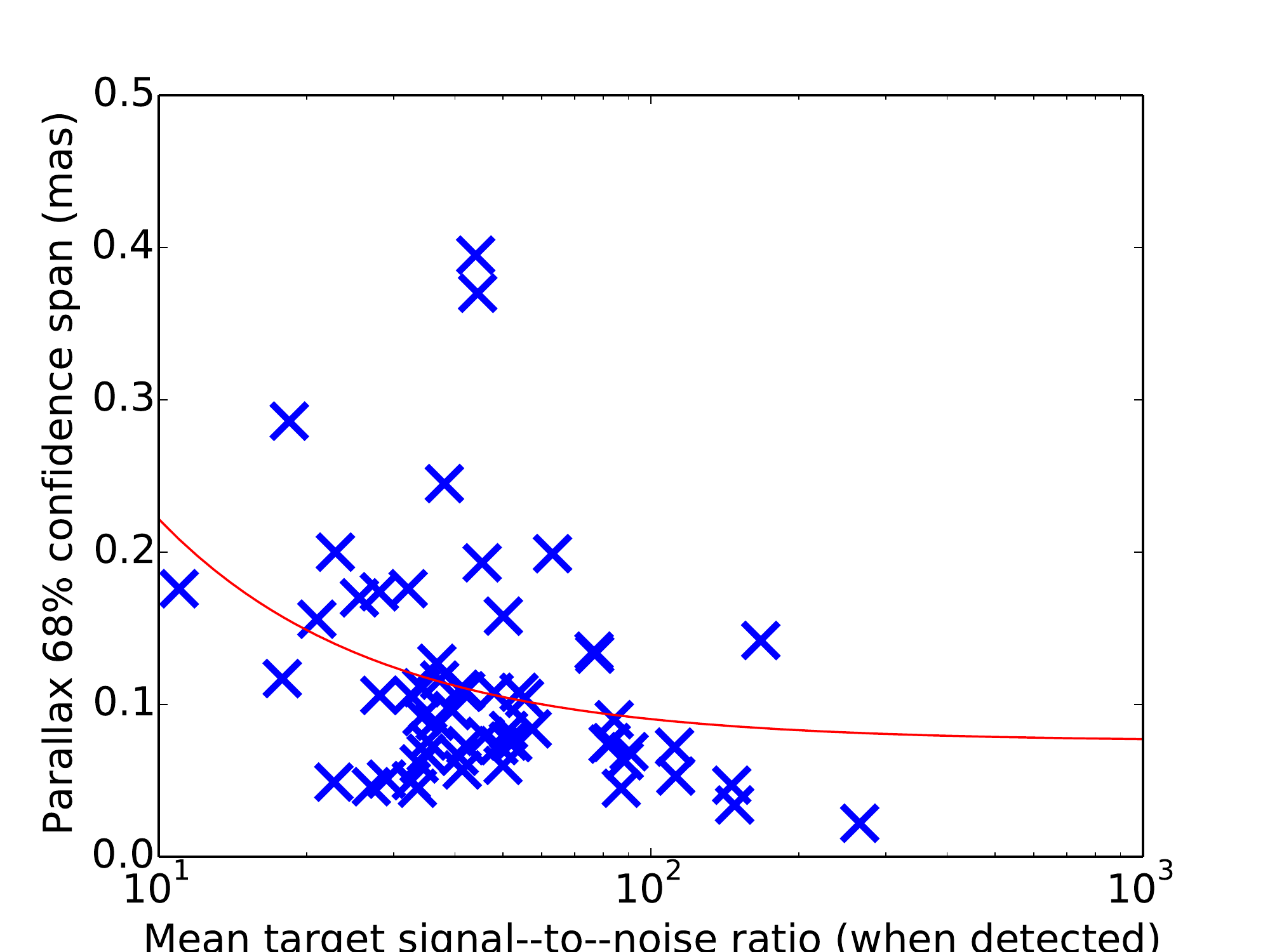} 
% Figure generated using code in the dropbox file, under accuracyplots/
\caption{\label{fig:errvstargetsnr} Parallax uncertainty for \psrpi\ pulsars plotted against the average signal--to--noise ratio achieved on the target (disregarding non--detections, if there were any).  The red line shows a best-fit linear regression.}
\end{figure}

\begin{figure}
\includegraphics[width=0.48\textwidth]{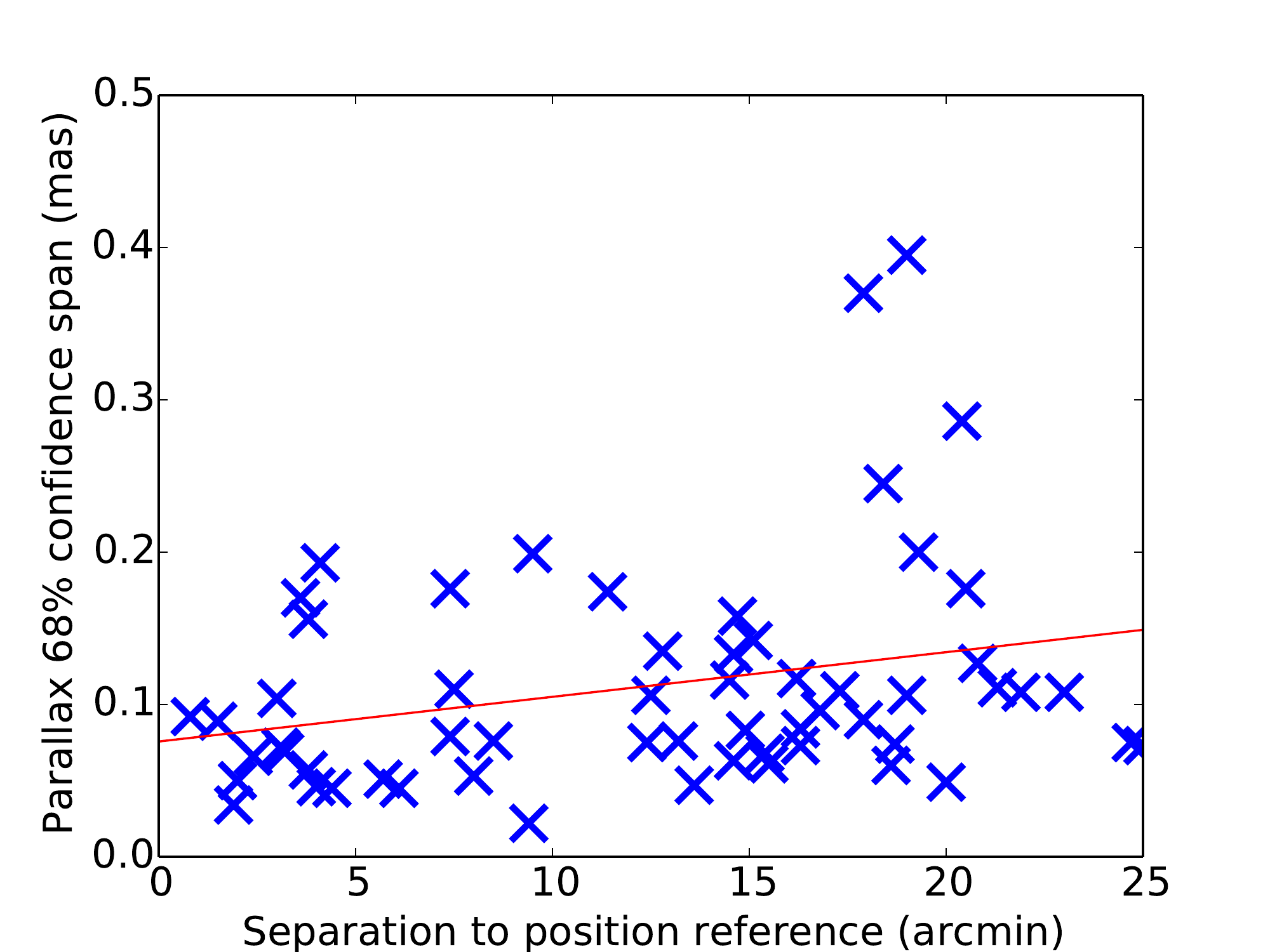} 
% Figure generated using code in the dropbox file, under accuracyplots/
\caption{\label{fig:errvsseparation} Parallax uncertainty for \psrpi\ pulsars plotted against angular separation to the in--beam position reference source.  The red line shows a best-fit linear regression.}
\end{figure}

\begin{figure}
\includegraphics[width=0.48\textwidth]{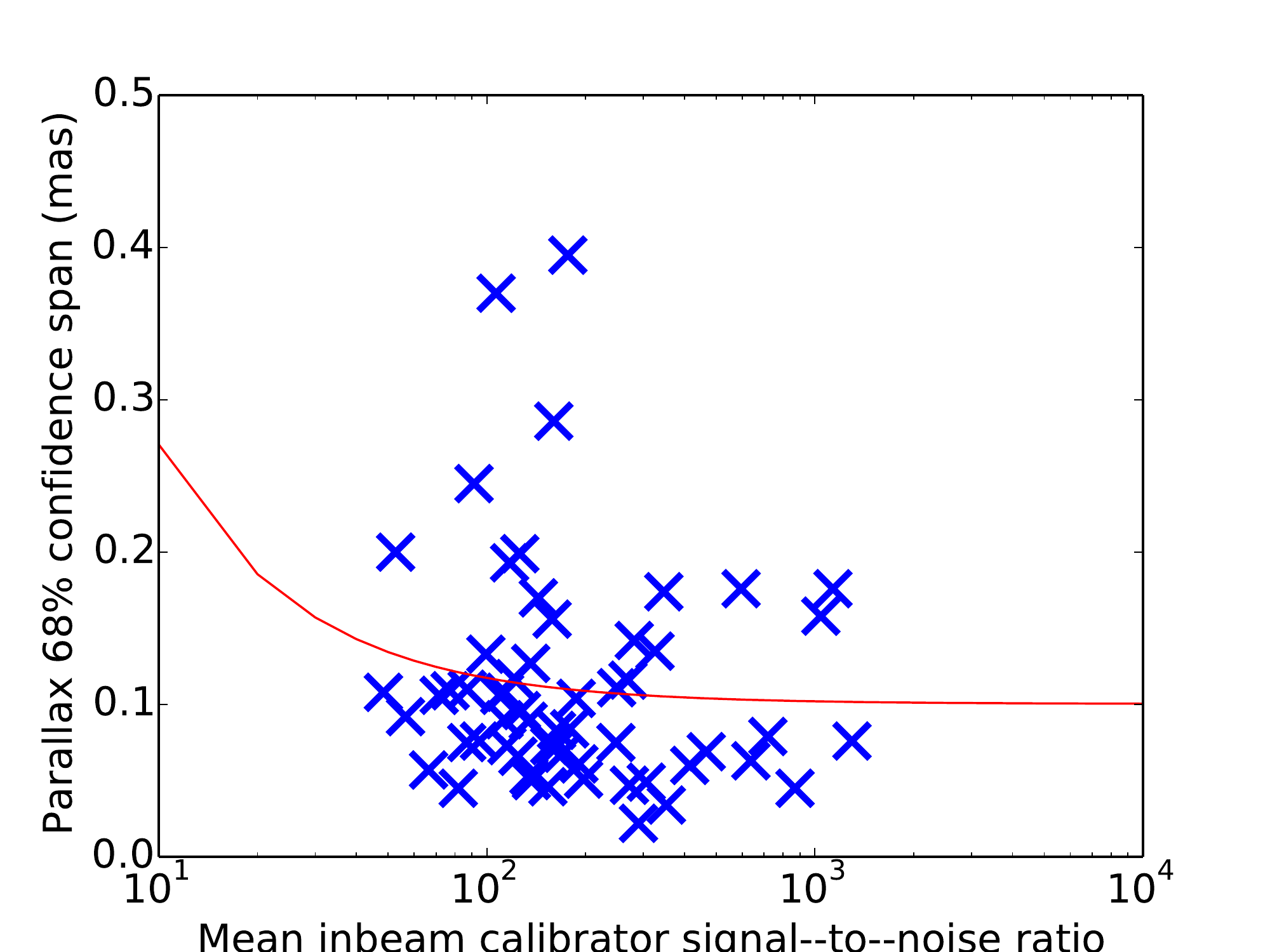} 
% Figure generated using code in the dropbox file, under accuracyplots/
\caption{\label{fig:errvscalsnr} Parallax uncertainty for \psrpi\ pulsars plotted against the average signal--to--noise ratio achieved on the inbeam calibrator source(s).  The red line shows a best-fit linear regression.}
\end{figure}

As discussed in Section~\ref{sec:errorbudget}, the astrometric accuracy achieved for a given target is expected to be influenced by the target brightness, the calibrator brightness, the target--calibrator separation, and the average observing conditions (principally the magnitude of the ionospheric gradients and the observing elevation).  In Figures~\ref{fig:errvstargetsnr} -- \ref{fig:errvscalsnr}, we explore this hypothesis by examining the parallax uncertainty achieved for \psrpi\ pulsars as a function of each potential influence separately, before attempting to find an parameterized function that predicts the achieved accuracy given the known observing parameters.  In all plots, we have excised PSR~J2317+1439, where the parallax uncertainty is artificially inflated by the non-detections of the pulsar discussed in Section~\ref{sec:j2317}, and the red line shows a best-fit linear regression.  The parallax uncertainty shown is the width of the 68\% confidence interval in milliarcseconds (i.e., the addition of the uncertainties in the positive and negative directions).

Figure~\ref{fig:errvstargetsnr} shows the parallax uncertainty plotted against the average signal--to--noise ratio (S/N) achieved on the target.  Because the brightest pulsars achieve very high S/N, a log scale is used for the x-axis. As expected, the faintest targets tend to have higher uncertainties, but systematic errors dominate in most cases.

Figure~\ref{fig:errvsseparation} shows the parallax uncertainty plotted against separation to the position reference source.  A weak trend towards larger uncertainties at larger separations is seen, but with a large scatter ($r^2 = 0.08$, where $r$ is the correlation coefficient).

Figure~\ref{fig:errvscalsnr} shows the parallax uncertainty plotted against the average S/N achieved on the inbeam calibrator, or the quadrature addition of the S/N if multiple calibrators were used.  Because the brightest calibrators achieve very high S/N, a log scale is used for the x-axis. The pulsars with the faintest calibrators tend to have higher uncertainties, but calibrator brightness is no more dominant than calibrator--target separation.

Finally, we use a multiparameter estimation including calibrator S/N, elevation--weighted calibrator--target separation, and target S/N to attempt to predict the parallax accuracy achieved in a \psrpi\ observing setup.  The predicted parallax error $\pi_\mathrm{p}$ in mas is given by:

\begin{equation}
\label{eq:predictedpx}
\pi_\mathrm{p} = \sqrt{\left( \frac{A}{\mathrm{S/N}_{t} \times \sqrt{N_{\mathrm{obs}}}} \right)^2 + \
                     \left( \frac{B}{\mathrm{S/N}_c} \right)^2 + \left(C \times \Delta\theta\right)^2}
\end{equation}
where $A$, $B$, and $C$ are constants which we fit from our dataset and find values $A=9.0$, $B=4.5$, $C=0.0028$, $\mathrm{S/N}_t$ is the average signal--to--noise ratio on the target, $N_{\mathrm{obs}}$ is the number of observations, $\mathrm{S/N}_c$ is the average signal--to--noise ratio on the in--beam calibrator, and $\Delta\theta$ is the angular separation of the pulsar and in-beam calibrator in arcminutes divided by the sine of the average observing elevation.  

\begin{figure}
\includegraphics[width=0.48\textwidth]{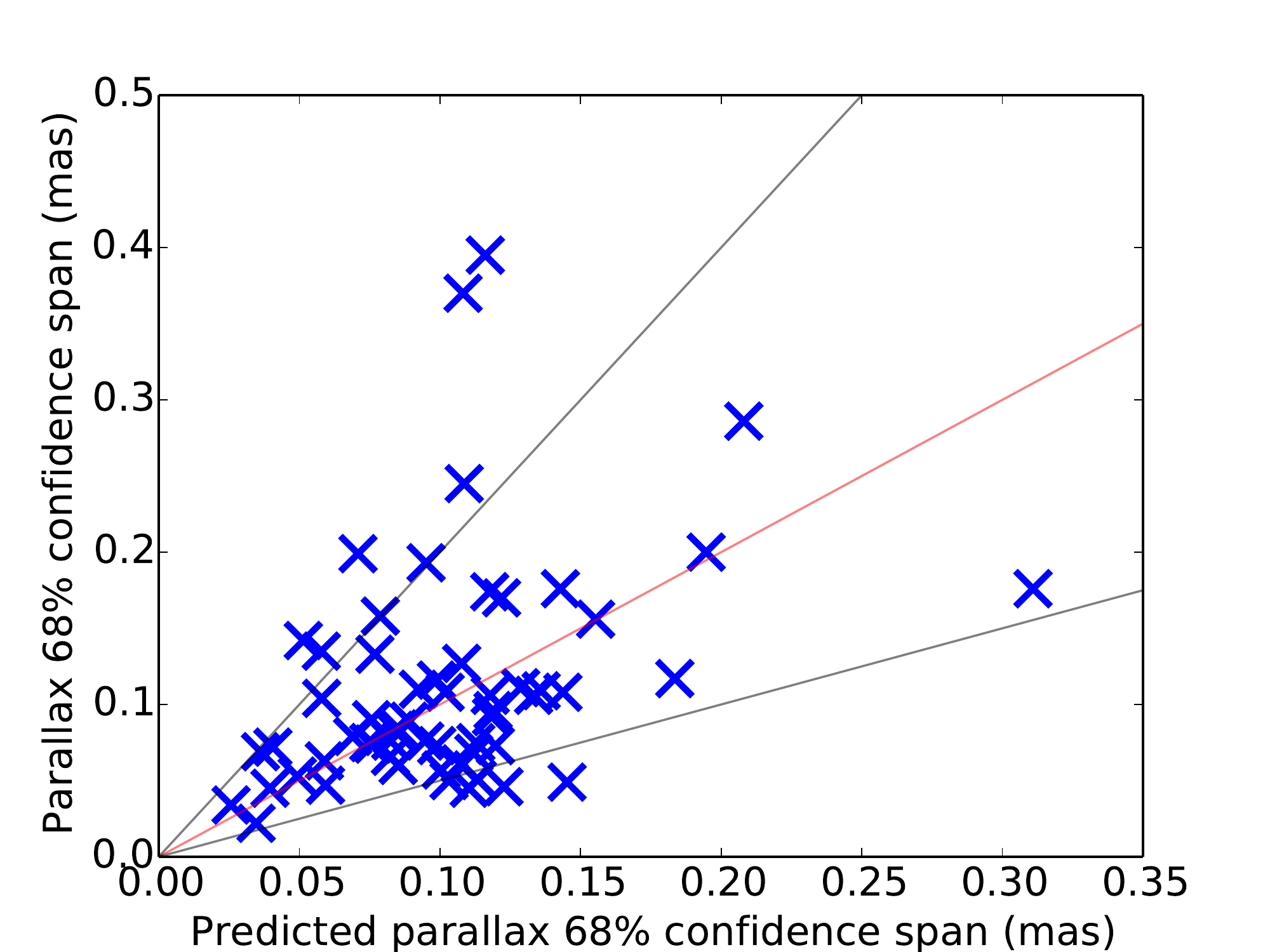} 
% Figure generated using code in the dropbox file, under accuracyplots/
\caption{\label{fig:predictedvsactualerr} Parallax uncertainty for \psrpi\ pulsars plotted against the predicted parallax uncertainty based on calibrator S/N, target S/N, and calibrator--target separation given in Equation~\ref{eq:predictedpx}.  The red line shows a 1:1 relationship, while the grey lines show $\pi_\mathrm{obs} = 0.5\times\pi_\mathrm{p}$ and $\pi_\mathrm{obs} = 2\times\pi_\mathrm{p}$.}
\end{figure}

Figure~\ref{fig:predictedvsactualerr} shows the actual parallax uncertainty $\pi_\mathrm{obs}$ vs the predicted value given by Equation~\ref{eq:predictedpx}, with the red line highlighting the expected 1:1 relationship.  Over 80\% of the pulsars fall within the range $0.5\times\pi_\mathrm{p} < \pi_\mathrm{obs} < 2\times\pi_\mathrm{p}$, indicated by the grey lines on the plot.

The values of the constants fitted in Equation~\ref{eq:predictedpx} can be used with caution to estimate the probable astrometric accuracy of a future 1600~MHz VLBA astrometric campaign of comparable duration (8 epochs).  For instance, in the limit of a very bright target and calibrator, then for an observation with typical observing elevation of 45\degrees, the expected parallax uncertainty is 4 $\mu$as per arcminute of separation to the in-beam calibrator.  Alternatively, given a target accuracy, these results can be used to estimate the characteristics of the in-beam calibrator that will be required.  For a bright target, if the desired parallax accuracy is 20 $\mu$as, the in-beam calibrator should be separated by no more than 5 arcminutes from the target, and should be at least $\sim$20 mJy (in order to achive the necessary signal--to--noise ratio of 225).  Obviously, extending the number of observations in the campaign could be undertaken to lower these expected limits.

\section{Conclusions}
\label{sec:conclusions}
We have presented the largest sample of VLBI astrometric results for radio pulsars assembled to date, obtaining a significant ($>$2$\sigma$) parallax for 53 pulsars using the VLBA.  Under moderately unfavourable observing conditions (relatively close to solar maxima, where ionospheric disturbances are more prevalent), we obtain a median parallax accuracy of $\sim$40 $\mu$as, meaning that precise distances can be obtained for pulsars out to $\sim$2.5 kpc, and reasonable constraints a factor of 2--3 further.  Observations with the VLBA at higher sensitivity (the standard continuum recording rate is now 4 times higher, doubling the sensitivity of comparable observations) and in more favourable ionospheric conditions should be capable of measuring a parallax-based distance for almost any sufficiently bright pulsar in the northern sky.

Comparisons of \psrpi\ distances to those predicted by the NE2001 and YMW16 Galactic electron density distribution models show that distance predictions based on dispersion measure are less accurate than claimed, although the biased nature of the \psrpi\ sample makes it difficult to quantify the level at which the DM--based distance uncertainties are typically underestimated.  It is clear, however, that results for nearby pulsars and pulsars at high Galactic latitudes should be treated with particular caution.

Extending the comparison of pulsar timing astrometry to VLBI with two additional new pulsars, we reinforce that timing proper astrometry can yield underestimated errors, particularly for pulsars at low ecliptic latitude.

Finally, we use the ensemble of \psrpi\ results to estimate the typical astrometric accuracy that could be obtained at 1600~MHz with the VLBA, where a suitable in-beam calibrator can almost always be found in regions where Galactic scattering is not too intense.  For a typical astrometric program like \psrpi\ with 8 observing epochs spread over 18 months and clustered near the parallax extrema, the parallax accuracy attainable in the limit of a sufficiently bright target and a brighth and stable calibrator is around 4 \uas\ per arcminute of separation.  Since the typical calibrator--target separation is of order 10 arcminutes, this implies that high quality parallax distances can be obtained at 1600~MHz out to several kpc.  With a smaller calibrator--target separation (either through good fortune, or higher sensitivity enabling the use of weaker calibrators), precise distances can be obtained up to $\sim$10 kpc.

\acknowledgements
The authors thank Allison Matthews for providing scripts used for plotting proper motion comparisons.  The Long Baseline Observatory is a facility of the National Science Foundation operated under cooperative agreement by Associated Universities, Inc.
This work made use of the Swinburne University of Technology software correlator, developed as part of the Australian Major National Research Facilities Programme and operated under licence \citep{deller11a}.  This research has made use of NASA's Astrophysics Data System
Bibliographic Services and the SIMBAD database, operated at
\hbox{CDS}, Strasbourg, France.
A.T.D. received support from an Australian Research Council Future Fellowship (FT150100415).
S.C., J.M.C., and T.J.W.L. acknowledge support from the NANOGrav Physics Frontiers Center (NSF award 1430284).
Pulsar research at the Jodrell Bank Centre for Astrophysics and the observations using the Lovell Telescope are supported by a consolidated grant from the STFC in the UK. 
Y.Y.K.\ was supported by the Russian Science Foundation grant 16-12-10481.

\bibliographystyle{apj}
\bibliography{psrpi,vpsr}

\end{document}